\documentclass[usenames,dvipsnames,prodmode,acmtocs]{acmsmall}

\usepackage[hyperref]{xcolor}
\usepackage{ifpdf}
\usepackage[pdfborder={0 0 0}]{hyperref}
\usepackage[mode=multiuser]{fixme}
\fxsetup{author=Nicolas}
\FXRegisterAuthor{jl}{ajl}{Julia}
\FXRegisterAuthor{np}{anp}{Nicolas}
\FXRegisterAuthor{gm}{agm}{Gilles}
\FXRegisterAuthor{gt}{agt}{Ga\"el}

\usepackage{ifthen}
\newboolean{showcomments}
\setboolean{showcomments}{false}
\ifthenelse{\boolean{showcomments}}
{ 
  \newcommand{\note}[2]{\fxnote*[author=Status]{#1}{#2}}
}{
  \newcommand{\note}[2]{#2}
}

\fxusetheme{color}
\definecolor{fxtarget}{rgb}{1,0,0}

\usepackage{amsmath}

\usepackage[T1]{fontenc}
\usepackage{microtype}
\usepackage{textcomp}

\usepackage{graphicx}
\usepackage{multirow}
\usepackage[scriptsize]{subfigure}
\usepackage[utf8]{inputenc}
\usepackage{xspace}

\usepackage{amssymb}
\usepackage{verbatim}

\newcommand{\mita}[1]{\mbox{\it{{#1}}}}

\newcommand{\fname}[1]{\url{#1}} % FIXME: Is it a good def. for fname?
\newcommand{\hl}[1]{\emph{#1}}

\begin{document}

% Page heads
\markboth{N. PALIX et al.}{Faults in Linux 2.6}

% Title portion
\title{Faults in Linux 2.6}
\author{
NICOLAS PALIX \affil{Grenoble - Alps University/UJF, LIG-Erods}
GAËL THOMAS,
SUMAN SAHA,
CHRISTOPHE CALVÈS,
GILLES MULLER and
JULIA LAWALL \affil{Inria/UPMC/Sorbonne University/LIP6-Regal}
}

%%
%% Load LaTeX macros generated
%% from the database.
%% Run figures/get_numbers.sh to update them
%%
%% File generated with ./get_numbers.sh
%% Wed Nov 27 17:24:05 CET 2013
\newcommand{\pctoverall}{\note{Checked 2013-11-27}12.5\%}

\newcommand{\nbreports}{\note{Checked 2013-11-27}5,169}

\newcommand{\nbfaults}{\note{Checked 2013-11-27}3,146}

\newcommand{\nbfaultsinlast}{\note{Checked 2013-11-27}855}

\newcommand{\notefirst}{\note{Faults in 2.6.0: 751}}

\newcommand{\noteoverall}{\note{Overall min 615 in linux-2.6.24}}

\newcommand{\notefinal}{\note{Faults in 3.0: 845}}

\newcommand{\notedensity}{\note{Density in 2.6.0: 0.2071 vs in 3.0 0.0897}}

\newcommand{\notelocccorr}{\note{Checked 2013-11-27}0.998}

\newcommand{\nbinitialreports}{\note{Checked 2014-03-23}51,449}

\begin{abstract}
  In August 2011, Linux entered its third decade. Ten years before,
Chou \emph{et al.}\ published a study of faults found by applying a
static analyzer to Linux versions 1.0 through 2.4.1.  A major result of
their work was that the {\tt drivers} directory contained up to 7 times
more of certain kinds of faults than other directories.  This result
inspired numerous efforts on improving the reliability of driver code.
Today, Linux is used in a wider range of environments, provides a
wider range of services, and has adopted a new development and release
model.  What has been the impact of these changes on code quality?

To answer this question, we have transported Chou \emph{et al.}'s
experiments to all versions of Linux 2.6, released between 2003 and
2011. We find that Linux has more than doubled in size during this period,
but the number of faults per line of code has been decreasing. And the
fault rate of {\tt drivers} is now below that of other directories, such as
{\tt arch}.  These results can guide further development and
research efforts for the decade to come. To allow updating these results as
Linux evolves, we define our experimental protocol and make our checkers
available.

%%% Check size with 'make count'

%%% Local Variables:
%%% mode: LaTeX
%%% TeX-master: "faults-in-linux-2.6-tocs"
%%% coding: utf-8
%%% TeX-PDF-mode: t
%%% ispell-local-dictionary: "american"
%%% End:

\end{abstract}




\acmformat{Nicolas Palix, Gaël Thomas, Suman Saha, Christophe Calvès,
  Gilles Muller and Julia Lawall. 2013. Faults in Linux 2.6}

\acmVolume{32}
\acmNumber{2}
\acmArticle{4}
\acmYear{2014}
\acmMonth{6}
\doi{2619090}

% \begin{bottomstuff}
%   % This work was partially supported by the Agence Nationale de la
%   % Recherche (France) under the contract ANR-09-BLAN-0158-01 and the
%   % Danish Research Council for Technology and Production Sciences.
%   % \jlnote{Not sure that this should be in a submission.}

%   Author's addresses: N. Palix, LIG-Erods, Domaine Universitaire de
%   Saint-Martin-d’Hères, BP 53, 38041 Grenoble cedex 9, France; Gaël
%   Thomas, Gilles Muller and Julia Lawall, LIP6, BP 169, 4 place
%   Jussieu, 75252 Paris Cedex 05, France; Suman Saha, Maxwell Dowrkin
%   33 Oxford Street, Room 309, Cambridge, MA 02138, USA.
% \end{bottomstuff}

\maketitle

\section{Introduction}

The Linux operating system is widely used, on platforms ranging from
embedded systems, to personal computers, to servers and supercomputers.  As
an operating system (OS) with a traditional monolithic kernel, Linux is
responsible for the security and integrity of the interactions between
the software and the underlying hardware.  Therefore, its correctness is
essential.  Linux also has a large developer base, as it is open source, and
is rapidly evolving.  Thus, it is
critical to be able to continually assess and control the quality of its
code.

Over 10 years ago, in 2001, Chou {\em et al.}\ published a study of
the distribution and lifetime of certain kinds of faults\footnote{ {
    Chou {\em et al.}\ used the terminology ``errors.''  In the
    software dependability literature \cite{ieee_guide}, however, this
    term is reserved for incorrect states that occur during execution,
    rather than faults in the source code, as were investigated by
    Chou {\em et al.}~and are investigated here.}} in OS code,
focusing mostly on the x86 code in the Linux
kernel~\cite{Engler:empirical:SOSP01}, in versions up to 2.4.1.  The
ability to collect fault information automatically from such a large
code base was revolutionary at the time, and this work has been highly
influential.  Indeed, their study has been cited over \note{Checked in
  Oct. 2013}{580} times, according to Google Scholar, and has been
followed by the development of a whole series of strategies for
automatically finding faults in systems code
\cite{Aiken:paste07,Li:fse05,sparse,smatch,wheeler06:flawfinder}.  The
statistics reported by Chou {\em et al.}\ have been used for a variety
of purposes, including providing evidence that driver code is
unreliable \cite{Herder:DSN09,Swift:TOCS06}, and evidence that certain
OS subsystems are more reliable than
others~\cite{Depoutovitch:EuroSys10}.

% supercomputers: \footnote{\scriptsize http://www.nccs.gov/computing-resources/jaguar/}

% 367 citations according to google scholar
% 86 citations according to acm

Linux, however, has changed substantially since 2001, and thus it is
worth examining the continued relevance of Chou \emph{et al.}'s
results.  In 2001, Linux was a relatively young OS, having first been
released only 10 years earlier, and was primarily used by specialists.
Today, well-supported Linux distributions are available, targeting
servers, embedded systems, and the general public
\cite{Fedora,Ubuntu}.  Linux code is changing rapidly, and only 30\%
of the code in the Linux 2.6.33 kernel is more than five years old
\cite{lwn374622,lwn374574}. \npnote*{Updated. Reviewer 2 asked for a
  clarification about dates.}{As of 2014, Linux supports 29
  architectures, up from 13 in Linux 2.4.1 (2001) and 24 in Linux 3.0
  (2011), and the developer base has grown commensurately.}  The
development model has also changed substantially.  Until Linux 2.6.0,
which was released at the end of 2003, Linux releases were split into
stable versions, which were installed by users, and development
versions, which accommodated new features.  \npnote*{Updated.
  Reviewer 2}{Since Linux 2.6.0 this distinction has disappeared;
  releases in the 2.6 and 3 series occur every three months}, and new
features are made available whenever they are ready.  Finally, a
number of fault finding tools have been developed that target Linux
code.  Patches are regularly submitted for faults found using
checkpatch \cite{Checkpatch}, Coccinelle \cite{Padioleau:eurosys08},
Coverity \cite{coverity}, smatch \cite{smatch} and Sparse
\cite{sparse,Sparse-web}.

In this paper, we transport the experiments of Chou {\em et al.}\ to
the versions of Linux 2.6, in order to reevaluate their results in the
context of the recent state of Linux development. Because Chou {\em et
  al.}'s fault finding tool and checkers were not released, and their
results were released on a local web site that is no longer available,
it is impossible to exactly reproduce their results on recent versions
of the Linux kernel.\footnote{Chou {\em et al.}'s work did lead to the
  development of the commercial tool Coverity, but using it requires
  signing an agreement not to publish information about its results
  ({\tt
    http://scan.co\-ver\-i\-ty.\-com/\-po\-li\-cy.\-html\-\#li\-cense}).}
To provide a baseline that can be more easily updated as new versions
are released, we propose an experimental protocol based on the open
source tools Coccinelle \cite{Padioleau:eurosys08}, for automatically
finding faults in source code, and Herodotos \cite{palix:aosd10}, for
tracking these faults across multiple versions of a software project.
We validate this protocol by replicating Chou {\em et al.}'s
experiments as closely as possible on Linux 2.4.1 and then apply our
protocol to all versions of Linux 2.6 released between December 2003
and May 2011.  \npnote*{Reviewer 2 asks for explanation about
  3.0}{Finally, the next Linux version, Linux 3.0 (released July
  2011), is also covered as a potential starting point for future
  studies. Note however that the change in version numbering is
  aesthetic and to celebrate the 20$^{th}$ anniversary of
  Linux\footnote{See \url{https://lkml.org/lkml/2011/5/29/204} and
    \url{https://lkml.org/lkml/2011/7/21/455}}, and so the difference
  between 2.6.39 and 3.0 is in principle no greater or lesser than the
  difference between any other 2.6 version and its successor.} To
ensure the perenity and reproducibility of our work, our tools and
results are available in the appendices\npnote*{New data set will be
  available shortly.}{\footnote{One may have to open the
    \emph{detailed view} to see the appendices. The version 2 of the
    appendices corresponds to the data and tools of the preliminary
    version of this work \cite{Palix:asplos11}. Version 3 updates the
    data and tools used for this paper.}} of a research
report~\cite{palix:inria-00509256} hosted in an open archival
repository.%PALIX:2010:INRIA-00509256:1}.

%% ``3. Competition. Coverity's Intellectual
%% Property may include elements that would assist competitors in creating
%% or improving products competitive to Coverity's tools. You agree that by
%% accepting access to the Service you commit not to distribute or share
%% details of the service or its analysis with any entity without prior
%% authorization from Coverity.''

The contributions of our work are as follows:

\begin{itemize}
\item We provide a repeatable methodology for finding faults in Linux code,
  based on open source tools, and a publicly available archive containing
  our complete results.

\item We show that the faults kinds considered 10 years ago by Chou
  {\em et al.}\ are still relevant, because such faults are still
  being introduced and fixed, in both new and existing files.  These
  fault kinds vary in their impact, but we have seen many patches for
  these kinds of faults submitted to the Linux kernel mailing list
  \cite{lkml} and have not seen any receive the response that the
  fault was too trivial to fix.

\item We show that while the rate of introduction of such faults continues
  to rise, the rate of their elimination is rising slightly faster,
  resulting in a kernel that is becoming more reliable with respect to
  these kinds of faults.  This is in contrast with previous results
  for earlier versions of Linux, which found that the number of faults
  was rising with the code size.

\item We show that the rate of the considered fault kinds is falling in the
  {\tt drivers} directory, which suggests that the work of Chou {\em et
    al.} and others has succeeded in directing attention to driver code.
  The directories {\tt arch} (HAL) and {\tt fs} (file systems) now show a
  higher fault rate, and thus it may be worthwhile to direct
  research efforts to the problems of such code.

\item We show that the lifespan of faults in Linux 2.6 is comparable
  to that observed by Chou \textit{et al.} for earlier versions. Chou
  \textit{et al.} observed a lifespan of 2.5 or 1.8 years according to
  how faults not yet fixed at the end of the considered period are
  taken into account.  For Linux 2.6, we observe an average fault
  lifespan between 2.5 and 3.0 years. The Linux 2.6 median lifespan
  ranges between 1.5 and 2.2 years and thus has increased since the
  period considered by Chou \emph{et al.}, where the reported median
  was 1.25 years. However, the median age of faults has decreased by 6
  months over the Linux 2.6 versions.  Moreover, we find that fault
  kinds that are more likely to have a visible impact during execution
  have a much shorter average lifespan, of as little as one year.

%   The new development model, in
%   which all releases are intended for users, means that users benefit
%   immediately when a fault is fixed. \jlfatal*{Fixme}{Not clear what is the point
%     of the last sentence.}\np{Agree. Moreover, there was fixes in the
%     previous series and this new development model also implies that
%     users get bugs ``more easily''.}

\item Although fault-finding tools are now being used regularly in Linux
  development, they seem to have only had a small impact on the kinds of
  faults we consider.  Research is thus needed on how such tools can be
  better integrated into the development process.  Our
  experimental protocol exploits previously collected information about false
  positives, reducing one of the burdens of tool use, but we propose that
  approaches are
  also needed to automate the fixing of faults, and not just the fault finding
  process.
\end{itemize}

% All of our results, as well as the
% scripts used and the source code of the Coccinelle and Herodotos tools are
% available on the open access archive HAL \cite{HAL}.

The rest of this paper is organized as follows.  Section~\ref{meth}
briefly presents our experimental protocol based on Coccinelle and
Herodotos. Section~\ref{sec:backgr-evol-linux} gives some background
on the evolution of Linux. Section~\ref{sec:baseline} establishes a
baseline for our results, by comparing our results for Linux 2.4.1
with those of Chou {\em et al.}  Section~\ref{sec:linux-2.6} presents
a study of Linux~2.6, considering the kinds of code that contain
faults, the distribution of faults across Linux code, and the lifetime
of faults.  Section~\ref{sec:fault-factors} correlates the faults with
other factors related to the Linux development model such as the
activity of the developers, the structure of the code and its
evolution.  Section~\ref{sec:towards-new-code} evaluates the interest
and applicability of the study to new
code. %, and effect of the use of
%fault-finding tools.  Section~\ref{sec:tools} considers how our
% experimental protocol eases the extension of the results to new versions of
% Linux. 
Section~\ref{sec:limitations} presents some limitations of our
approach.  Finally, Section~\ref{sec:related} describes related work and
Section~\ref{sec:concl} presents our conclusions.
Throughout the paper, our main findings are shown in {\em italics}.

%%% Local Variables:
%%% mode: LaTeX
%%% TeX-master: "faults-in-linux-2.6-tocs.tex"
%%% coding: utf-8
%%% TeX-PDF-mode: t
%%% ispell-local-dictionary: "american"
%%% End:

\section{Experimental protocol}
\label{meth}

In the laboratory sciences there is a notion of experimental protocol,
giving all of the information required to reproduce an experiment.  We
believe that there is a need for such a protocol in the study of
software as well. For a study of faults in operating systems code,
such a protocol should include the definition of the fault finding
tools and checkers, as each of these elements substantially affects
the results. \npnote*{Reviewer 2 asks for a forward ref.}{In this
  section, we first present our checkers (Section~\ref{sec:checkers}),
  then describe how we calcuate the fault rate
  (Section~\ref{sec:relev-site-find}), and finally describe the tools
  (Section~\ref{sec:coccinelle}) that we have used in the fault
  finding and validation process.} All of our results, as well as the
scripts used and the source code of the tool used, Coccinelle and
Herodotos, are available on the open access archive
HAL~\cite{palix:inria-00509256} and on the publicly available project
website~\cite{10years_web}.

\subsection{Fault finding checkers}
\label{sec:checkers}

The exact definitions of the checkers used by Chou {\em et al.} are
not publicly available.  Based on the descriptions of Chou {\em et
  al.}, we have implemented our interpretations of their
\textbf{Block}, \textbf{Null}, \textbf{Var}, \textbf{Inull},
\textbf{Range}, \textbf{Lock}, \textbf{Intr}, \textbf{LockIntr},
\textbf{Float}, and \textbf{Size} checkers.  We omit the \textbf{Real}
checker, related to the misuse of {\tt realloc}, and the
\textbf{Param} checker, related to dereferences of user-level
pointers, as in both cases, we did not have enough information to
define checkers that found any faults. \npnote*{Reviewer 1 asks for a
  clarification of that point.}{For instance, to perform memory
  reallocation, there is no \texttt{krealloc} function in Linux 2.4.1,
  as there is in recent kernel versions; We tried looking for other
  memory reallocation function, \emph{e.g.}
  \texttt{skb\_realloc\_headroom}, but the checker did not find any
  fault.}  In the description of each checker, the initial citation in
italics is the description provided by Chou {\em et al.}

% our web page:
% \verb+http://coccinelle.lip6.fr/osdi10+.\footnote{This data will soon be
%   available from the HAL archive.}

\paragraph*{\textbf{Block}} \emph{``To avoid deadlock, do not call
  blocking functions with interrupts disabled or a spinlock held.''}
Implementing this checker requires knowing the set of functions that
may block, the set of functions that disable interrupts, and the set
of functions that take spinlocks.  These functions vary across Linux
versions.  Identifying them precisely requires a full interprocedural
analysis of the Linux kernel source code, including a precise alias
analysis, as these operations may be performed via function pointers.
To our knowledge, Chou {\em et al.}'s tool {\tt xgcc} did not provide
these features in 2001, and thus we assume that these functions were
identified based on their examination of the source code and possibly
heuristics for collecting functions with particular properties.  We
take the same approach, but add a simple interprocedural analysis,
based on the iterative computation of a transitive closure through the
call graph.  This iterative analysis implies that our {\bf Block}
checker automatically takes into account the new blocking functions
that are added in each version.

To identify blocking functions, we consider two kinds of functions as the
starting point of our interprocedural analysis.  First, we observe that basic
memory allocation functions, such as the kernel function {\tt
  kmalloc}, often take as argument the constant {\tt GFP\_KERNEL} when they
are allowed to block until a page becomes available.  Thus, we consider
that a function that contains a call with {\tt GFP\_KERNEL} as an argument
may block.
Second, we observe that blocking is directly caused by calling the function
{\tt schedule}.  Given this initial list of blocking functions, we then
iteratively augment the list with the names of functions that call
functions already in the list without first explicitly releasing locks or
turning on interrupts, until reaching a fixed point.

To identify functions that turn off interrupts and take locks, we rely
on our knowledge of a set of commonly used functions for these
purposes, listed in the appendix. \textbf{BlockIntr} checks for the
case where a blocking function is called while interrupts are
disabled.  However, we observe that blocking with interrupts turned
off is not necessarily a fault, and indeed core Linux scheduling
functions, such as {\tt interruptible\_sleep\_on}, call {\tt schedule}
with interrupts turned off. We have taken this issue into account when
checking for false positives. Analogously, we consider the case where
a blocking function is called while holding a spinlock, which is
always a fault.  We refer to this checker as {\bf BlockLock} to
highlight the different design.

\paragraph*{\textbf{Null}} \emph{``Check potentially \texttt{NULL}
  pointers returned from routines.''} To collect a list of the
functions that may return {\tt NULL}, we follow the same iterative
strategy as for the {\bf Block} checker, with the starting point of
the iteration being the set of functions that explicitly return {\tt
  NULL}. Once the transitive closure is computed, we check the call
sites of each collected function to determine whether the returned
value is compared to \texttt{NULL} before it is used.

\paragraph*{\textbf{Var}} \emph{``Do not allocate large stack
  variables ($>$ 1K) on the fixed-size kernel stack.''} Our checker
looks for local variables that are declared as large arrays,
\emph{e.g.}, 1,024 or more elements for a \texttt{char} array.  Many
array declarations express the size of the array using macros, or even
local variables, rather than explicit constants.  Because Coccinelle
does not expand macros, and indeed some macros may have multiple
definitions across different architectures, we consider only array
declarations where the size is expressed as an explicit constant.

\paragraph*{\textbf{Inull}} \emph{``Do not make inconsistent
  assumptions about whether a pointer is \texttt{NULL}.''}  We distinguish
two cases: \textbf{IsNull}, where a null test on a pointer is followed by a
dereference of the pointer, and \textbf{NullRef}, where a dereference of a
pointer is followed by a null test on the pointer.  The former is always an
error, while the latter may be an error or may simply indicate overly
cautious code, if the pointer can never be \texttt{NULL}.  Still, at least
one \textbf{NullRef} fault has been shown to allow an attacker to obtain
root access \cite{spencer-exploit}. This kind of code is
  \emph{unstable}~\cite{Wang:2013:TOS:2517349.2522728}, implying that
  compiler optimisations can introduce vulnerabilities.

\paragraph*{\textbf{Range}}
\emph{``Always check bounds of array indices and loop bounds derived
  from user data.''}  We recognize the functions {\tt memcpy\_fromfs},
{\tt copy\_from\_user} and {\tt get\_user} as giving access to user
data.  Possible faults are cases where a value obtained using one of
these functions is used as an array index, with no prior test on its
value, and where some value is checked to be less than a value
obtained using one of these functions, as would occur in validating a
loop index.

\paragraph*{\textbf{Lock and Intr}}
\emph{``Release acquired locks; do not double-acquire locks
  (\textbf{Lock}).''}  \emph{``Restore disabled interrupts
  (\textbf{Intr}).''}  In early versions of Linux, locks and
interrupts were managed separately: typically interrupts were disabled
and reenabled using {\tt cli} and {\tt sti}, respectively, while locks
were managed using operations on spinlocks or semaphores.  In Linux
2.1.30, however, functions such as {\tt spin\_lock\_irq} were
introduced to combine locking and interrupt management.  Thus, our {\bf
  Lock} checker is limited to operators that only affect locks
(spinlocks and, from Linux 2.6.16, mutexes), our {\bf Intr} checker is
limited to operators that only disable interrupts, and for the
combined operations, we introduce a third checker, {\bf LockIntr}.
The considered functions are listed in the appendix.

\paragraph*{\textbf{Free}} \emph{``Do not use freed memory.''} Like
the {\bf Null} checker, this checker first iteratively collects
functions that always apply either {\tt kfree} or some collected
function to some parameter, and then checks each call to either {\tt
  kfree} or a collected function for a use of the freed argument after
the call.

\paragraph*{\textbf{Float}} \emph{``Do not use floating point in the
  kernel.''}  Most uses of floating point in kernel code are in
computations that are performed by the compiler and then converted to
an integer, or are in code that is part of the kernel source tree, but
is not actually compiled into the kernel.  Our checker only reports a
floating point constant that is not a subterm of an arithmetic
operation involving another constant, and thus may end up in the
compiled kernel code.

%% \paragraph*{Real}
%% The {\bf Real} checker is described by Chou {\em et al.}\ as \emph{``Do
%%   not leak memory by updating pointers with potentially \texttt{NULL}
%%   realloc return values.''}  We were not able to identify a generic
%% {\tt realloc} function in the versions of Linux considered by Chou
%% {\em et al.}.  Because there is not sufficient information available
%% to be confident of computing comparable results, we have chosen to
%% simply omit this checker.  This checker is furthermore responsible for
%% less than 1\% of the faults identified by Chou {\em et al.}\ in Linux
%% 2.4.1, so we expect that this omission has little impact on our
%% results.

%% \paragraph*{Param}
%% The {\bf Param} checker is described by Chou {\em et al.}\ as \emph{``Do not
%%   dereference user pointers.''}  As in the {\tt Range} checker, we focus on
%% calls to {\tt memcpy\_fromfs} and {\tt copy\_from\_user} ({\tt get\_user}
%% obtains an integer from user level, so it is not likely to be relevant
%% here).  A potential fault is then considered to be a dereference of a field
%% of a copied structure.  We did not find any faults using this checker, and
%% thus, for conciseness, we have omitted it from our graphs.

\paragraph*{\textbf{Size}} \emph{``Allocate enough memory to hold the
  type for which you are allocating.''}  Because our checker works at
the source code level without first invoking the C preprocessor, it is
not aware of the sizes of the various data types on a particular
architecture.  We thus focus on the information apparent in the source
code, considering the following two cases.  In the first case, one of
the basic memory allocation functions, {\tt kmalloc} or {\tt kzalloc},
is given a size argument involving a {\tt sizeof} expression defined
in terms of a type that is different from the type of the variable
storing the result of the allocation.  To reduce the number of false
positives, the checker ignores cases where one of the types involved
represents only one byte, such as {\tt char}, as these are often used
for allocations of unstructured data.  We consider as a fault any case
where there is no clear relationship between the types, whether the
allocated region is too large or too small.  In the second case, there
is an assignment where the right hand side involves taking the size of
the left hand side expression itself, rather than the result of
dereferencing that expression.  In this case, the allocated region has
the size of a pointer, which is typically significantly smaller than
the size intended.

\vspace{\baselineskip}

These faults vary in how easy they are to find in the source code, how
easy they are to fix once found, and the likelihood of a runtime
impact.  Table \ref{crash} summarizes these properties for the various
fault kinds, based on our observations in studying the code.  Faults
involving code within a single function are often easy for both
maintainers and tools to detect, and thus we designate these as
``Easy.''  Finding ``Hard'' faults requires an interprocedural
analysis to identify functions that have specific properties.
Interprocedural analysis requires more effort or expertise from a
maintainer, or more complexity in a tool.  Fixing a fault may require
only an easy local change, as in {\bf Size}, where the fix may require
only changing the argument of {\tt sizeof} to the type of the
allocated value.  Cases that require creating new error handling code,
such as {\bf Null}, or choosing between several alternative fixes
({\em e.g.}, moving a dereference or dropping an unnecessary null
test), such as {\bf NullRef}, are more difficult.  Instances of fault
kinds that entail more difficult fixes may benefit less from the use
of tools, as the tool user may not have enough expertise to choose the
correct fix.  Finally, we indicate a low impact when a crash or hang
is only likely in an \emph{exceptional} condition \npnote*{Add an
  example of the impact definition.}{(\textit{e.g.}, for \textbf{Null}
  or \textbf{IsNull} faults in error-handling code)}, and high when it
is likely in \emph{normal} execution.

\begin{table}
 \tbl{Assessment of the difficulty of finding and fixing faults, and the
  potential of a fault to cause a crash or hang at runtime}
{
\begin{tabular}{|p{15mm}|p{1cm}|p{1cm}|p{1cm}|}\cline{2-4}
\multicolumn{1}{p{15mm}|}{} &                Find&            Fix&             Impact\\\hline
Block&           Hard&            Hard&            Low\\
Null&            Hard&            Hard&            Low\\
Var&             Easy&            Easy&            Low\\
IsNull&          Easy&            Easy&            Low\\
NullRef&         Easy&            Hard&            Low\\
Range&           Easy&            Easy&            Low\\\hline
\end{tabular}
\hspace{10mm}
\begin{tabular}{|p{15mm}|p{1cm}|p{1cm}|p{1cm}|}\cline{2-4}
\multicolumn{1}{p{15mm}|}{} &                Find&            Fix&             Impact\\\hline
Lock&            Easy&            Easy&            High\\
Intr&            Easy&            Easy&            High\\
LockIntr&        Easy&            Easy&            High\\
Free&            Hard&            Easy&            High\\
Float&           Easy&            Hard&            High\\
Size&            Easy&            Easy&            High\\\hline
\end{tabular}}
\label{crash}
\end{table}

\subsection{Assessing the fault rate}
\label{sec:relev-site-find}

The maximum number of faults that code can contain is the number of
occurrences of code relevant to the fault, \textit{i.e.} where a given
kind of fault may appear.  For example, the number of {\bf Block}
faults is limited by the number of calls to blocking functions.  We
follow Chou \emph{et al.}\ and refer to these occurrences of relevant
code as {\em notes}.  Then,
\[
\mita{fault rate} = \mita{faults} / \mita{notes}
\]

We find the notes associated with each of our checkers as follows.  For
{\bf Block}, {\bf Null}, and {\bf Free}, a note is a call to one of the
functions collected as part of the transitive closure in the fault-finding
process.  For {\bf Var}, a note is a local array declaration.  For {\bf
  Inull} ({\bf IsNull} and {\bf NullRef}), a note is a null test of a value
that is derefenced elsewhere in the same function.  For {\bf Range} and for
{\bf Lock}, {\bf Intr}, or {\bf LockIntr}, a note is a call to one of the
user-level access functions or locking functions, respectively.  For {\bf
  Size}, a note is a use of {\tt sizeof} as an argument to one of the basic
memory allocation functions {\tt kmalloc} or {\tt kzalloc} when the
argument is a type, or a use of {\tt sizeof} where the argument is an
expression.  In the former case, as for the checker, we discard some cases
that are commonly false positives such as when the argument to {\tt sizeof}
is a one-byte type such as {\tt char}.  Finally, we do not
calculate the number of notes for {\bf Float}, because we consider that
every occurrence of a float in a context where it may be referenced in the
compiled code is a fault, and thus the number of notes and faults is the
same.

\subsection{Tools}
\label{sec:coccinelle}

%\note{Software version updated Oct. 2013}
Our experimental protocol relies on two open-source tools: Coccinelle
(v1.0.0-rc15), to automatically find potential faults and notes in the
Linux kernels \cite{Padioleau:eurosys08}, and Herodotos (v0.6.1-rc1),
to correlate the fault reports between versions~\cite{palix:aosd10}.
We store the resulting data in a PostgreSQL database (v9.1.5), and
analyze it using SQL queries.

Coccinelle performs control-flow based pattern searches in C code.  It
provides a language, the Semantic Patch Language (SmPL), for specifying
searches and transformations and an engine for performing them.
Coccinelle's strategy for traversing control-flow graphs is based on the
temporal logic CTL \cite{Brunel:POPL09}, while that of the tool used by
Chou {\em et al.} is based on automata.  There are technical
differences between these strategies, but we do not expect that they are
relevant here.

A notable feature of Coccinelle is that it does not expand preprocessor
directives.  We have only found this feature to be a limitation in the {\bf
  Var} case, as noted in Section~\ref{sec:checkers}. On the other hand,
this feature has the benefit of making the fault-finding
process independent of configuration information, and thus we can find
faults across the entire Linux kernel source tree, rather than being
limited to a single architecture.

To be able to understand the evolution of faults in Linux code, it is
not sufficient to find potential faults in the code base; we must also
understand the history of individual fault occurrences. To do so, we
must be able to correlate potential fault occurrences across multiple
Linux versions, even in the presence of code changes in the files, and
manage the identification of these occurrences as real faults and
false positives. For these operations, we use Herodotos.  To correlate
fault occurrences, Herodotos first uses {\tt diff} to find the changes
in each pair of successive versions of a file for which Coccinelle has
produced fault reports. If a pair of reports in these files occurs in
the unchanged part of the code, at corresponding lines, the reports
are automatically considered to represent the same fault, with no user
intervention.  Otherwise, in a pair of reports, if the report in the
older version occurs in the changed part of the code, then its status
is considered to be unknown, and the user must indicate, via an
interface based on the emacs ``org'' mode~\cite{org-mode}, whether the
pair to which it belongs represents the same fault or unrelated
ones. We have validated this approach by manually inspecting a
randomly chosen set of automatic correlations. None of the hundreds of
correlations inspected was suspicous among the more than 46,000
automatic correlations performed. Once the correlation process is
complete, a similar interface is provided to allow the user to
classify each group of correlated reports as representing either a
fault or a false positive.  Further details about the process of using
Herodotos are provided elsewhere~\cite{palix:aosd10}.

Once the fault reports are correlated and assessed for false
positives, we import their histories into a database, along with the
associated notes.  The database also contains information about Linux
releases such as the release date and code size, and information about
Linux files (size, number of modifications between releases) and
functions (starting and ending line numbers).  The complete database,
including both the reports and the extra information, contains 2 GB of
data.  To analyze the collected data, we wrote over \note{Still true
  in oct 2013}{2,100 lines} of PL/pgSQL and SQL queries and about 300
lines of R that extract and correlate information.

\paragraph*{Extending the results to new versions}
A benefit of our experimental protocol is that it makes it
easy to extend the results to new versions of Linux.  When a new version
of Linux is released, it is only necessary to run the checkers on the new
code, and then repeat the correlation process.  As our collected data
contains information not only about the faults that we have identified, but
also about the false positives, Herodotos automatically annotates both
faults and false positives left over from previous versions as such,
leaving only the new reports to be considered by the user.

%% \np{Applying these checkers on Linux 2.6.33 takes approximately two
%%   hours on a HP ProLiant server with two 3~GHz quad-core Xeon
%%   processors and 16~GB of memory.}

%% Figure \ref{correl} shows the percentage of potential fault occurrences that
%% Herodotos either automatically correlated or automatically determined to be
%% fixed, for each of the fault types and Linux versions considered in this
%% paper.

%% \jlfatal*{Fixme}{Some FPs are due to structures that have headers, and thus the alloc is
%%   expressed in terms of a sum of smaller things.  Others are due to the
%%   confusion between int and unsigned int, but this is probably not worth
%%   mentioning...}

%%% Local Variables:
%%% mode: LaTeX
%%% TeX-master: "faults-in-linux-2.6-tocs"
%%% coding: utf-8
%%% TeX-PDF-mode: t
%%% ispell-local-dictionary: "american"
%%% End:

\section{Evolution of Linux}
\label{sec:backgr-evol-linux}

%\np{389 dev for 2.6.11, >  1 000 dev. / release since 2.6.24}

%%  git log --no-merges --oneline v2.6.36..v2.6.37 --numstat -- fs | grep -v '[^ ] ' | sort -n

\begin{figure}
  \centering
  \includegraphics[width=0.8\linewidth,keepaspectratio]{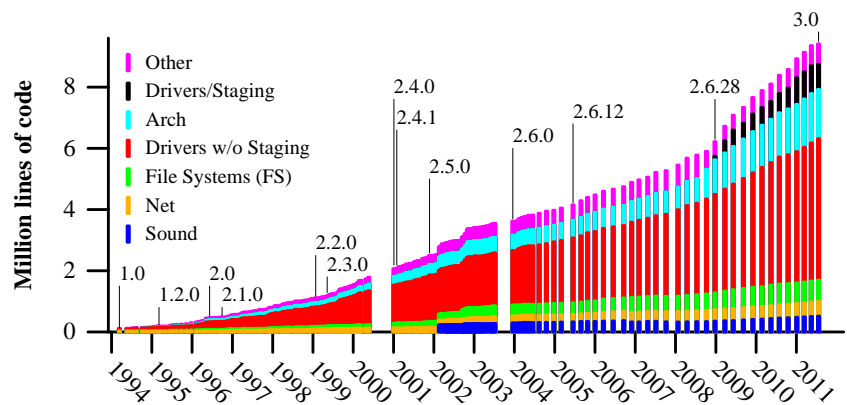}
  \caption{Linux directory sizes (in MLOC)}
  \label{fig:code-size}
\end{figure}

To give an overview of the complete history of Linux, we first
consider the evolution in code size of the Linux kernel between
version 1.0, released in March 1994, and version 3.0, released in July
2011, as shown in Figure~\ref{fig:code-size}.  Prior to Linux 2.6,
this figure shows the size of the development versions (with odd
version number), when available, as new code is added in these
versions, and this added code is then maintained in the subsequent
stable versions (with even version number). \jlnote*{Updated for
  clarity}{Starting with Linux 2.6, we show all major versions, all of
  which can contain new code.}  Code sizes are computed using David
A. Wheeler's SLOCCount (v2.26)~\cite{sloccount} and include only the
ANSI C code.  The code sizes are broken down by directory,
highlighting the largest directories: {\tt drivers/staging}, {\tt
  arch}, {\tt drivers}, {\tt fs} (file systems), {\tt net}, and {\tt
  sound}.  {\tt Drivers/staging} was added in Linux 2.6.28 as an
incubator for new drivers that are not yet mature enough to be used by
end users.  Code in {\tt drivers/staging} is not compiled as part of
the default Linux configuration, and is thus not included in standard
Linux distributions.  {\tt Sound} was added in Linux 2.5.5, and
contains sound drivers that were previously in the {\tt drivers}
directory.  The largest directory is {\tt drivers}, which, including
{\tt drivers/staging}, has made up 57\% of the source code since Linux
2.6.30.

\begin{figure}
  \centering
  \includegraphics[width=0.8\linewidth,keepaspectratio]{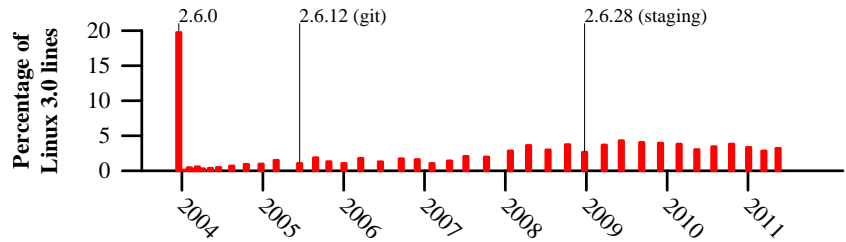}
  \caption{Distribution of the lines of code of Linux 3.0 by the Linux
    2.6 version in which they were introduced}
  \label{fig:line-age}
\end{figure}

Due to the large increase in size and the high level of churn,
relatively few lines of Linux 3.0 have been present in the Linux
kernel since Linux 2.6.0, as shown by
Figure~\ref{fig:line-age}. Indeed, as reported by the
\texttt{linetags} tool from the \texttt{gitdm}\footnote{The original
  version of \texttt{linetags} is available at
  \url{git://git.lwn.net/gitdm.git} and was developed by Jonathan
  Corbet, LWN. We made a modified version
  (\url{https://github.com/npalix/gitdm}) to support Linux versions
  prior to 2.6.12. This latter was applied on a Linux repository that
  aggregates linux-history and linux-2.6 thanks to a git graft point
  between commits 1da177e4c3f41524e886b7f1b8a0c1fc7321cac2 and
  e7e173af42dbf37b1d946f9ee00219cb3b2bea6a.}  toolsuite, only 20\% of
the lines of Linux 2.6.0 have been kept untouched througout the Linux
2.6 history. Except for Linux 2.6.0, every version of Linux 2.6 has
contributed as new code less than 5\% of the code found in Linux
3.0. Nevertheless, recent versions tend to introduce a larger part of
Linux 3.0 as compared to older versions.  Indeed, \hl{each version
  released after 2007 introduces more than 2\% of the Linux 3.0 code,
  {\em i.e.}, more than 55\% in total, whereas each earlier version,
  except 2.6.0, introduces less than 2\%.}  Linux 2.6.0 is an
exception as it represents in part the code that is inherited from the
Linux 2.5 era, and as we use it as the starting point for the origin
of lines. Most of the code that has remained unchanged since 2.6.0
comes from driver code for old hardware (\texttt{drivers}), old
hardware platforms (\texttt{arch}) and old network protocols that no
longer change (\texttt{net}), with respectively 65.6\%, 18.8\% and
12.2\% of the Linux 2.6.0 code that is still present in Linux 3.0.

%% smbfs move to staging 2116b7a473bf1c8d26998b477c294e7fe294921f
%% ceph factor out to libceph(net) 3d14c5d2b6e15c21d8e5467dc62d33127c23a644
%% 5db53f3 [LogFS] add new flash file system

\begin{figure}
  \centering
  \includegraphics[width=0.8\linewidth,keepaspectratio]{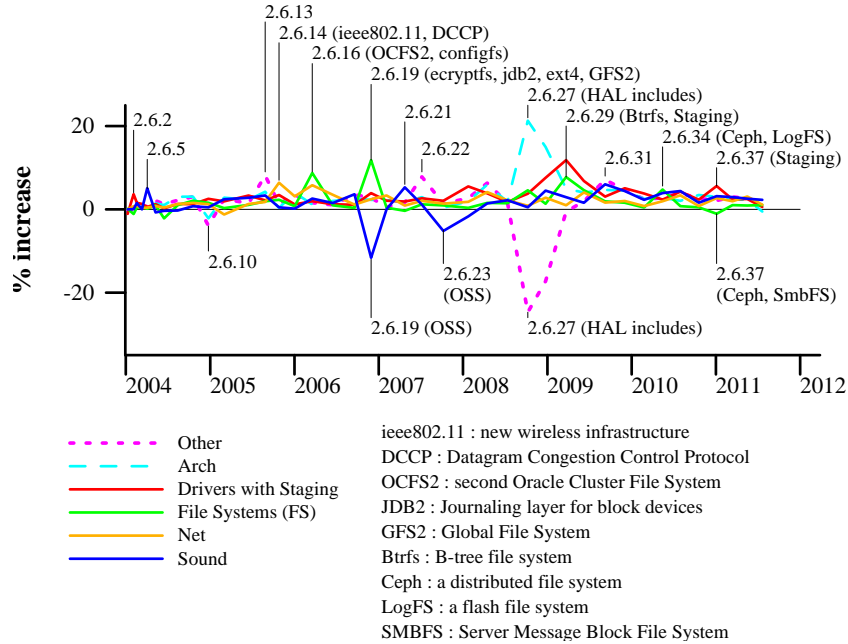}
  \caption{Linux directory size increase}
  \label{fig:code-increase}
\end{figure}

Figure~\ref{fig:code-size} shows that,
for most directories, the code growth has been roughly linear since Linux
1.0.  Some exceptions are highlighted in Figure~\ref{fig:code-increase},
which shows the percentage code size increase in each directory from one
version to the next.  We have marked some of the larger increases and
decreases.  Many of the increases involve the introduction of new services
(\textit{e.g.}, ieee802.11 for wireless support in version 2.6.14), and
new file systems (\textit{e.g.}, the Btrfs filesystem in version
2.6.29).  In Linux 2.6.19 and 2.6.23, old OSS drivers already supported by
ALSA were removed from {\tt sound}, decreasing its size.  In Linux 2.6.27,
{\tt arch} was reorganized, and received some large header files from {\tt
  include}, adding around 180,000 lines of C code. The most recent decrease
is in the \texttt{fs} directory of Linux 2.6.37, where the Ceph distributed
file system was reorganized and partly moved to the \texttt{net}
directory. Another decrease, at the same time, is due to the Server Message
Block File System (SMBFS), which is on its way out of the kernel, as it has
been superseded by the Common Internet File System (CIFS). Finally,
\texttt{staging} grew substantially in 2.6.29 and 2.6.37.  All in all,
these changes have resulted in code growth from 2~MLOC in 2001 to more than
9~MLOC in 2011.

%% \begin{figure}[tb!]
%%   \centering
%%   \includegraphics[width=\linewidth,height=\textheight,keepaspectratio]{figures_db/notes-evol}
%%   \caption{Notes per line of code through time}
%% \label{fig:notes-evol}
%% \end{figure}

%% \begin{figure}[tb!]
%%   \centering
%%   \includegraphics[width=0.95\linewidth,keepaspectratio]{figures_db/notes-evol-dir}
%%   \caption{Notes through time per directory}
%%   \label{fig:notes-evol-dir}
%% \end{figure}

\begin{figure}
  \centering
  \includegraphics[width=0.95\linewidth,keepaspectratio]{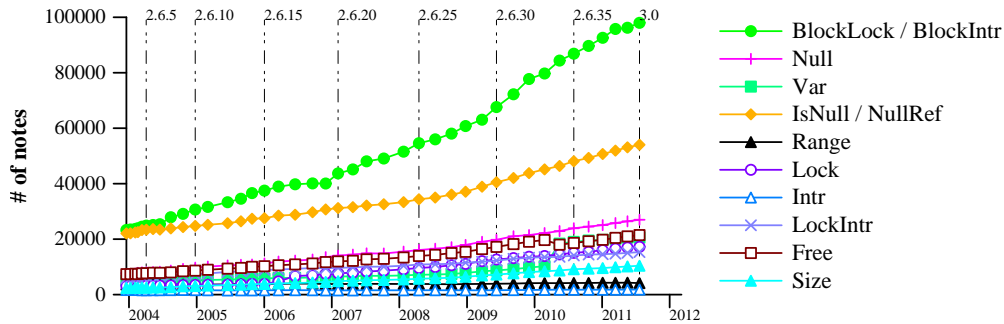}
  \caption{Notes through time per kind}
  \label{fig:notes-evol-kind}
\end{figure}

In our study, we are less interested in the absolute number of lines
of code than the amount of code relevant to our fault kinds,
\textit{i.e.}  notes.  As shown in Figure \ref{fig:notes-evol-kind},
the increase in code size has induced an similar increase in the
number of notes, as defined in Section~\ref{sec:relev-site-find}, in
almost all cases.  In fact, \hl{across all of Linux 2.6, the number of
notes per line of code is essentially constant, between 0.027 and
0.030 and the number of lines and the number of notes are highly
correlated: 0.998 by the Pearson's correlation
coefficient.}\footnote{This coefficient ranges from -1 to 1 inclusive,
  with 0 indicating that the two variables are not correlated, and an
  absolute value of 1, that the variable are correlated.}%  \jl{For the
%   correlation, I think that one has to know the formula that was used
%   in order to appreciate the number.  The formula that I know of is
%   Spearman's rho, which provides both a degree of correlation and a
%   degree of confidence.  But maybe there are other ways to compute
%   correlation?} \npnote*{Answer}{NP: ANSI SQL provides only
%   Pearson. Spearman is a \emph{rank} correlation coefficient. Seems a
%   better coefficient as the process is not linear. Do we change?}
% \jl{I have no idea.}

%%% Local Variables:
%%% mode: LaTeX
%%% TeX-master: "faults-in-linux-2.6-tocs"
%%% coding: utf-8
%%% TeX-PDF-mode: t
%%% ispell-local-dictionary: "american"
%%% End:

%\section{Validation of our experimental protocol}
\section{Faults in Linux 2.4.1}
\label{sec:baseline}

Linux 2.4.1 was the latest version of Linux considered by Chou
\emph{et al.}~\cite{Engler:empirical:SOSP01}.  To validate our
experimental protocol, we have used our checkers to find faults and
notes in this version, and we compare our results to those
  provided in their paper.  We focus on the results that are specific
to Linux 2.4.1, rather than those that relate to the history of Linux
up to that point, to avoid the need to study earlier versions that are
of little relevance today.

\subsection{What code is analyzed?}
\label{sec:codesize}

For the results of fault finding tools to be comparable, the tools
must be applied to the same code base.  Chou {\em et al.}\ focus only
on x86 code, finding that 70\% of the Linux 2.4.1 code is devoted to
drivers.  Nevertheless, we do not know which drivers, file systems,
\textit{etc}.\ were included.  To calibrate our results, we use
SLOCCount to obtain the number of lines of ANSI C code in the Linux
kernel and in the {\tt drivers} directory, considering the three
possibilities reported in Table~\ref{code_size}: all code in the Linux
2.4.1 kernel source tree (``All code''), the set of {\tt .c} files
compiled when using the default x86 configuration (``Min
x86''),\footnote{\scriptsize This configuration was automatically
  generated using \texttt{make menuconfig} without any modification to
  the proposed configuration. To collect the {\tt .c} files, we
  compiled Linux 2.4.1 according to this configuration using a Debian
  3.1 (Sarge) installation in a virtual machine, with \texttt{gcc}
  version 2.95.4 and \texttt{make} version 3.80.} and all 2.4.1 code
except the {\tt arch} and {\tt include} subdirectories that are
specific to non-x86 architectures (``Max x86'').  Max x86 gives a
result that is closest to that of Chou {\em et al.}, although the
proportion of driver code is slightly higher than 70\%.  This is
reasonable, because some driver code is associated with specific
architectures and cannot be compiled for x86.  Nevertheless, these
results show that we do not know the precise set of files used in Chou
{\em et al.}'s tests.

\begin{table}[h!t]
\tbl{Percentage of Linux code found in {\tt drivers} calculated
  according to various strategies}{
\begin{tabular}{|l||r|r|r|}
\cline{2-4}
\multicolumn{1}{c|}{}& All code & Min x86 & Max x86 \\\hline
Drivers LOC & 1,248,930 &  71,938 & 1,248,930 \\
Total LOC   & 2,090,638 & 174,912 & 1,685,265 \\\hline
Drivers \%  & 59\% & 41\% & 74\% \\\hline
\end{tabular}}
\label{code_size}
\end{table}

%% \begin{table}
%% \begin{center}
%% {%\footnotesize
%%   \scriptsize
%% \begin{tabular}{|l||r|r|r||r|}
%% \cline{2-5}
%% \multicolumn{1}{c|}{}& All code & Min x86 & Max x86 & Chou {\em et al.} \\\hline
%% Drivers LOC & 1,248,930 &  71,938 & 1,248,930 & 1,150,000 \\
%% Total LOC   & 2,090,638 & 174,912 & 1,685,265 & 1,650,000 \\\hline
%% Drivers \%  & 59\% & 41\% & 74\% & 69\%\\\hline
%% \end{tabular}}
%% \end{center}
%% \caption{The percentage of Linux code found in {\tt drivers} calculated
%%   according to various strategies.
%% ``Min x86'' refers to the set of .c files compiled using the default
%%   x86 configuration provided with the Linux kernel.
%% ``Max x86'' refers to all code except non-x86 arch and include code}
%% \label{code_size}
%% \end{table}

% arch is 285117 (1)
% arch/i386 is 23369 (2)
% include is 226716 (3)
% include/asm* is 152607 (4)
% include/asm-i386 is 8982 (5)
% total - (1) + (2) - (4) + (5)

In our experiments, we consider the entire kernel source code, and not just
the code for x86, as every line of code can be assumed to be relevant to
some user.

\subsection{How many faults are there?}
For the entire Linux 2.4.1 kernel, using the checkers described in
Section \ref{sec:checkers}, we obtain about 700 reports, of which we
have determined that 569 represent faults and the remainder represent
false positives.  Chou {\em et al.}'s checkers find 1,025 faults in
Linux 2.4.1.  They have checked 602 of these reports; the remainder
are derived from what they characterize as low false positive
checkers.  We have checked all of the reports included in our study.

%% checked 2011/01/15, without BlockIntr
%% SELECT count(*) FROM "public"."full_reports" WHERE "version_name" =
%% 'linux-2.4.1' and standardized_name != 'BlockIntr' [ and status = 'BUG' ]

%% checked 2010/12/22: 729 - 569

%% to find the number of bugs and the number of reports in 2.4.1:
%% select * from study_bugs_and_reports_number where version_name='linux-2.4.1';

Table \ref{tab:comparative_count} compares the number of faults found per
checker.  In most cases, we find fewer faults.  This may be due to
different definitions of the checkers, or different criteria used when
identifying false positives.  In the case of {\bf Var}, we find fewer
faults because we consider only cases where the size is explicitly expressed as a
number.
In a few cases, we find more faults.  For
example, Chou {\em et al.}'s {\bf Inull} checker can be compared to our
{\bf IsNull} and {\bf NullRef} checkers.  We find fewer {\bf IsNull} faults
than their {\bf Inull} faults, but far more {\bf NullRef} faults.  We also
find slightly more {\bf Free} faults.  This may derive from considering a
larger number of files, as we have found that only one of our {\bf Free}
faults occurs in a file that is compiled using the default x86
configuration.  Results from Chou {\em et al.}'s checkers were available at
a web site interface to a database, but Chou has informed us that this
database is no longer available.  Thus, it is not possible to determine the
precise reasons for the observed differences.

% \np{New experiment to refine the var checker. Check for big structure
%   allocation}

% 500 and 625 checked on July 26, 2010

\begin{table}[thp]
\tbl{Comparative fault count}
{
\begin{tabular}{|l|l|l|l|}
\hline
\multirow{2}{*}{Checker}& \multicolumn{2}{c|}{Chou {\em et al.}} & \multirow{2}{*}{Our results} \\
     & checked & unchecked & \\\hline
Block & 206 & 87 & N/A \\
BlockLock & N/A & N/A & 43 \\
BlockIntr & N/A & N/A & 102 \\
Null & 124 & 267 & 98 \\
Var & 33 & 69 & 13
\\\hline
Inull & 69 & 0 & N/A \\
IsNull & N/A & N/A & 36 \\
NullRef & N/A & N/A & 221
\\\hline
Range & 54 & 0 & 11 \\
Lock & 26 & 0 & 5 \\
Intr & 27 & 0 & 2 \\
LockIntr & N/A & N/A & 6 \\
Free & 17 & 0 & 21
\\\hline
Float & 10 & 15 & 8 \\
%Real & 10 & 1 & N/A \\
%Param & 7 & 0 & 0 \\
Size & 3 & 0 & 3 \\\hline\hline
Total & 569 & 438 & 569 \\\hline
\end{tabular}}
\label{tab:comparative_count}
\end{table}

\subsection{Where are the faults?}
\label{buckets}

In this section, the faults are classified using the same criterion
as Chou \emph{et al.}, \textit{i.e.}, by subsystem and function size.

\paragraph{By subsystem}
\label{sec:subsystem}

\begin{figure}
  \centering
\begin{tabular}{@{}c@{}}
%  \subfigure[Number of faults \np{broken down per} directory and category]{
  \subfigure[Number of faults broken down by directory and category]{
    \includegraphics[width=\linewidth,height=\textheight,keepaspectratio]{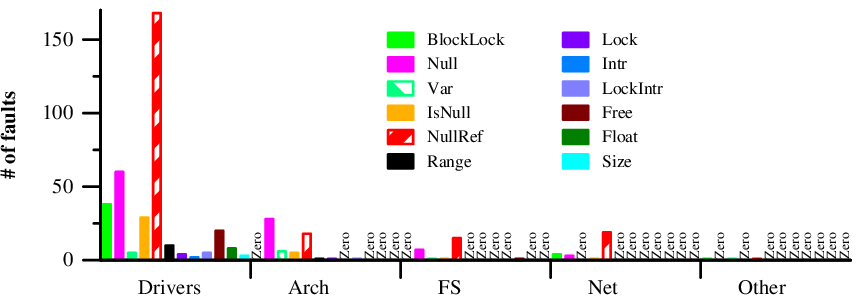}
    \label{fig:where-count-241}
  }
%% \\
%%   \subfigure[Rate of faults per directory and category]{
%%     \includegraphics[width=\linewidth,height=\textheight,keepaspectratio]{figures_db/rate-per-dir-4-241}
%%     \label{fig:where-rate-241}
%%   }
\\
%  \subfigure[Rate of faults compared to other directories]{
  \subfigure[Rate of faults compared to all other directories]{
    \includegraphics[width=\linewidth,height=\textheight,keepaspectratio]{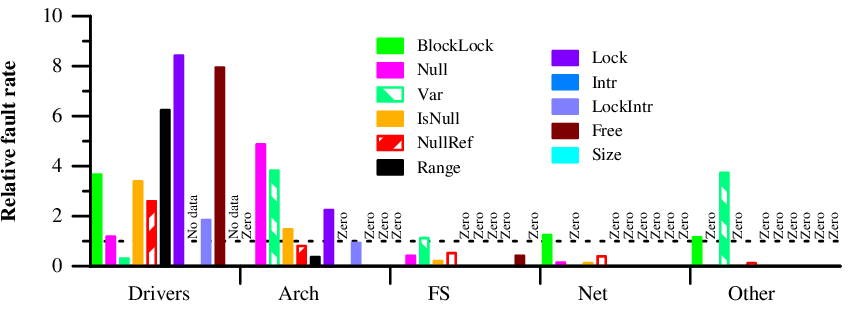}
    \label{fig:where-cmp-rate-241}
  }
\end{tabular}
\caption{Faults in Linux 2.4.1}
\label{fig:faults241}
\end{figure}

Chou {\em et al.}\ find that the largest number of faults is in the {\tt
  drivers} directory and that the largest number of these faults are in the
categories {\bf Block}, {\bf Null}, and {\bf Inull}, with around 180, 95,
and 50 faults in {\tt drivers}, respectively.\footnote{\scriptsize These
  numbers are approximated from the provided graphs.}  As shown in Figure
\ref{fig:where-count-241}, we also observe that the largest number of
faults is in the {\tt drivers} directory, with the largest number of these
faults also being in {\bf BlockLock}, {\bf Null}, and {\bf Inull} ({\bf IsNull}
and {\bf NullRef}), although in different proportions.

A widely cited
result of Chou {\em et al.}\ is that the {\tt drivers} directory contains
almost 7 times as many of a certain kind of faults ({\bf Lock}) as all
other directories combined.  They computed this ratio using the following
formula for each directory $d$, where $d$ refers to the directory $d$
itself and $\overline{d}$ refers to all of the code in all other
directories:
\[
\mita{fault rate}_d / \mita{fault rate}_{\overline{d}}
\]

\noindent
Figure \ref{fig:where-cmp-rate-241} shows the values of the same formula,
using our results.  We obtain a similar ratio with a relative rate of over
8 for \textbf{Lock} in {\tt drivers}, as compared to all other directories
combined.  We also find that the {\tt drivers} directory has a rate of {\bf
  Free} faults that is almost 8 times that of all other directories
combined. Chou {\em et al.}, however, found a fault rate of only around
1.75 times that of all other directories combined in this case.  With both
approaches, however, the absolute number of {\bf Free} faults is rather
small.  Like Chou {\em et al.}, we also observe a high fault rate in the
{\tt arch} directory for the {\bf Null} checker, in both cases about 4.8
times that of all other directories combined.  Finally, unlike Chou {\em et
  al.}, we observe a high rate of {\bf Var} faults in both {\tt arch} and
{\em other}.  In the {\tt arch} case, all of the {\bf Var} faults found are
for architectures other than x86.  Indeed, overall for {\tt arch}, we find
60 faults, but only 3 (all {\bf Null}) in the x86 directory.

% Numbers checked July 26, 2010

% Pour arch:
% select b.study_dirname, count(b.correlation_id) from
% full_bug_correlations b join file_names fn on b.file_name=fn.file_name
% where b.standardized_name !=
% 'BlockIntr' and linux_le(b.correlation_birth_version,  'linux-2.4.1') and
% linux_ge(b.correlation_last_version,  'linux-2.4.1') and
% b.study_dirname='arch' group by b.study_dirname;

% Pour x86:
% select count(b.correlation_id) from full_bug_correlations b join
% file_names fn on b.file_name=fn.file_name where b.standardized_name !=
% 'BlockIntr' and
% linux_le(b.correlation_birth_version,  'linux-2.4.1') and
% linux_ge(b.correlation_last_version,  'linux-2.4.1') and
% (fn.type_name='x86' or fn.type_name='i386') and fn.family_name='arch';

\paragraph{By function size}
\label{sec:function-size}

\begin{figure}[pt]
  \centering

  \subfigure[Functions with notes - 4 linear buckets (Chou et al.'s methodology)]{
    \includegraphics[width=1\linewidth,keepaspectratio]{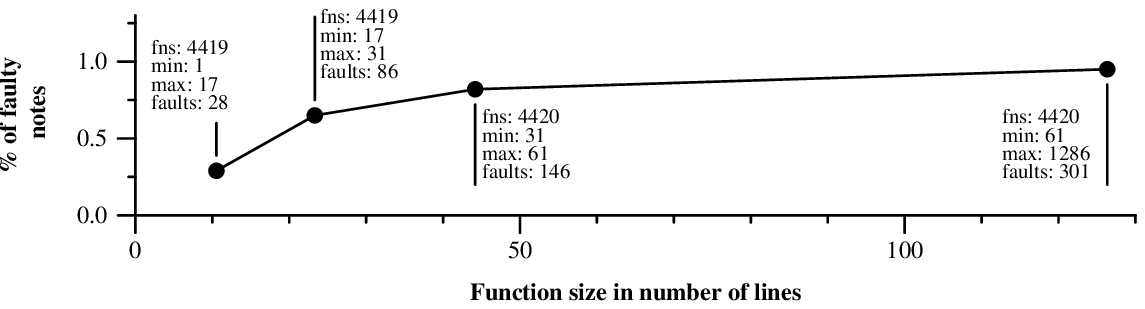}
  \label{fig:err-by-fct-size-4-241-lin4b-noted_fcts}
  }

  \subfigure[Functions with notes - 4 logarithmic buckets]{
    \includegraphics[width=1\linewidth,keepaspectratio]{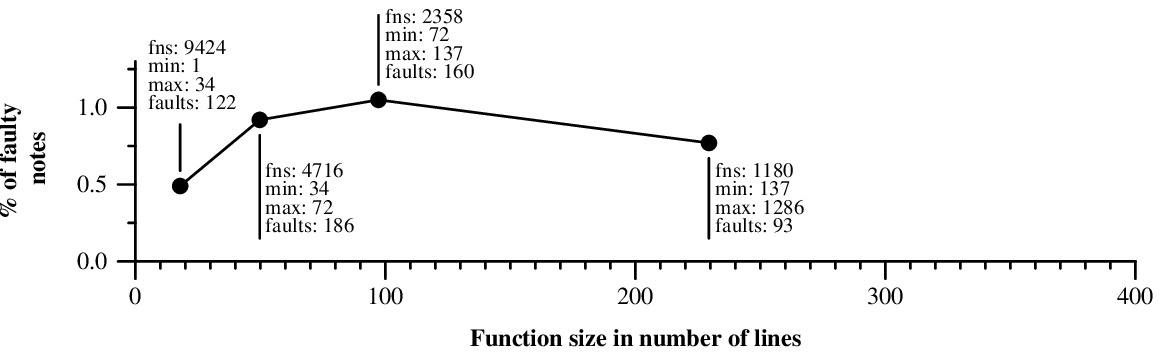}
  \label{fig:err-by-fct-size-4-241-log4b-noted_fcts}
  }

  \subfigure[All functions and functions with notes - 6 logarithmic buckets]{
    \includegraphics[width=1\linewidth,keepaspectratio]{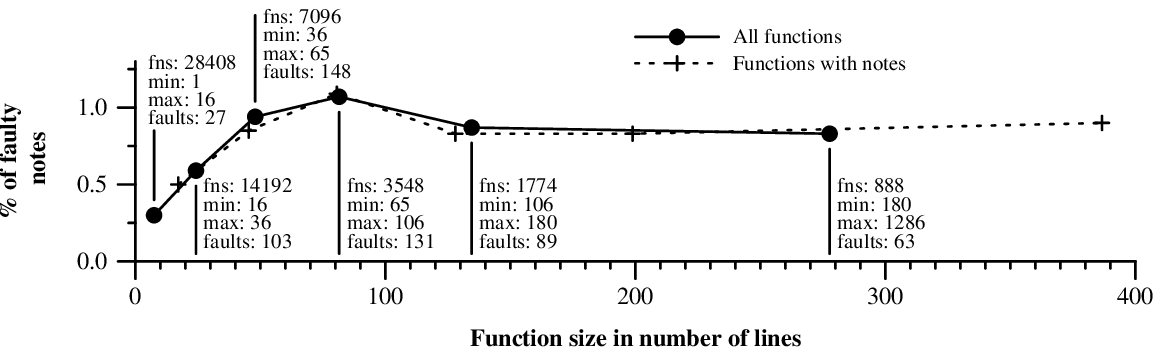}
  \label{fig:err-by-fct-size-4-241-log6b-all_fcts}
  }
  \begin{minipage}{.8\linewidth}\scriptsize
    \npnote*{}{At each point, \textbf{fns} is the number of
      functions affected to the bucket, \textbf{min}, respectively
      \textbf{max} is the size of smallest, respectively largest,
      function of the bucket, and \textbf{faults} is the number of faults
      present in the functions assigned to the bucket.}
  \end{minipage}

  \caption{Fault rate by function size in 2.4.1}
\label{fig:err-by-fct-size-4-241}
\end{figure}

Figure~\ref{fig:err-by-fct-size-4-241} shows the percentage of faulty
notes according to the size of the functions. Following Chou
\textit{et al.}, we only consider functions with notes and divide
these functions into 4 buckets, each containing the same number of
functions.  Like Chou \textit{et al.}, we find that the largest
functions have the highest fault rate, as shown in
Figure~\ref{fig:err-by-fct-size-4-241-lin4b-noted_fcts}.

%% For log4-all_fcts
% Using this logarithmic distribution of the functions inside the bucket
% allows to discover that about 30,000 functions of less than 17 lines
% of code have only a fault rate of 0.31. The next bucket, which
% represents about 15,000 functions, has a fault rate of 0.59. The third
% bucket, a fault rate of 1.08, and the fourth and last bucket a fault
% rate of 0.89. So, large functions have indeed a higer fault rate for
% the third first buckets but for the lastest bucket, the largest
% functions, between 88 and 1,286 lines of codes, have lower fault rate
% than the previous bucket.

\begin{figure}[pt]
  \centering
  \includegraphics[height=3cm,keepaspectratio]{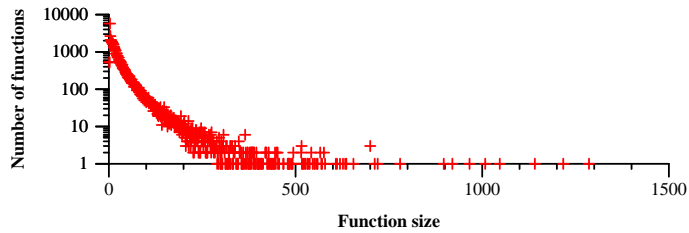}
  \caption{Function size distribution}
  \label{fig:fct-size-distrib}
\end{figure}

%% For log4-noted_fcts
Linux, however, contains more small functions than large ones, as
illustrated by Figure~\ref{fig:fct-size-distrib}, which shows the
distribution of the functions according to their size. Thus using
same-sized buckets tends to group a wide range of larger functions
into a single bucket.  We thus decided to consider a logarithmic
distribution of the functions into buckets, where the first bucket
contains the smallest half of the functions, the next bucket contains
the smallest half of the remaining functions,
\textit{etc}.\footnote{Note that at bucket boundaries, functions with
  the same length may appear in different buckets.} \hl{Using this
logarithmic bucketing strategy,
Figure~\ref{fig:err-by-fct-size-4-241-log4b-noted_fcts}, shows that it
is actually the functions of the third bucket, having between 72 and
137 lines, that have the highest fault rate, at 1.05, although the
largest functions continue to have a higher fault rate, at 0.77, than
the smallest ones, at 0.49.}
%% bucket 2 has fault rate 0.92

To study the relation between function size and fault rate in more
detail, Figure~\ref{fig:err-by-fct-size-4-241-log6b-all_fcts}
increases the number of buckets to 6, and considers all functions, not
just those with notes. The curve with only the functions containing
notes is given with a dotted line for reference. The same trend as
found in Figure~\ref{fig:err-by-fct-size-4-241-log4b-noted_fcts} is
observed for small, mid-size, and large functions.  \hl{This bucketing
  strategy, however, indicates the probability of a fault based solely
  on the function size, which represents a simpler criterion. As the
  two curves of Figure~\ref{fig:err-by-fct-size-4-241-log6b-all_fcts}
  are almost identical, the two strategies can be considered
  equivalent.}

% 2013-10-23
% study_rate_per_fct_size_restrict_s_all_fcts with 4 exp. buckets
%
% bucket	average_fct_size	min_fct_size	max_fct_size	nb_fncs_per_bucket	rate_in_percentage	number_of_bugs	number_of_notes
% 0	7.90	1	17	29816	0.31	32	10257
% 1	27.03	17	42	14908	0.59	118	19939
% 2	59.53	42	88	7454	1.08	215	19982
% 3	157.77	88	1286	3728	0.89	196	22049

%%%%%%%%%%%%%%%%%%%
%%%%%%%%%%%%%%%%%%%

\subsection{How old are the faulty files?}
\label{sec:file-age}

\begin{figure}
  \centering
  \includegraphics[width=.8\linewidth,keepaspectratio]{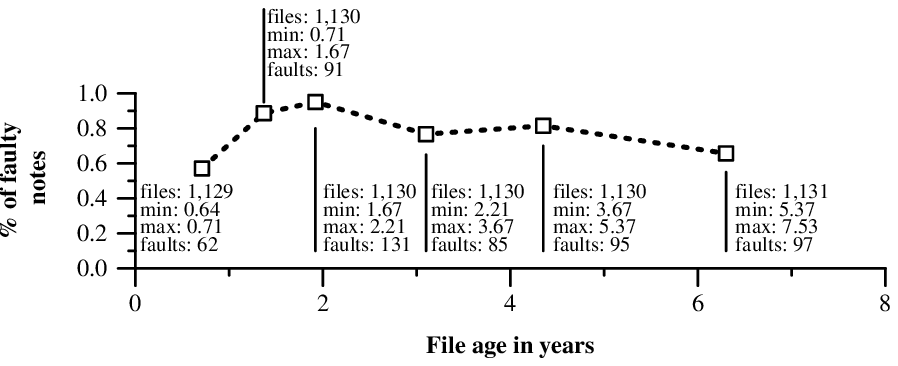}

  \begin{minipage}{.8\linewidth}\scriptsize
    At each point, \textbf{files} refers to the number of files
    considered to compute the point, \textbf{min} and \textbf{max}
    refer to the minimal or maximal age of the files in the bucket,
    and \textbf{faults} indicates the number of faults that occur in
    the bucket files.
  \end{minipage}

  \caption{Fault rate by file age in 2.4.1}
  \label{fig:rate-by-age-4-241}
\end{figure}

To explore the relation between the Linux 2.4.1 fault rate and file age,
we first consider the distribution of files according to their age, as
shown in Table~\ref{tab:file-age-distrib-241}. In Linux 2.4.1, there
were few very recently added files; most of the files had been part of
the Linux kernel for between 1 and 3 years. There is also an
almost equal number of files for each of the numbers of years above
3. We thus decided to distribute the files equally among the buckets,
as done by Chou \textit{et al.}, \npnote*{Reviewer 2 found bucket
  description confusing}{\emph{i.e.} we first order the files
  according to their age and then assign each file to a bucket such
  that each bucket has almost the same number of files, as illustrated
  Figure~\ref{fig:rate-by-age-4-241}.} However, we again increase the
number of buckets to 6, to emphasize the relation between the faults
and the file age for the youngest files, without disturbing the
observation of other files, either young or old.

\begin{table}[h!tp]
\tbl{File age distribution for Linux 2.4.1}
{  \centering
  \begin{tabular}{| l |c|c|c|c|c|c|c|c|} \hline
    Age (in years) & 0 & 1 & 2 & 3 & 4 & 5 & 6 & 7 \\ \hline
    Number of files
    & 1330
    & 1892
    &   526
    &   1217
    &    541
    &     627
    &    325
    &    322 \\ \hline
  \end{tabular}
}
  \label{tab:file-age-distrib-241}
\end{table}

Chou \emph{et al.}  find that younger files have a higher fault rate,
of up to 3\% for the {\bf Null} checker.  We also find fault rates of
more than 3\% for the {\bf Null} checker in 4 out of 6 buckets, being
the first three buckets and the last one. As illustrated by
Figure~\ref{fig:rate-by-age-4-241}, moderately young files (between 8
months and 26 months) have the highest fault rates, over
0.89\%. However, the youngest files (under 8 months) have the lowest
fault rate, 0.57\%. For files older than 2.5 years, we observe lower
fault rates, above 0.82\%. Interestingly, middle aged files, of about
3 years, and the oldest files, of over 5 years have a lower fault rate
than files between 4 and 5 years old.

In conclusion, our methodology gives comparable results with Chou
\textit{et al.} about the relation of faults and the age of
files. However, by using two more buckets, we have been able to
emphasize particularities of the newest and oldest files.

\subsection{How are faults distributed?}

%table_linux_2_4_1_without_BlockLock_BlockIntr : [
%   [ 1 , 215 ],
%   [ 2 , 80 ],
%   [ 3 , 24 ],
%   [ 4 , 11 ],
%   [ 5 , 2 ],
%   [ 6 , 2 ],
%   [ 7 , 3 ],
%   [ 9 , 1 ]
% ];
 
\begin{figure}
  \centering
  \includegraphics[width=.8\linewidth,keepaspectratio]{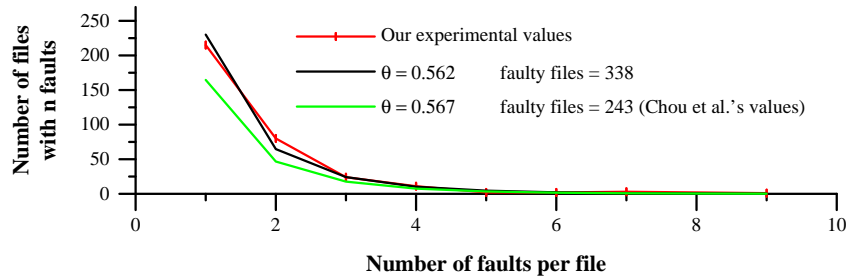}
  \caption{Comparison of the error distributions}
\label{fig:thetas}
\end{figure}

Chou {\em et al.}~plot the numbers of faults against the percentage of
files containing each number of faults and find that for all of the
checkers except {\bf Block}, the resulting curve fits a log series
distribution, as determined by $\chi^2$ test, with a $\theta$ value of
0.567 and a degree of confidence (p-value) of 0.79 (79\%). We observe
a $\theta$ value of 0.562 and a p-value of 0.234 without {\bf Block}.
We conjecture that this low p-value is due to the fact that two files
have 5 and 6 faults while three files have 7 faults each; such an
increase for larger values is not compatible with a log series
distribution.  Nevertheless, the values involved are very small, and,
as shown in Figure~\ref{fig:thetas}, the curve obtained from our
experimental results fits well with the curve obtained using the
corresponding calculated $\theta$ value.  The curve obtained from Chou
{\em et al.}'s value of $\theta$ is somewhat lower, because they found
a smaller number of faulty files, probably due to having considered
only the x86 architecture.

\subsection{Assessment}

In this section, we have seen that our checkers do not find the same
number of faults in Linux 2.4.1 code as the checkers of Chou {\em et
  al.}  We recall that Chou {\em et al.}'s checkers are not very
precisely described, and thus we expect that most of the differences
are in how the checkers are defined.  Furthermore, we do not consider
exactly the same set of files.  Nevertheless, the distribution of
these faults among the various directories is roughly comparable, and
their distribution among the files is also comparable. We thus
conclude that our checkers are sufficient to provide a basis for
comparison between Linux 2.6 and the previous versions studied by Chou
\emph{et al}.

\section{Faults in Linux 2.6}
\label{sec:linux-2.6}

We now assess the extent to which the trends observed for
Linux 2.4.1 and previous versions continue to apply in Linux 2.6, and study
the points of difficulty in kernel development today.  We consider what has
been the impact of the increasing code size and the addition of new
features on code quality, and whether drivers are still a major
problem.

%
% Numbers are provided by figures/numbers.tex
% and can be updated with figures/get_numbers.sh
%
Concretely, we study a period of over 7 years, beginning with the
release of Linux 2.6.0 at the end of 2003 and ending in July 2011 with
the release of Linux 3.0. \jlnote*{Updated. Reviewer 1 and 3 made a
  remark about reports.}{For the complete set of Linux 2.6 kernels and for Linux
  3.0, using the checkers described in Section \ref{sec:checkers}, we
  obtain {\nbinitialreports} reports, resulting in {\nbreports}
  reports \emph{after correlation}, of which we have manually
  determined that {\nbfaults} represent faults and the rest represent
  false positives.}

%% Last checked the 2013-06-11
%%
%% SELECT count (*) FROM "public"."full_correlations"
%% WHERE
%% [ "status" = 'BUG' AND ]
%% "correlation_birth_version" != 'linux-2.4.1' AND
%% standardized_name != 'BlockIntr' and
%% standardized_name != 'BlockRCU' and
%% standardized_name != 'LockRCU' and
%% standardized_name != 'DerefRCU'

\subsection{How many faults are there?}
\label{sec:what}

% Percentages checked the 2013-06-21
% We count BlockIntr but not RCU related faults
% \notefirst
% \noteoverall
% \notefinal
% \notedensity

\begin{figure}[tp]
  \centering
  \begin{tabular}{@{}c@{}}
  \subfigure[Faults]{
    \begin{minipage}{.80\linewidth}
      \includegraphics[width=\linewidth,keepaspectratio]{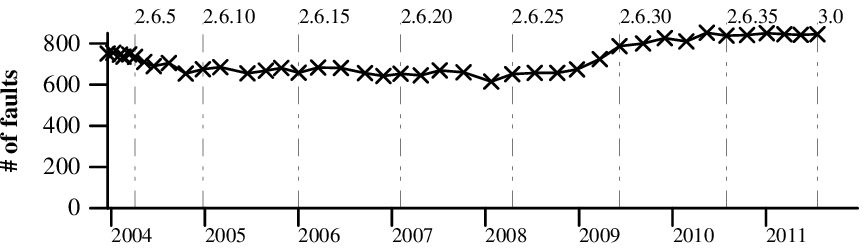}
      \vspace{.5em}
    \end{minipage}
  \label{fig:evol-count}
  }
\\[10mm]
  \subfigure[Fault density (Faults per 1KLOC)]{
    \begin{minipage}{.80\linewidth}
    \includegraphics[width=\linewidth,keepaspectratio]{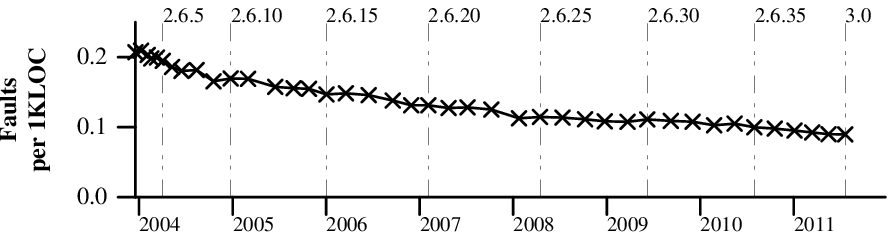}
      \vspace{.5em}
    \end{minipage}
  \label{fig:evol-density}
  }
\\[10mm]
  \subfigure[Introduction and elimination of faults]{
    \begin{minipage}{.80\linewidth}
    \includegraphics[width=\linewidth,keepaspectratio]{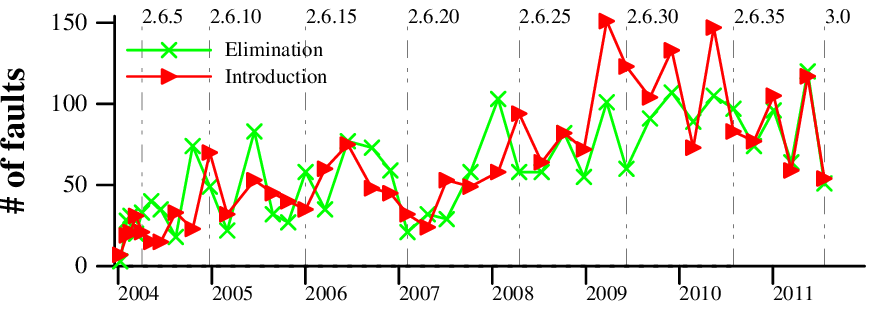}
      \vspace{.5em}
    \end{minipage}
  \label{fig:evol-birth-and-death}
  }
  \end{tabular}
\caption{Faults in Linux 2.6.0 to 3.0}
\label{fig:evol-linux-faults}
\end{figure}

We first analyze the relation between the code growth and the total
number of faults in Linux 2.6.  As shown in
Figure~\ref{fig:evol-count}, the number of the faults considered has
held roughly steady over this period, with an overall increase of only
\pctoverall.  Indeed there was a decrease of 18\% from 2.6.0 to the
minimum value in the considered time period, which occured in 2.6.24.
This latter decline is quite remarkable given that the code size
increased by more than 50\% during these 25 versions
(Figure~\ref{fig:code-size}).  However, since the minimum of 2.6.24,
the number of faults has somewhat increased, including a jump of 17\%
between 2.6.28 and 2.6.30. As a result, we find in Linux 3.0 37\% more
faults than in 2.6.24, and {\pctoverall} more than in 2.6.0.  \hl{Still,
due to the overall increase of 160\% in code size from Linux 2.6.0 to
Linux 3.0, the rate of faults per line of code has significantly
decreased, by 57\%, as shown in Figure \ref{fig:evol-density}, and for
version 3.0, we find fewer than 90 faults per MLOC.} These observations
are quite different from those for versions up through Linux 2.4.1: in
those versions, there was a code size increase of over 17 times
between Linux 1.0 and Linux 2.4.1 and an increase in the number of the
faults considered of over 33 times~\cite{Engler:empirical:SOSP01}.
Finally, Figure \ref{fig:evol-birth-and-death} reveals the reason for
our observations about 2.6. \hl{Faults are still introduced, indeed at a
growing rate, but in many versions even more faults are eliminated.}

\begin{figure}[pt]
  \centering
  \begin{tabular}{@{}c@{}}
    \subfigure[Fewer than 100 faults]{
    \begin{minipage}{.9\linewidth}
      \includegraphics[width=\linewidth,height=\textheight,keepaspectratio]{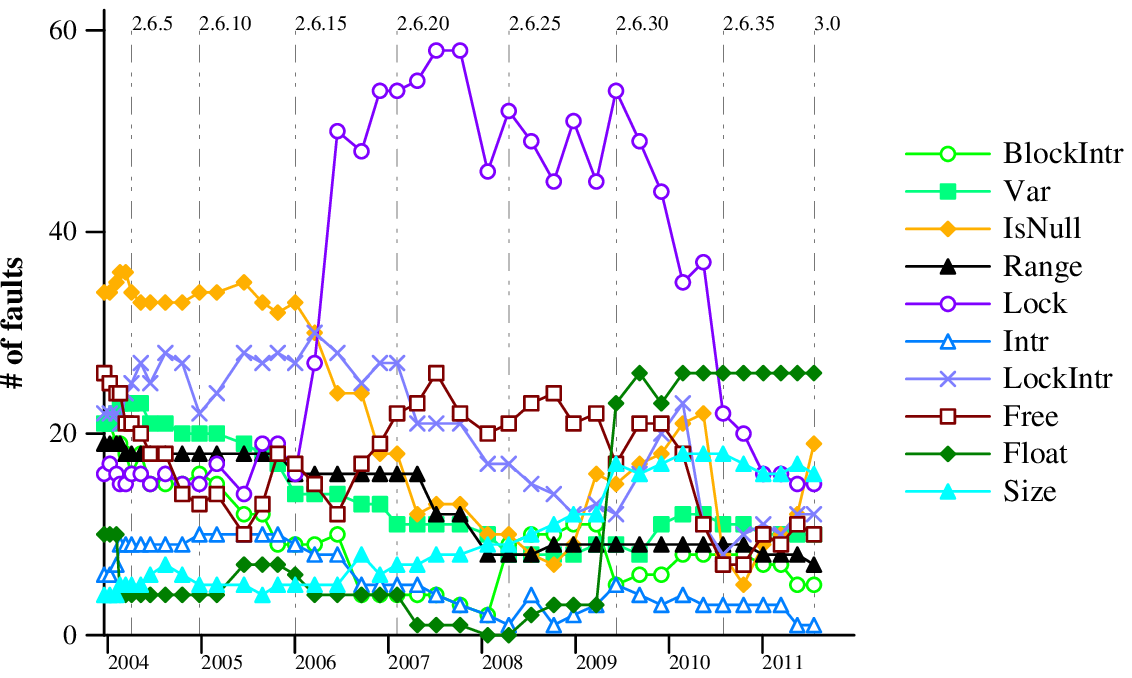}
      \vspace{.5em}
    \end{minipage}
      \label{fig:evol-low}
    }
\\[10mm]
%     \subfigure[Increasing faults]{
%     \begin{minipage}{.9\linewidth}
%       \includegraphics[width=\linewidth,height=\textheight,keepaspectratio]{figures/count-evol-inc}
%       \vspace{.5em}
%     \end{minipage}
%       \label{fig:evol-inc}
%     }
% \\[10mm]
%     \subfigure[Decreasing faults]{
%     \begin{minipage}{.9\linewidth}
%       \includegraphics[width=\linewidth,height=\textheight,keepaspectratio]{figures/count-evol-dec}
%       \vspace{.5em}
%     \end{minipage}
%       \label{fig:evol-dec}
%     }
% \\[10mm]
    \subfigure[More than 100 faults (BlockLock, NullRef and Null)]{
    \begin{minipage}{.9\linewidth}
      \includegraphics[width=\linewidth,height=\textheight,keepaspectratio]{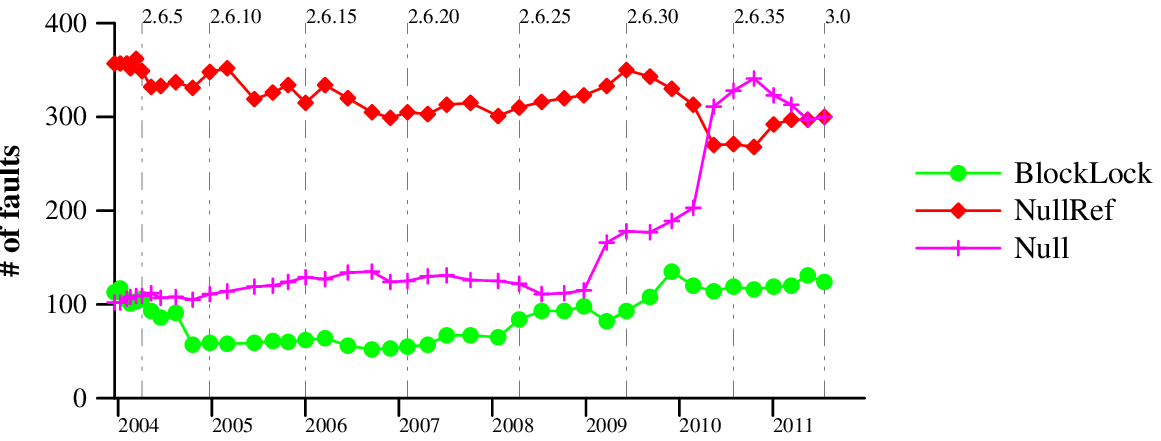}
      \vspace{.5em}
    \end{minipage}
      \label{fig:evol-over100}
    }
  \end{tabular}
  \caption{Faults through time}
  \label{fig:evol}
\end{figure}

\npnote*{Figures 11(a) and 11(b) were merged}{Figure~\ref{fig:evol}
  shows the number of each kind of fault found in Linux 2.6, separated
  for readability by their order of magnitude into those that have
  fewer than 100 faults at their maximum, shown in Figure~\ref{fig:evol-low},
  and the others, shown in Figure~\ref{fig:evol-over100}.}

%\note{Checked 2013-06-24}
\hl{For many fault kinds, the number of faults is essentially constant
over the considered period.}  Three notable exceptions are {\bf Lock}
in Linux 2.6.16 and 2.6.17, {\bf Null} in Linux 2.6.29 and 2.6.34, and
{\bf Float} in Linux 2.6.30, as shown in Figure~\ref{fig:evol}. In Linux 2.6.16,
the functions {\tt mutex\_lock} and {\tt mutex\_unlock} were
introduced to replace mutex-like occurrences of the semaphore
functions {\tt down} and {\tt up}.  9 of the 11 {\bf Lock} faults
introduced in Linux 2.6.16 and 23 of the 25 {\bf Lock} faults
introduced in Linux 2.6.17 were in the use of {\tt mutex\_lock}.  In
Linux 2.6.29, the {\tt btrfs} file system was introduced, as seen in
Figure~\ref{fig:code-increase}.  31 {\bf Null} faults were added with
this code.  7 more {\bf Null} faults were added in {\tt
  drivers/staging}, which more than tripled in size at this time.  31
other {\bf Null} faults were also added in this version.  Next, in
Linux 2.6.30 there was a substantial increase in the number of Comedi
drivers \cite{comedi-drv} in {\tt drivers/staging}.  All of the 21
{\bf Float} faults introduced in this version were in two Comedi
files.  These faults are still present in Linux 3.0.  Recall, however,
that {\tt staging} drivers are not included in Linux
distributions. Finally, in Linux 2.6.34, 75 \textbf{Null} faults, out
of 119, were introduced in the \texttt{arch} and \texttt{fs}
directories.

In the \textbf{IsNull} category, the number of faults decreases across
the studied period. Nevertheless two spikes in the last two years can
be observed. Indeed, 22 faults were introduced between 2.6.28 and
2.6.34, of which 7 were introduced in 2.6.29. These faults were
introduced almost entirely in \texttt{drivers}. In Linux 2.6.35, the
number of faults dropped to only 8, but it quickly grew up to 19
faults in 3.0 through the introduction of 18 new faults (note that
some faults were corrected at the same time, implying some fault
turnover). In each version, the faults were introduced mainly in
\texttt{drivers}. So, between 2.6.28 and 2.6.34, 40 \textbf{IsNull}
faults were introduced, of which 10 were in the \texttt{staging}
directory and 32 in other drivers.

%% \jlfatal*{Fixme}{Between 2.6.6 and 2.6.7 there was a drop in bad\_lock2 (Block) of 7, all
%%   in one file.  Between 2.6.7 and 2.6.8 there was an increase of 8 for the
%%   same pattern, but in multiple files.  Between 2.6.8 and 2.6.9 there was a
%%   drop in bad\_lock1 of 10, affecting only one file.  This accounts for the
%%   down up down effect on the left side of the Block graph.  Perhaps this is
%%   too detailed, however.}

%% \jlfatal*{Fixme}{
%% Likewise, in 2.6.26, 7 bad\_lock2 faults were corrected in various files.
%% In 2.6.26, 6 new faults were added in a single file.}

%% Total 2.6 bugs for Lock is 1112
%% Total 2.6 bugs for Block is 1567

\begin{figure}[tb!]
  \centering
  \includegraphics[width=\linewidth,height=\textheight,keepaspectratio]{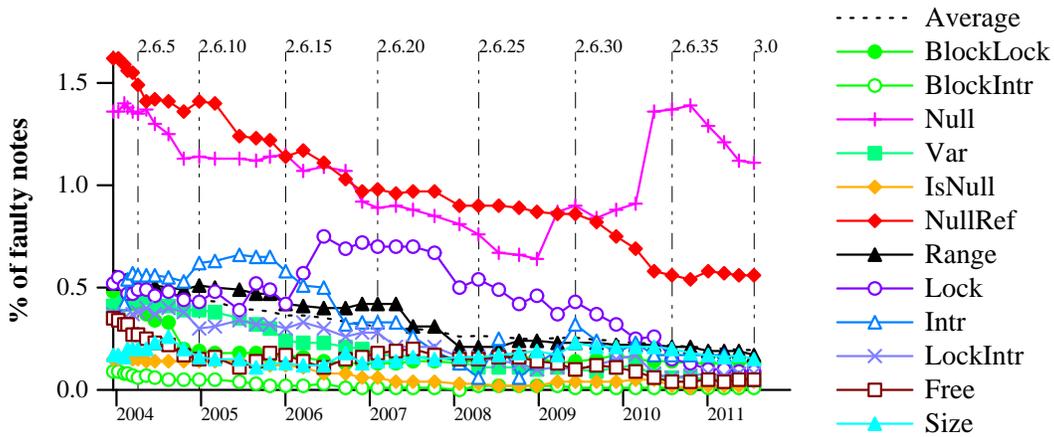}
  \caption{Fault rate per fault kind}
  \label{fig:evol-kind}
\end{figure}

\hl{As shown in Figure~\ref{fig:evol-kind}, the fault rate, {\em i.e.},
the ratio of observed faults to the number of notes, for the
considered fault kinds confirms the increase in reliability} (Float is
omitted, as described in Section~\ref{sec:relev-site-find}). As the
number of notes increases roughly with the size of the Linux kernel
while the number of faults is relatively stable, the fault rate tends
to decline.  The main increases, in {\bf Lock} and {\bf Null}, are due
to the introduction of {\tt mutex\_lock} and the {\tt btrfs} file
system, respectively, as mentioned previously.

\subsection{Where are the faults?}
\label{where}

The presence of a high rate of faults in a certain kind of code may
indicate that this kind of code overall needs
more attention.  Indeed, Chou {\em et al.}'s work motivated studies
of many kinds of driver faults, going beyond the fault kinds they
considered.  Many
properties of the Linux kernel have, however, changed since 2001, and so we
reinvestigate what kind of code has the highest rate of
faults, to determine whether attention should now be placed elsewhere.

%in the following, over half means around 50%
% For example
% linux-2.6.33      | drivers       | 3901868 | 7894327 |0.49426227213542079014

\begin{figure}[tp!]
  \centering
  \includegraphics[width=\linewidth,height=\textheight,keepaspectratio]{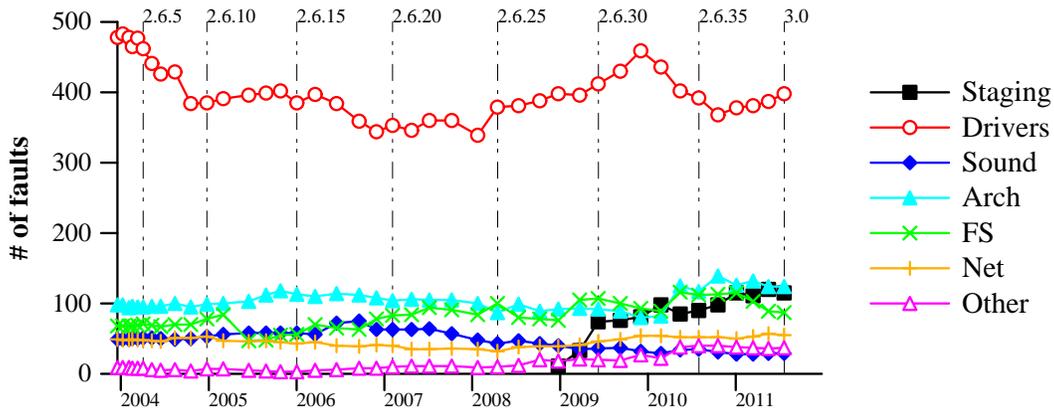}
   \caption{Faults per directory}
  \label{fig:count-evol-dir}
\end{figure}

\begin{figure}[tp!]
  \centering
  \includegraphics[width=\linewidth,height=\textheight,keepaspectratio]{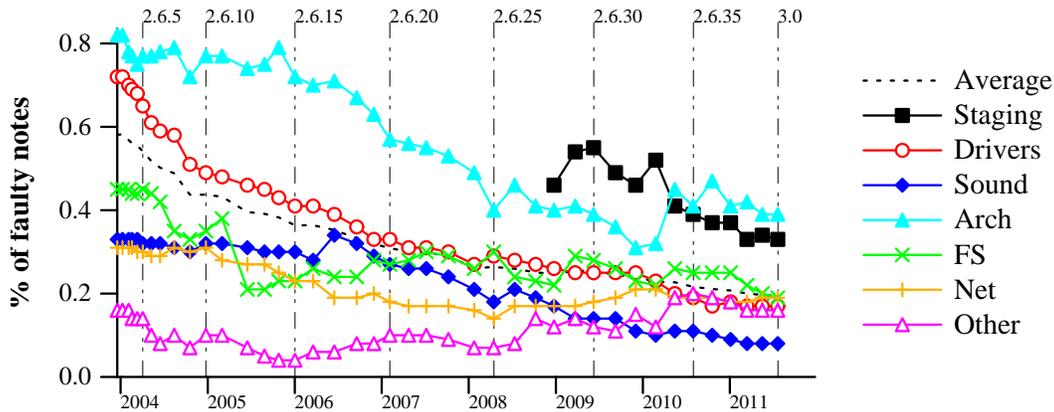}
  \caption{Fault rate per directory}
  \label{fig:evol-dir}
\end{figure}

\hl{As shown in Figure~\ref{fig:count-evol-dir}, the largest number of
faults is still in {\tt drivers}, which indeed makes up over half of
the Linux kernel source code.  The second-largest number of faults is
in {\tt arch}, accompanied by {\tt fs} and {\tt drivers/staging} in
recent versions.  In contrast to the case of Linux 2.4.1, however, as
shown in Figure~\ref{fig:evol-dir}, {\tt drivers} no longer has the
largest fault rate, and indeed since Linux 2.6.19 its fault rate has
been right at the average.}  There was not a large increase in the
number of {\tt drivers} notes at that time, so this decrease is
indicative of the amount of attention drivers receive in the peer
reviewing process.  {\tt Arch} on the other hand has many faults and
relatively little code, and so it has the highest fault rate
throughout most of Linux 2.6.  Around 30\% of the {\tt arch} faults
are {\bf Null} faults, although there appears to be no pattern to
their introduction. Over 90\% of the {\tt arch} faults are outside of
the {\tt x86}/{\tt{i386}} directories, with many of these faults being
in the {\tt ppc} and {\tt powerpc} code.

% Dir: cifs, ocfs2, btrfs
% select version_name , count(report_id) from full_reports where
% study_dirname = 'fs' and type_name = DIR and status = 'BUG' group
% by version_name, release_date order by release_date;

The largest numbers of faults in {\tt fs} are in {\tt cifs},
\texttt{ocfs2} and \texttt{btrfs}. For the \texttt{cifs} filesystem,
there are over 40 faults in Linux 2.6.0. But the number of faults
suddenly drops from 52 to 16 between 2.6.11 and 2.6.12 and only two
such faults remain in 3.0. For the {\tt ocfs2} filesystem, there are
10-14 faults per version starting in Linux 2.6.16 until the versions
2.6.39 and 3.0, where there are respectively only 4 and 5 faults
remaining. For the {\tt btrfs} filesystem, there are respectively 36
and 38 faults in Linux 2.6.29 and 2.6.30. Again, there are
respectively only 8 and 6 faults in the two last versions. All of
these faults are in recently introduced file systems: {\tt cifs} was
introduced in Linux 2.5.42, {\tt ocfs2} in Linux 2.6.16, and {\tt
  btrfs} in Linux 2.6.29.

{\tt Drivers/staging}, introduced in Linux 2.6.28, also has a
high fault rate, exceeding that of {\tt arch}.  This directory is thus
receiving drivers that are not yet mature, as intended.  The
introduction of {\tt drivers/staging}, however, has no impact on the
fault rate of {\tt drivers}, as {\tt drivers/staging} accommodates
drivers that would not otherwise be accepted into the Linux kernel
source tree.  Such drivers benefit from the expertise of the Linux
maintainers, and are updated according to API changes with the rest of
the kernel.

%% checked 2010/12/22: Around 30%

%% arch faults: 320 => at revision 338
%% SELECT count("standardized_name") FROM "public"."Bug ages" WHERE
%% "osdi_dirname" = 'arch' AND "min" != '01-30-2001' AND "standardized_name"
%% != 'Real'

%% arch Null faults: 99 => at revision 99
%% SELECT count("standardized_name") FROM "public"."Bug ages" WHERE
%% "osdi_dirname" = 'arch' AND "min" != '01-30-2001' AND "standardized_name"
%% = 'Null'

%% checked again 2011/01/15: Around 30%

%% arch faults: 315 without BlockIntr, gael database
%% SELECT count(*) FROM "public"."full_bug_correlations" WHERE "study_dirname"
%% = 'arch' AND "status" = 'BUG' AND "correlation_birth_version" !=
%% 'linux-2.4.1' and "standardized_name" != 'BlockIntr' and
%% standardized_name != 'BlockRCU' and standardized_name != 'DerefRCU'

%% arch NULL faults: again 99
%% SELECT count(*) FROM "public"."full_bug_correlations" WHERE "study_dirname"
%% = 'arch' AND "status" = 'BUG' AND "correlation_birth_version" !=
%% 'linux-2.4.1' and "standardized_name" = 'Null'

\begin{figure}[tp!]
  \centering
  \includegraphics[width=\linewidth,height=\textheight,keepaspectratio]{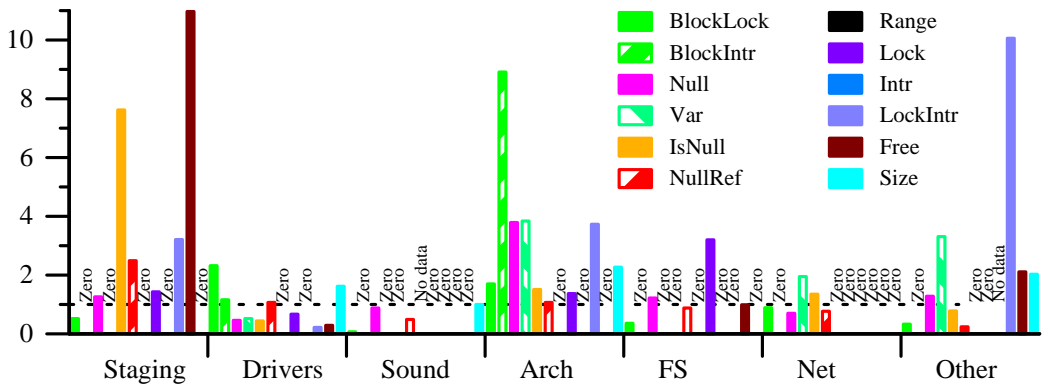}
  \caption{Fault rates compared to all other directories (Linux 3.0)}
  \label{fig:comparative-rate-per-dir}
\end{figure}

For Linux 2.4.1, we observed that {\tt
  drivers} had a much higher fault rate for certain kinds of faults
than other directories.  Figure \ref{fig:comparative-rate-per-dir}
shows the comparison of fault rates for Linux 3.0. \hl{It is more common
that {\tt drivers/staging}, {\tt arch}, or {\it other} has the highest
fault rate, indicating that the drivers that are intended for use in
the Linux kernel are no longer the main source of faults.} Indeed,
there are about 11 times more faulty notes in \texttt{drivers/staging}
than in all other directories. The maximum for \texttt{arch} and
\textit{other} is respectively 9 for \textbf{BlockIntr}, and 10 for
\textbf{LockIntr}.  In comparison, {\tt drivers} has 2.3 more faulty
{\bf BlockLock} notes than all other directories in Linux 3.0. This
worst case of the \texttt{drivers} directory is more than 4 times
smaller than the three other maximum values mentioned above.

For \texttt{staging}, dereference of a NULL pointer (\textbf{IsNULL})
and missing memory frees (\textbf{Free}) are the most common errors,
with respectively 8 and 11 times more errors in \texttt{staging} than
in the rest of the kernel.  For \texttt{arch}, the highest ratio is
for \textbf{BlockIntr} which is 9 times higher than in the rest of the
kernel. \texttt{Arch} has also high ratios for \textbf{Null},
\textbf{Var} and \textbf{LockIntr}, with a factor of 4 in these cases.
In the case of \texttt{fs}, the most problematic point is
\textbf{Lock} with a ratio of 3.2.  \texttt{Sound} and \texttt{net}
have the smallest fault rates.  In a preliminary version of this work, we compared
fault rates between directories in Linux 2.6.33 (Figure 11 of
\cite{Palix:asplos11}).  The two highest rates for {\tt sound} found
in that study are no longer visible in the results for Linux 3.0,
indicating the {\tt sound} has improved compared to the other
directories.

%arch has 167 total faults and 66 in Null
%I had noted this, but not it seems to be only 77 faults in all???

%% \jlfatal*{Fixme}{fs has Null, NullRef, and Lock faults in 2.6.33.  The most common
%%   subdirectories btrfs, introduced in 2.6.29, and then ocfs2, introduced in
%%   2.6.16.}

%% The fault rate significantly increases in only two directories during Linux
%% 2.6: {\tt sound} in Linux 2.6.17 and {\tt fs} in Linux 2.6.29.  The former
%% is again due to the introduction of {\tt mutex\_lock} and the latter is
%% again due to the introduction of {\tt btrfs}.  In both cases, the fault
%% rates decline thereafter.

\begin{figure}[tb!]
  \centering
  \includegraphics[width=0.95\linewidth,keepaspectratio]{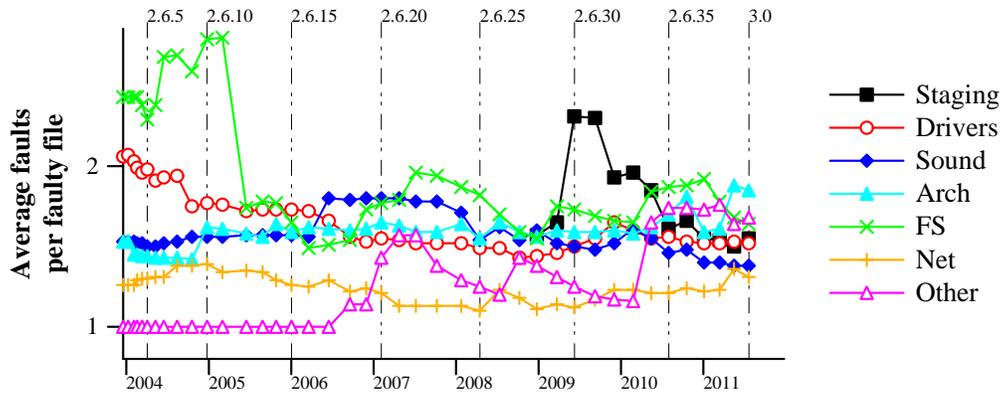}
  \caption{Faults per faulty file per directory}
  \label{fig:flt-file-evol-dir}
\end{figure}

Finally, in Figure \ref{fig:flt-file-evol-dir}, we consider the number
of faults per file containing at least one fault.  The highest
average number of faults per faulty file is for {\tt fs} in the
versions prior to 2.6.12, at 2.8 faults. In this case, there was a
single file with many {\bf NullRef} faults, as many as 45 in Linux
2.6.11.  In later versions, the highest average is for {\tt
  drivers/staging}, for which the average was over 2 in Linux 2.6.30.
At that point, a large number of drivers had recently been introduced
in this directory.  Many of these faults have been corrected and the
rate of entry of new drivers has slowed, and thus the average has
dropped to around 1.5, close to that of other directories.  {\tt
  Sound} had a relatively high number of faults per faulty file
starting in Linux 2.6.16 with the introduction of {\tt mutex\_lock};
faulty functions often contain more than one {\tt mutex\_lock}, and
thus a single omitted {\tt mutex\_unlock} may result in multiple {\bf
  Lock} reports.

%% checked 2010/12/22: as many as 45

%% check the 45
%% select count(r.report_id), f.file_name from reports r, ``Bug ages'' c, files f where r.correlation_id=c.correlation_id and r.file_id=f.file_id and f.version_name='linux-2.6.11' and c.standardized_name='NullRef' group by file_name order by count;

% values checked July 26, 2010

%%% Local Variables:
%%% mode: LaTeX
%%% TeX-master: "faults-in-linux-2.6-tocs"
%%% coding: utf-8
%%% TeX-PDF-mode: t
%%% ispell-local-dictionary: "american"
%%% End:

\subsection{How long do faults live?}
\label{long}

Eliminating a fault in Linux code is a three step process.  First, the
fault must be detected, either manually or using a tool.  Then, it
must be corrected, and a patch submitted to the appropriate
maintainers.  Then, the patch must be accepted by a hierarchy of
maintainers, ending with Linus Torvalds. Finally, there is a delay of
up to 3 months until the next release. It has thus been found that
most patches are integrated within 3-6
months~\cite{DBLP:conf/msr/JiangAG13} into a Linux release,
\textit{i.e.} it takes one or two releases to integrate patches. The
lifespan of a fault, modulo this three/six-month delay, is an
indication of the efficiency of the fault-elimination process.

\paragraph*{Fault survival analysis}
\label{sec:fault-surv-analys}

A challenge in measuring the lifetime of faults present within a time
period is that some faults may have been introduced before the time
period began, while others may last after the time period
ends. \npnote*{Updated to include Linux 3.0}{In this section, we
  consider the probability that a fault, that is observed at least
  once between Linux 2.6.0 and Linux 3.0, reaches a certain age.} A
fault that was introduced before the considered time period
(\textit{i.e.} before Linux 2.6.0) is said to have its lifetime left
censored, while a fault that persists after the considered time period
(\textit{i.e.} after Linux 3.0) is said to have its lifetime right
censored. The problem of censoring is addressed by a survival analysis
with the Kaplan-Meier
estimator~\cite{kaplan1958nonparametric}. Survival analysis gives the
probability that an observation (\textit{e.g.}, the death age of
faults) reaches a given value. The Kaplan-Meier estimator corrects the
bias introduced by the partial observations (\textit{i.e.}, the
observations of faults where only a lower bound of the age is known
because either the fault introduction or the fault removal was
unobserved), making the computed average lifespan unbiased as compared
to the average lifespan of the raw observations. To refine the
estimator, the approach allows approximating partial observations by
an interval. For complete observations, the exact value of the fault
age at death is used.

% Max. The median lifespan of the faults is 2.15 years.
% Right. The average lifespan of a fault is 2.27 years
% {\textpm} 21.5 days, but half of the faults live less than 1.26
% years.
\npnote*{Changed for clarity}{To handle the censored data of Linux 2.6.0,
  as we did not know the exact lifespan of the faults, we bound the
  lifespan by an interval and thus set a lower-bound and an
  upper-bound on the lifespan of the Linux 2.6.0 faults.} The
lower-bound is always the observed lifespan, {\em i.e.}, we assume
that the fault was introduced where we first observe it, in Linux
2.6.0. For the upper-bound, we considered three alternatives: 1)
minimize the upper bound by setting it equal to the lower-bound (all
Linux 2.6.0 faults are new), 2) maximize the upper bound by extending
the lifespan up to Linux 1.0 and 3) set the upper bound by extending
the lifespan up to the median point between Linux 1.0 and Linux 2.6.0,
which is in January 1999, near the release of Linux 2.2.0.

\begin{figure}[tp]
  \centering
  \includegraphics[width=.8\textwidth]{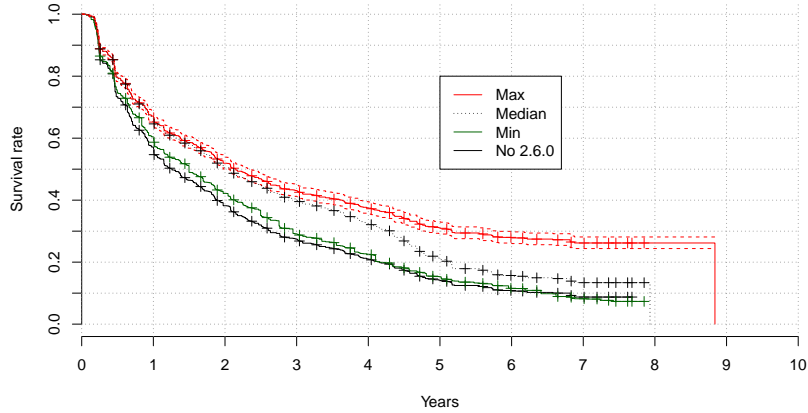}
  \caption{Kaplan-Meier survival analysis of a fault}
  \label{fig:kaplan-meier-min-max}
\end{figure}

Figure~\ref{fig:kaplan-meier-min-max} reports the survival analysis of
faults, \npnote*{R ref. added}{computed with R~\cite{R-project}},
according to four configurations: first, ignoring the faults of Linux
2.6.0 and then each of the three alternatives mentioned above. The
first configuration, illustrated in black at the bottom left, only
considers the faults introduced after Linux 2.6.0, as they are at worst
only right-censored. The second one, the solid green line in the
middle, considers that all Linux 2.6.0 faults are new, and thus gives
the lower bound of 0 as the estimation. The third one, in red at the
top, considers that the Linux 2.6.0 faults may have been introduced in
Linux 1.0, and thus gives the upper bound of the estimation. Finally,
the fourth one, in dotted black in the middle, gives a median
estimation where the Linux 2.6.0 faults may have been introduced at
any time between January 1999 and December 2003 (Linux 2.6.0). This
may extend the lifespan of the faults of Linux 2.6.0 by up to five
years with respect to the observed lifespan.

\npnote*{New.This adds explanations about
  Figure~\ref{fig:kaplan-meier-min-max}.}{The faults considered in the
  first configuration have an average lifespan of 2.3 years
  {\textpm}~22 days, and a median lifespan of 1.3 years. As shown in
  Figure~\ref{fig:kaplan-meier-min-max}, this is the lowest lifespan
  observed. The faults, for which the birth is observed, are thus
  fixed quicker than the set of all the faults, \textit{i.e.}, the set
  that includes the faults already present in Linux 2.6.0 (as
  considered in the remaining configurations). For the second and
  third configurations, respectively the min and max strategies, the
  average lifespan ranges from 2.4 to 3.7 years, and the median
  lifespan from 1.5 to 2.2 years. The strategy for left-censoring thus
  has an impact of up to about 1 year. For the median estimation
  (fourth configuration), the average lifespan of a fault is 2.9 years
  {\textpm} 19 days, and the median lifespan is 2.1 years. This
  strategy affects the average lifespan more than the median
  lifespan. We can thus conclude that \hl{the average lifespan is about
  between 2.5 and 3.0 years and the median lifespan is about 2
  years.}%  At the rate of a Linux release every 3 months, and
  % considering fixes are properly backported, it takes 8 major
  % releases to fix half of the faults in a stable branch.
}

\begin{figure}[tp]
  \centering
  \includegraphics[width=.8\textwidth]{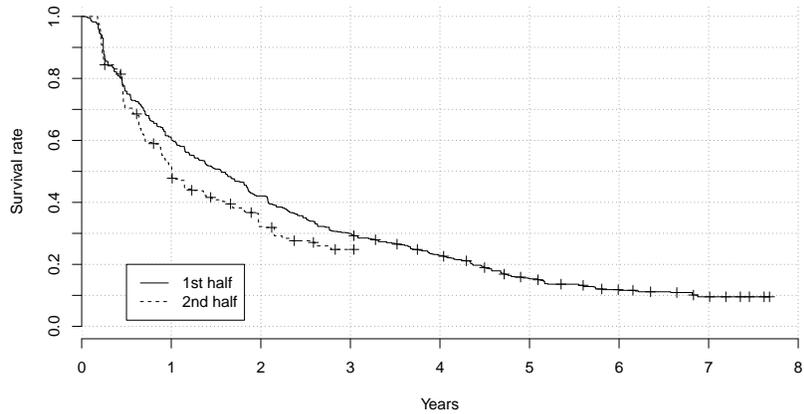}
  \caption{Kaplan-Meier survival analysis comparison}
  \label{fig:kaplan-meier-improvement}
\end{figure}

Figure~\ref{fig:kaplan-meier-improvement} compares the lifespans of
faults introduced in the first half of the versions (with the first
configuration, \textit{i.e.}  ignoring the Linux 2.6.0 faults) with
the lifespans of faults introduced in the last half of the
versions. We observe an average lifespan of 2.41 years and 2.60 years, for
respectively the first and the last half of the versions.  These
results show that \hl{the average lifespan has increased. However, the
median lifespan has decreased from 1.56 to 1.00 years.}

In Figures~\ref{fig:kaplan-meier-min-max}
and~\ref{fig:kaplan-meier-improvement}, each cross corresponds to a
right-censored fault. The 0.95 confidence band is given as a pair of dotted
lines in Figure~\ref{fig:kaplan-meier-min-max}, for the upper-bound
estimation. This band is rather tight and is similar for all of the curves
(not shown for the sake of clarity).

\paragraph*{Lifespan per directory and fault kind}

\begin{figure}[t!p]
  \centering
  \begin{tabular}{@{}c@{}}
  \subfigure[Per directory]{
  \begin{minipage}{.8\linewidth}
      \includegraphics[width=\linewidth,keepaspectratio]{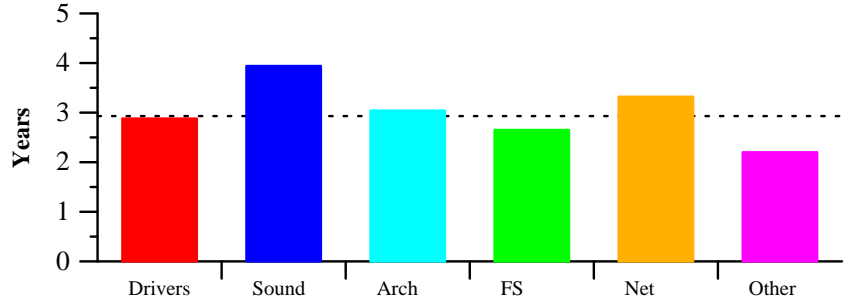}
     \vspace{.5em}
    \end{minipage}
   \label{fig:avg-ages-per-dir}
  }
%% \\
%%   \subfigure[Per fault kind]{
%%   \includegraphics[width=0.8\linewidth,keepaspectratio]{avg-ages}
%%   \label{fig:avg-ages-per-kind}
%%   }
\\[10mm]
  \subfigure[Per finding and fixing difficulty, and impact likelihood]{
    \begin{minipage}{.8\linewidth}
    \includegraphics[width=\linewidth,keepaspectratio]{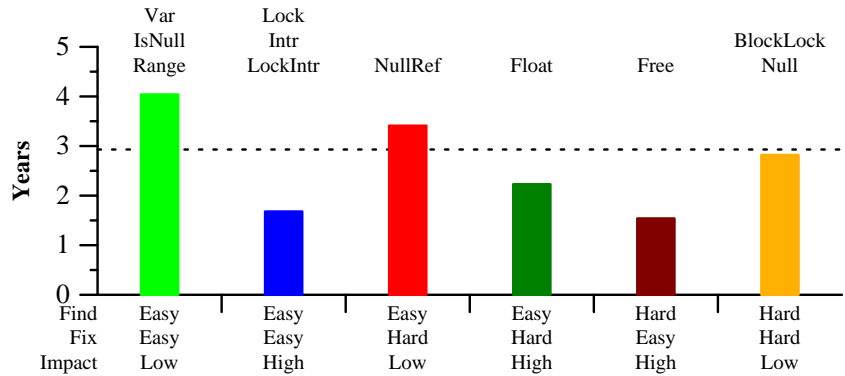}
    \vspace{.5em}
    \end{minipage}
  %   \begin{tabular}{|l|l|l||l|}
%   \hline
%   Find&Fix&Impact&Checkers\\
%   \hline
%   Easy&Easy&Low & Var, IsNull, Range \\
%   Easy&Easy&High & Lock, Intr, LockIntr, Size \\
%   Easy&Hard&Low & NullRef \\
%   Easy&Hard&High & Float \\
%   Hard&Easy&High & Free \\
%   Hard&Hard&Low & Block, Null \\\hline
%   \end{tabular}
  \label{fig:avg-ages-per-kind-difficulty}
  }
  \end{tabular}
  \caption{Average fault lifespans (without staging)}
  \label{fig:avg-ages}
\end{figure}

Figure \ref{fig:avg-ages} presents the average lifespan of faults
across Linux 2.6, by directory and by fault kind.  We omit {\tt
  drivers/\-staging} because it was only introduced recently. The
lifespan averages are reported by the Kaplan-Meier estimator with the
median strategy.
The average fault lifespan across all files of Linux 2.6 is 2.93
years, as indicated by the horizontal dotted line in
Figure~\ref{fig:avg-ages}. The lifespans vary somewhat by directory.
As shown in Figure \ref{fig:avg-ages-per-dir}, \hl{the average lifespan of
faults in the {\tt drivers} directory is the same as the average
lifespan of all faults, and indeed is less than the average lifespan
of faults in the {\tt sound} directory.}  {\tt Sound} faults now have
the longest average lifespan.  {\tt Sound} used to be part of {\tt
  drivers}; it may be that the sound drivers are no longer benefiting
from the attention that other drivers receive. In
Figure~\ref{fig:evol-dir} of Section~\ref{where}
(page~\pageref{fig:evol-dir}), we showed that the fault rates of
\texttt{fs} and \texttt{arch} are now worse than that of
\texttt{drivers}. Here, we find similarly that \hl{faults in {\tt arch}
have a longer lifespan than faults in {\tt drivers}. However, the
faults in \texttt{fs} have the second smallest lifespan.}% This is
% certainly due to the fact that faults in a filesystem could corrupt
% user data, and thus that they must be quickly fixed. 

For the fault kinds, Figure \ref{fig:avg-ages-per-kind-difficulty}
shows that the average lifespans correspond roughly to our assessment
of the difficulty of finding and fixing the faults and their
likelihood of impact (Table \ref{crash}).  In particular, all of the
fault kinds we have designated as having high impact, meaning that the
fault is likely to have an observable effect if the containing
function is executed, have the lowest average lifespans. Moreover,
\hl{among the high impact faults, the ones that are easily fixed have the
lowest average lifespans. On the other hand, the ease of finding the
faults has little impact on their lifespan}, showing that developers
are willing to invest in tracking down any faults that cause obvious
problems, and are prone to quickly accept simple fixes.

\begin{figure}[t!p]
  \centering
  \begin{tabular}{@{}c@{}}
    \subfigure[Cumulative number]{
      \begin{minipage}{1.0\linewidth}
        \includegraphics[width=0.9\linewidth,keepaspectratio]{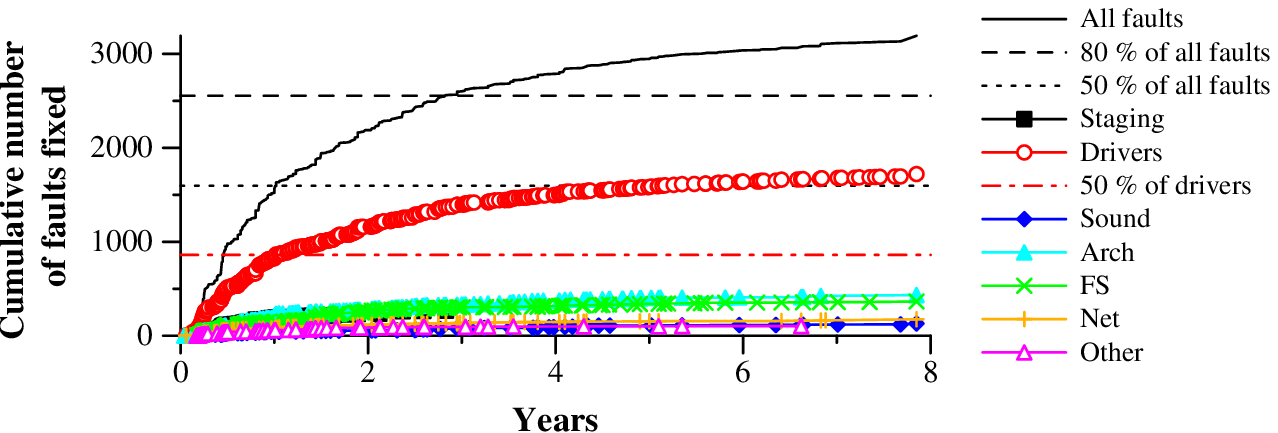}
        \vspace{.5em}
      \end{minipage}
      \label{fig:cumul-time-to-fix}
    }
    \\[10mm]
    \subfigure[Percentage]{
      \begin{minipage}{1.0\linewidth}
        \includegraphics[width=0.9\linewidth,keepaspectratio]{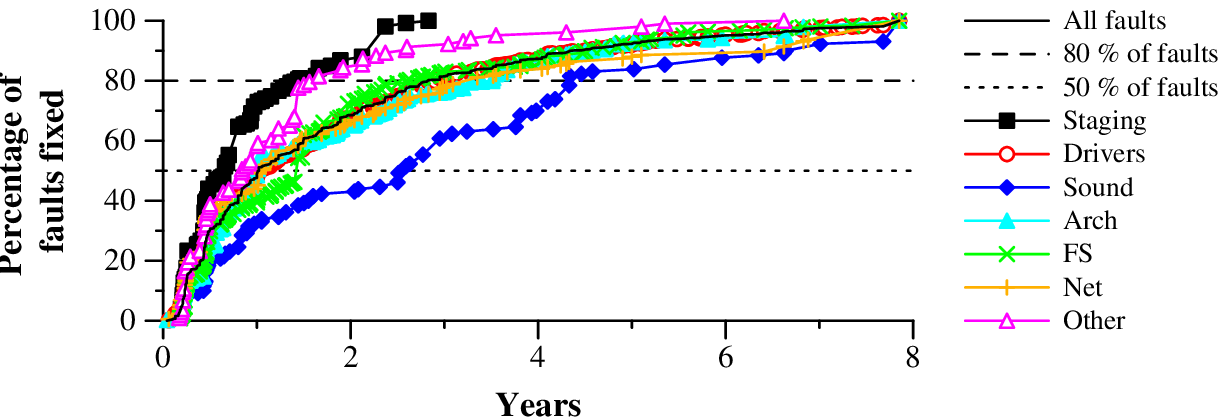}
        \vspace{.5em}
      \end{minipage}
      \label{fig:pct-time-to-fix}
    }
  \end{tabular}
  \caption{Fixed faults per amount of elapsed time}
  \label{fig:time-to-fix}
\end{figure}

Figure \ref{fig:time-to-fix} examines fault lifespans in more detail,
by showing the number of faults (Figure \ref{fig:cumul-time-to-fix})
and the percentage of faults (Figure \ref{fig:pct-time-to-fix}) that
have been fixed in less than each amount of time. In these figures, we
include the faults already present in Linux 2.6.0 and the faults still
remaining in Linux 3.0. \texttt{Staging} is also added back to compare
it with the other directories, but as it was introduced only two and a
half years before 3.0, no \texttt{staging} fault can have a lifetime
longer than 2.5 years. If we consider all of the faults, half of them
were fixed within just above one year, but it took about 3 years to
fix 80\% of them. This global trend is observed for almost all of the
directories. Indeed, half of the faults were fixed in around one year
for each subsystem, except \texttt{sound} where 2.5 years were
required to fix half of the faults. In 2010, we did a similar study,
reported in a preliminary version of this paper~\cite{Palix:asplos11},
and found for \texttt{staging} that half of the faults were fixed
within 6 months. We now observe that the time is almost 8
months. These drivers may thus have lost the attention of code
reviewers as compared to the time at which \texttt{staging} was
created.

%% Checked the 2013-06-12
% SELECT
% info."study_dirname","life_in_years","cumulative_corrected_bugs","percentage_of_corrected_bugs", info.min
% FROM 
%  ( SELECT "study_dirname","life_in_years","cumulative_corrected_bugs","percentage_of_corrected_bugs",
%   min(abs("percentage_of_corrected_bugs" - 50)) as  min
%   FROM "public"."study_fig_cumulative_bugs_per_life_dir"
%   GROUP BY "study_dirname"
%   ,"life_in_years","cumulative_corrected_bugs","percentage_of_corrected_bugs"
%   ORDER BY min) as info,
%  (SELECT "study_dirname",
%   min(abs("percentage_of_corrected_bugs" - 50)) as  min
%   FROM "public"."study_fig_cumulative_bugs_per_life_dir"
%   GROUP BY "study_dirname"
%   ORDER BY min
% ) as mid
% WHERE info."study_dirname" = mid.study_dirname
% AND info.min = mid.min

\paragraph*{Origin of faults}

\begin{figure}[p!]
  \centering
 \includegraphics[height=.93\textheight,keepaspectratio]{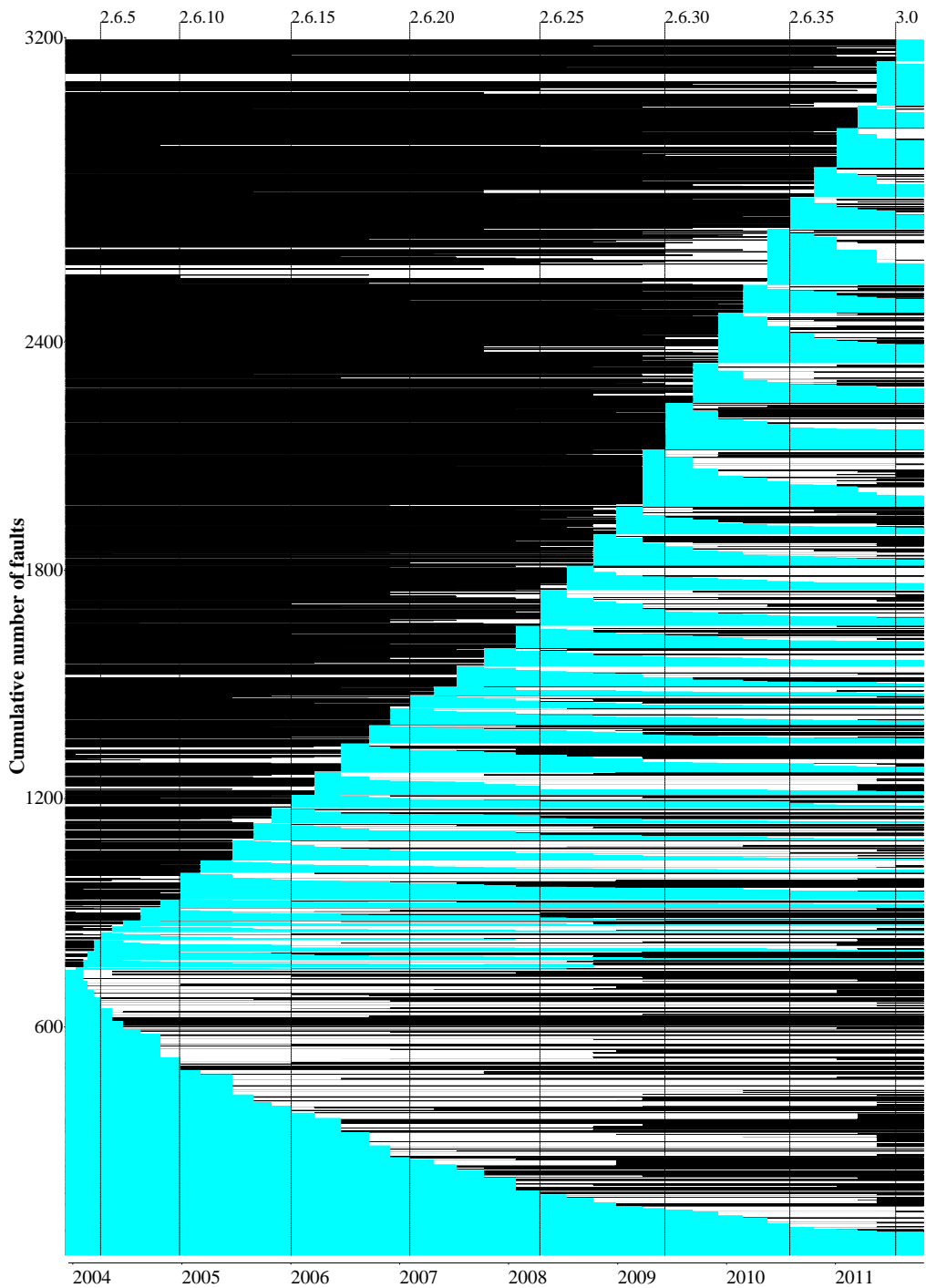} %%width=\linewidth
  \caption{Lifetime of faults.  Each row represents a separate fault.  Blue
    (grey) lines indicate the periods where the fault is present.  Black
    lines indicate the period where the file containing
    the fault does not exist and white lines indicate the period where
    the file does exist but the fault does not.
  }
  \label{fig:lifetime}
\end{figure}

\npnote*{Reviewer 2 asked for sorting algorithm}{Figure
  \ref{fig:lifetime} shows the lifetime of each of the faults found in
  our study.\footnote{The details of this figure may be more apparent
    on a printout than on a monitor.}  The faults are ordered by birth
  date, with faults of Linux 2.6.0 at the bottom of the graph and new
  faults of Linux 3.0 at the top. For each version, the faults are
  further ordered by their death age.} The 751 faults in Linux 2.6.0
are marked as introduced at that point, even though they may have been
introduced earlier.  \hl{Of the faults introduced and eliminated within
the period considered, 35\% of the faults introduced in or after Linux
2.6.0 were introduced with the file and 12\% of the faults eliminated
before Linux 3.0 were eliminated with the file. These percentages have
remained stable as compared to the ones of our previous
study~\cite{Palix:asplos11}.}

%% SELECT count("correlation_id") FROM "public"."full_bug_correlations"
%% WHERE "correlation_birth_version" = 'linux-2.6.0'
%% => 753 (2013-06-18)

%% SELECT count("correlation_id") FROM "public"."full_bug_correlations"
%% WHERE "correlation_birth_version" = 'linux-2.6.0'
%% AND standardized_name != 'BlockRCU'
%% and standardized_name != 'DerefRCU'
%% and standardized_name != 'LockRCU'
%% => 751 (2013-06-18)

%% SELECT fb.standardized_name, count("correlation_id")
%%  FROM "public"."full_bug_correlations" fb, bug_categories b
%%  WHERE "correlation_birth_version" = 'linux-2.6.0'
%%  AND fb.standardized_name = b.standardized_name
%%  GROUP BY fb.standardized_name, b.standardized_name
%%  ORDER BY b.standardized_name_id

%% Overall bugs
%% SELECT count("correlation_id") FROM "public"."full_bug_correlations" 
%% 3229

%% New with file
%% SELECT count("correlation_id") FROM "public"."full_bug_correlations" 
%% WHERE "correlation_birth_version" = "file_birth_version"
%% 1157 - 35%

%% SELECT count("correlation_id") FROM "public"."full_bug_correlations" 
%% WHERE "correlation_death_version" = "file_death_version"
%% 389 - 12%

% Checked 15 jan 2011
% Graph gr/occ.jgr
% Working on; Linux-2.6
% 1344; file(s) affected by; 2383; bug(s)
%
% 1.77; bug(s) per file
% 875; bug(s) (36%) introduced by a new file.
% 308; bug(s) (12%) removed by a file deletion.
% 1; version(s) for the shortest bug life.
% 34; version(s) for the longest bug life.
% 9.47; version(s) in average for a bug life.
% 14; days for the shortest bug life.
% 2260; days for the longest bug life.
% 643; days in average for a bug life.
% 1469 ; 61.64 % ; drivers
%  315 ; 13.22 % ; arch
%  291 ; 12.21 % ; fs
%  138 ; 5.79 % ; net
%  110 ; 4.62 % ; sound
%   18 ; 0.76 % ; kernel
%   12 ; 0.50 % ; mm
%    8 ; 0.34 % ; tools
%    7 ; 0.29 % ; lib
%    5 ; 0.21 % ; security
%    3 ; 0.13 % ; virt
%    3 ; 0.13 % ; block
%    2 ; 0.08 % ; include
%    1 ; 0.04 % ; Documentation
%    1 ; 0.04 % ; ipc

\begin{figure}[t!p]
  \centering
  \includegraphics[width=.9\linewidth,keepaspectratio]{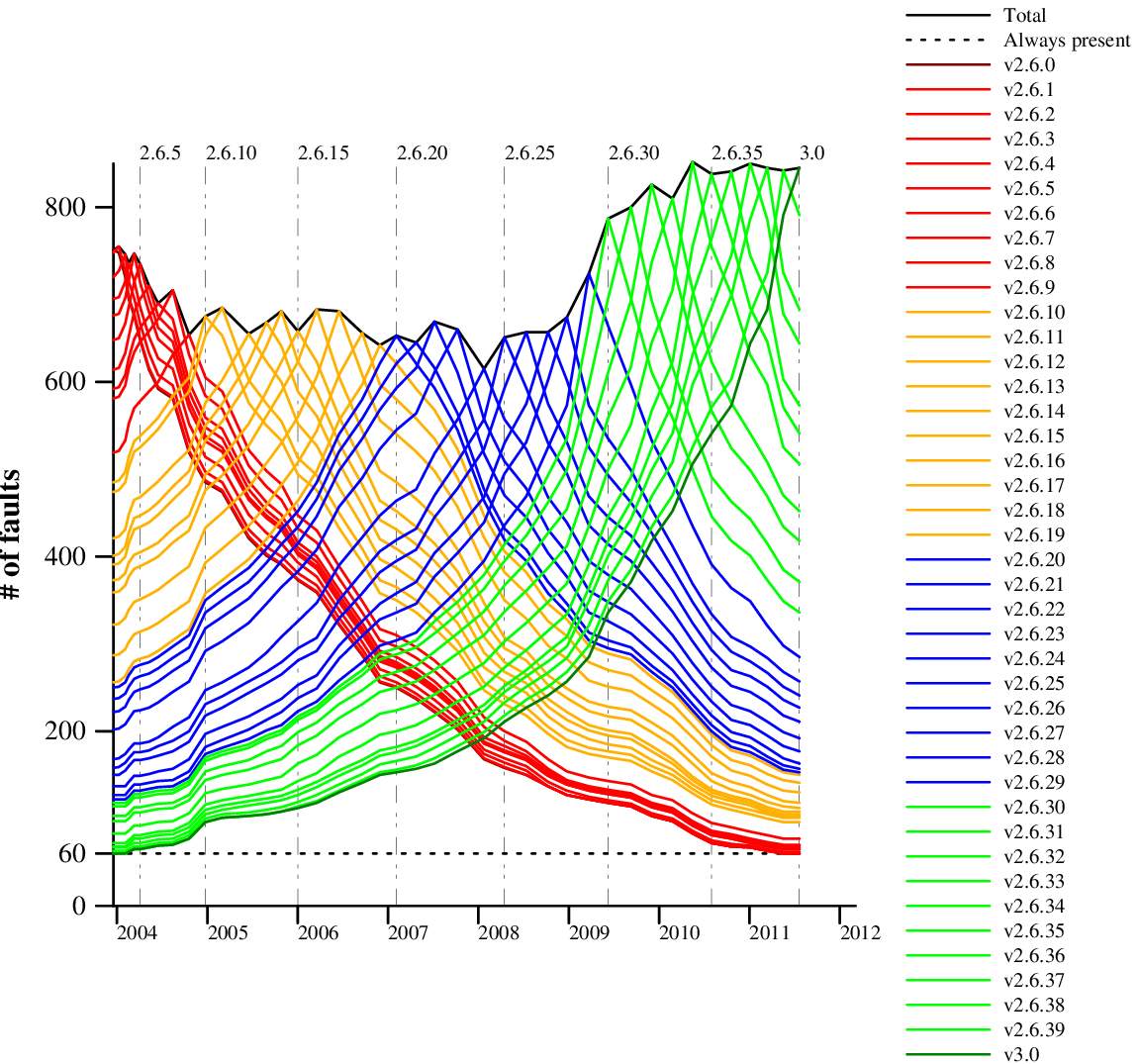}
  \caption{Lifetime of faults across versions}
  \label{fig:lifetime_across}
\end{figure}

While we have seen that the total number of faults is essentially
constant across the versions, Figure \ref{fig:lifetime_across} shows
that since Linux 2.6.27, a significantly larger number of faults have
been introduced.  {\bf Null} and {\bf NullRef} faults predominate,
with for example 51\% of the added faults in Linux 2.6.27 being {\bf
  NullRef} faults, most of which were introduced in various drivers.
In Linux 2.6.30 and Linux 2.6.32, 43\% and 26\% of the introduced
faults, respectively, were in {\tt drivers/staging}.  In each case,
about half of the introduced faults were fixed within a few versions.

\npnote*{Rephrased for clarity}{Figure \ref{fig:lifetime_across} shows
  the number of faults in each version (highest point of each line)
  that are also present in each of the previous and successive
  versions.  For instance about 200 faults of Linux 2.6.0 are still
  present in Linux 2.6.23, as seen by following the Linux 2.6.23 curve
  backward until Linux 2.6.0 was released. Linux 2.6.23 furthermore
  has more than 600 faults, and about 200 of these faults remain until
  Linux 3.0, as seen by following the curve forward up to the Linux
  3.0 release.}  Except for the increase at Linux 2.6.29 and 2.6.30,
as previously noted, the height of all of the lines is fairly similar,
indicating that the rate of introductions and eliminations of faults
across the versions is relatively stable.  These facts indicate a
maturity in the Linux code and its development model. However, the
angle of the lines is becoming sharper indicating the improvement of
fault correction, as previously reported in
Figure~\ref{fig:kaplan-meier-improvement}.

\paragraph*{Use of tools}

\begin{figure}[t!p]
  \[\includegraphics[width=\linewidth,keepaspectratio]{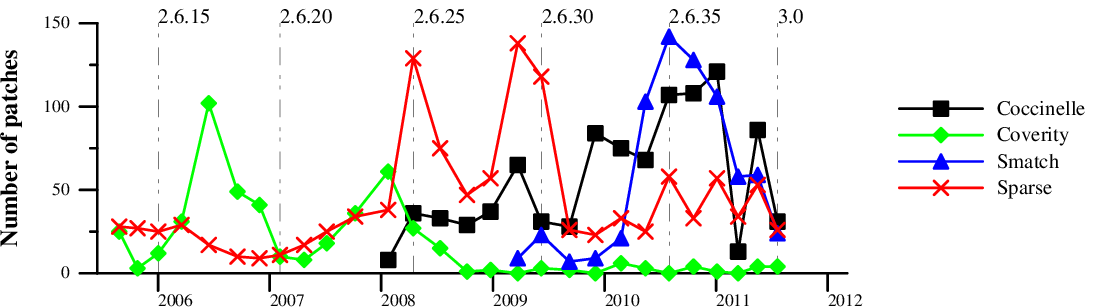}\]
\caption{Tool usage since the introduction of git (Linux 2.6.12)}
\label{fig:tool-usage}
\end{figure}

In principle, since the work of Chou {\em et al.}, it has been
possible to find all of the considered faults using tools.  Figure
\ref{fig:tool-usage} shows the number of patches in each Linux version
that mention one of the fault-finding tools Coccinelle
\cite{Padioleau:eurosys08} (used in this paper), Coverity
\cite{coverity} (the commercial version of Chou {\em et al.}'s {\tt
  xgcc} tool), Sparse \cite{sparse,Sparse-web}, and Smatch
\cite{smatch}.  As developers are not obliged to mention the tools
they use, these results may be an
underestimation. \npnote*{Moved}{However, \hl{we observe a Pearson's
  correlation coefficient of 0.641 between the number of tool-based patches and
  the elimination of faults}, which are reported respectively in
  Figure~\ref{fig:tool-usage} and
  Figure~\ref{fig:evol-birth-and-death}.}

The use of the various tools is somewhat variable across the different
versions.  In the case of Coverity, the main use occurred between 2006 and
2009, when its application to open-source software, including Linux, was
funded by the US Department of Homeland Security~\cite{coverity-dhs-end}.
Sparse usage shows several large peaks, at 2.6.25 and at 2.6.29-30, and
smaller ones at 2.6.35, 2.6.37 and 2.6.39. For 2.6.25, 32 developers used
Sparse to produce 129 patches, but half of these patches were produced by
only 3 developers.  For versions 2.6.29 and 2.6.30, a single developer was
responsible for 140 of the 253 Sparse-related patches.  This developer
fixed faults across the entire kernel.  Each of the three small peaks comes
from the use of Sparse by a single developer, who produced between 10 and
25 patches each time, among about 20 other developers. This developer was
different for each peak. The tools have also been used to find a wider
range of faults than those considered in this paper.  For example, in Linux
2.6.24, only about half of the patches that mention Coverity relate to the
kinds of faults we consider, particularly {\bf Null}, {\bf IsNull}, {\bf
  NullRef}, and {\bf Free}.  Overall, despite the variability in usage, the
results show a willingness on the part of the Linux developers to use
fault-finding tools and to pay attention to the kinds of faults that they
find.

\jlnote*{New paragraph about self-impact.}{Around the time of our
  preliminary study of faults in Linux 2.6 \cite{Palix:asplos11}, we
  contributed a number of patches using Coccinelle based on our findings.
  These include patches on faults found in released versions, but also
  faults introduced between versions, as found using the head of the Linux
  git repository.  While these patches made it possible to validate our
  results, it also meant that our own patches may interfere with the
  results on each version reported here.  During the period of Linux 2.6.30
  and Linux 3.0, which contains the largest number of uses of Coccinelle,
  we find that just under 22\% of the patches that refer to Coccinelle, by
  ourselves or others, fix the kinds of faults considered in this paper.
  Most of these faults are related to \texttt{NULL} pointers.  In contrast,
  the patches submitted using Coccinelle typically address memory leaks and
  modernizing various APIs, which are not considered here.  Thus, we
  consider that our contributions to the Linux kernel have only had a small
  impact on the results presented in this paper.}

%% Hannes Eder had 140 sparse patches in 2.6.29 and 2.6.30 combined
%% He had 144 sparse patches overall
%% Harrison had 38 sparse patches in 2.6.25 which is the most of any single
%% person, but is small as compared to the total of 126.
%% He had 102 sparse related patches overall.

%% Probably no point to mentioned the following.  Coverity is a commercial
%% tool, and if no one is paying for its use any more, it is not surprising
%% that it is not used.
%  \np{ Finally, we observe that Coverity was mainly used
%   between 2006 and 2009, and is relatively less referenced in recent
%   years. We suppect the reason is twofold: first of all, Coverity got
%   a contract \cite{coverity-dhs} to analyse the LAMP architecture from
%   2006 to 2009 \cite{coverity-dhs-end}, secondly, it is suggesting a
%   reluctance of open source developers to rely on closed source tools
%   having restrictive licenses.-- Should we say it was with the Department
% of Homeland Security ?}

%%% Local Variables:
%%% mode: LaTeX
%%% TeX-master: "faults-in-linux-2.6-tocs"
%%% coding: utf-8
%%% TeX-PDF-mode: t
%%% ispell-local-dictionary: "american"
%%% End:

\section{Code quality predictability in Linux 2.6}
\label{sec:fault-factors}

In the software-engineering community, substantial work has been done
on identifying metrics that predict code quality
\cite{Munson:ICSM98,Nagappan:ICSE05,bird2009does-icse}. Among these,
we consider human-related factors and how the code is structured and
evolves.

\subsection{Human factors}
\label{sec:human-factors}

We first evaluate the developer activity in the Linux project, and
then how developer commitment is related to the introduction of
faults.

\paragraph*{Developer activity}

\begin{figure}[tp!]
  \centering
  \subfigure[Authors per directory]{
    \includegraphics[width=.9\linewidth,keepaspectratio]{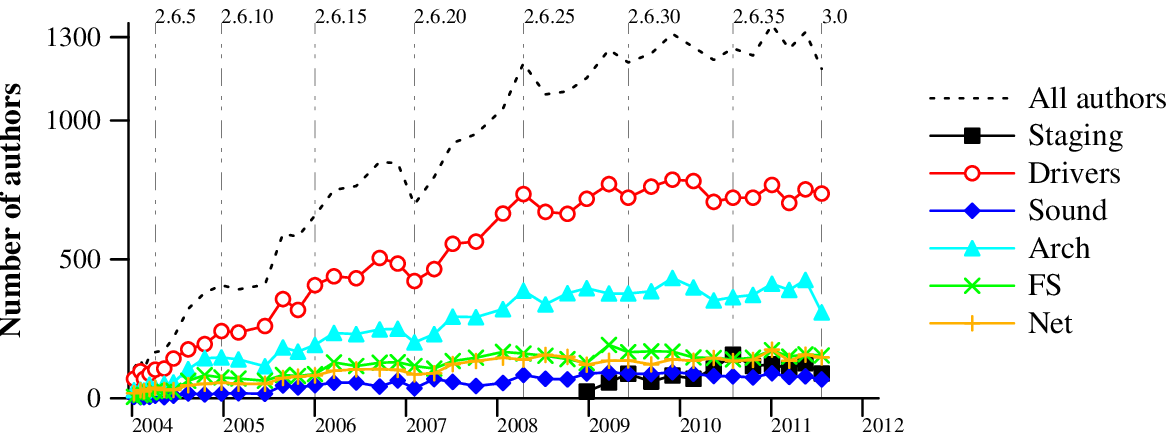}
    \label{fig:author}
  }
  \subfigure[Committers per directory]{
    \includegraphics[width=.9\linewidth,keepaspectratio]{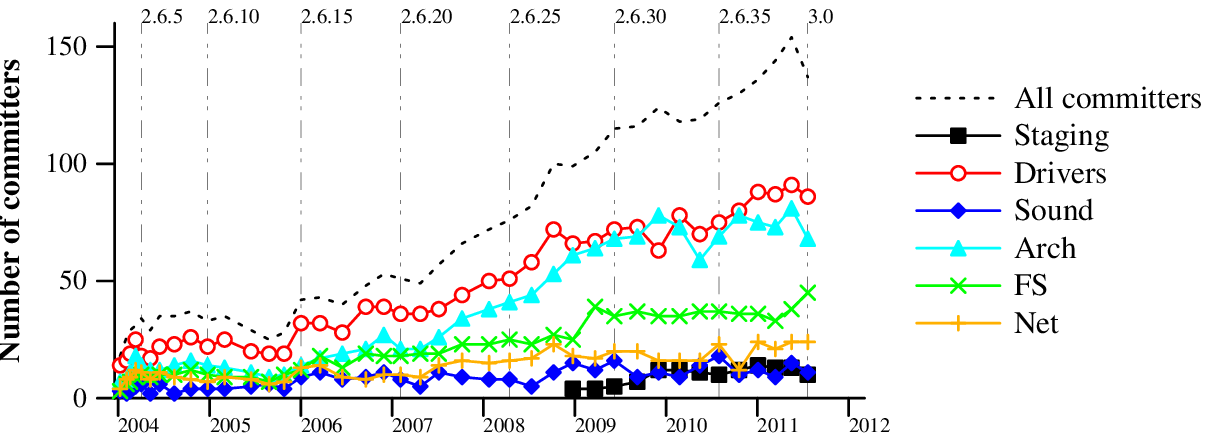}
    \label{fig:committer}
  }
  \caption{Authors and committers per version}
  \label{fig:author-committer}
\end{figure}

In the Linux development model, anyone (an {\em author}) can submit a
patch to a \emph{maintainer} and to relevant mailing lists for public
review. If the patch is accepted, it is then picked up by a
\emph{committer} (often the \emph{maintainer}), who {\em commits} it
into his git repository. The committer's git repository is then
propagated to Linus Torvalds, who makes a release. The number of patch
authors is thus an indicator of the number of participants in the
Linux development process, and the number of committers is an
indicator of the amount of manpower that is actively available to
begin the integration of patches into a release.
Figure~\ref{fig:author-committer} shows the number of authors and
committers associated with the patches included in each version, both
in total and broken down by directory.

\hl{For {\tt drivers}, the numbers of authors and committers are rising at
the same rate, roughly at the rate of the increase in the code size.
For {\tt arch} and {\tt fs}, however, where we have previously noted a
higher fault rate (Figures \ref{fig:evol-dir} and
\ref{fig:comparative-rate-per-dir}), the number of authors is rising
significantly more slowly than the number of
committers.} \npnote*{Reviewer 3}{The lower number of authors may
  suggest that potential authors are not able to develop adequate
  expertise to keep up with the number of new architectures and file
  systems (Figure~\ref{fig:code-increase}). The fewer authors engage,
  the less review there is on code.}  Finally, the small number of
{\tt sound} authors may explain the previously observed long life of
{\tt sound} faults (Figure~\ref{fig:avg-ages-per-dir}).

\paragraph*{Developer commitment}

The more a developer works on the Linux project, the more expertise
the developer may be assumed to acquire. We thus study developer
commitment as an approximation of developer expertise.

\begin{figure}[t!p]
  \centering
  \includegraphics[width=.9\linewidth,keepaspectratio]{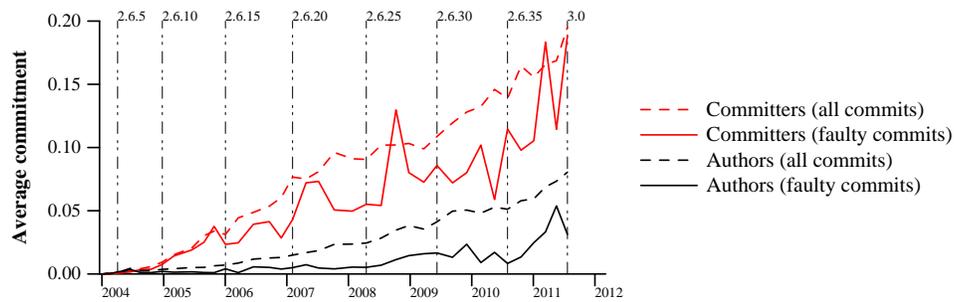}
  \caption{Developer commitment}
  \label{fig:exp-evol}
\end{figure}

To quantify the commitment of \emph{authors} and \emph{committers}, we
take the product of the number of commits integrated into the Linux
kernel and the period of time (in days) over which the commits have
occurred. \jlnote*{Changed. Reviewer asked for clarification.}{In
  doing so, a developer with many recent commits is considered to have
  the same degree of commitment as a developer with fewer commits over
  a longer period of time. For instance, an author who gets 12 commits
  integrated during one year, will have the same degree of commitment
  as some one who gets 24 commits integrated in 6 months.
  Figure~\ref{fig:exp-evol} gives the evolution of the average
  commitment for each release and for authors and commiters, showing
  both the value obtained by considering all the commits of that
  release, and for all the faulty ones, \emph{i.e.}, the commits that
  introduce one of the faults we found}.

Not surprisingly committer commitment is higher than author
commitment. Both forms of commitment progress almost constantly over
time. This indicates that there is little turnover among the
committers and authors. However, \hl{for faulty commits, the average
commitment of committers varies greatly even if it is constantly
greater than the average commitment of authors. Finally, the average
author commitment in the case of faulty commits is significantly lower
that the average author commitment for all commits. This indicates
that new authors with few commits are more likely to propose faulty
patches. As the average committer commitment is also generally lower
for faulty commits than the average committer commitment for
all commits, these faulty patches are applied.}

% SELECT author_name from authors join (select "author_id" FROM
% "public"."history" order by "author_expertise"  desc limit 1)
% commitment using (author_id)

\begin{figure}[t!p]
  \centering
  \includegraphics[width=.9\linewidth,keepaspectratio]{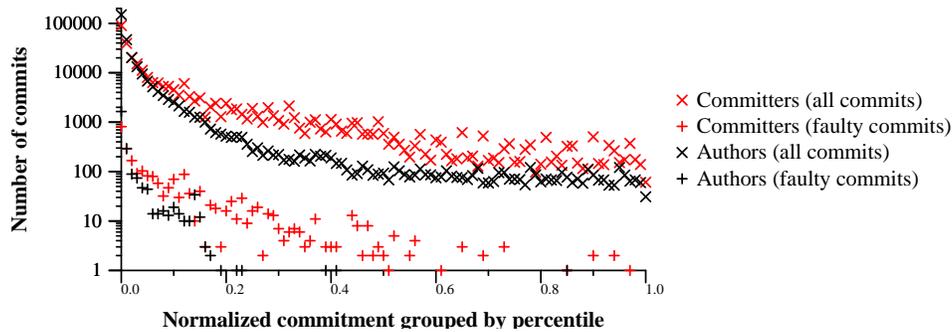}
  \caption{Developer and maintainer distribution}
  \label{fig:exp-distrib}
\end{figure}

Figure~\ref{fig:exp-distrib} shows the distribution of the commits
according to the commitment of the authors and the committers. Two
commitment values are defined for each commit: one for its author, the
other for its committer. These commitments are normalized with respect
to the maximal value across the entire period studied, \textit{i.e.},
the computed commitment of Linus Torvalds for the latest commit of
Linux 3.0. Finally, each commitment value is rounded and assigned to a
percentile of the normalized scale. Commits belonging to the same
percentile are grouped and counted. So, each point represents a set of
commits with similar author commitment, respectively committer
commitment. The size of the set is given on the y-axis.

Once again, for commits in general, the commitment of committers is
higher than that of authors, which gives us some confidence in this
metric.\hl{ Concerning faulty commits, more than 99.5\% of them are
proposed by authors who rank below 15\%. However, the commitment of
the committers is more spread across the entire scale.}

%%% Local Variables:
%%% mode: LaTeX
%%% TeX-master: "faults-in-linux-2.6-tocs"
%%% coding: utf-8
%%% TeX-PDF-mode: t
%%% ispell-local-dictionary: "american"
%%% End:

\newcommand{\lightred}{light/red}

\subsection{Code structure and evolution}
\label{sec:distrib}

We consider three metrics: code churn, file age, and function size,
and their relation to the number of faults.  We also evaluate the
quality of Linux code in light of the conjecture that open source code
is more reliable because it can be examined by many people.

% In the software engineering community, substantial work has been done on
% identifying metrics that predict code quality.  We consider three
% possible metrics: code churn, file age, and function size.  We also
% evaluate the quality of Linux code in light of the conjecture that open
% source code is more reliable because it can be examined by many people.

\paragraph*{Churn}
 
\begin{figure}[t!p]
  \centering
  \includegraphics[width=.8\linewidth,keepaspectratio]{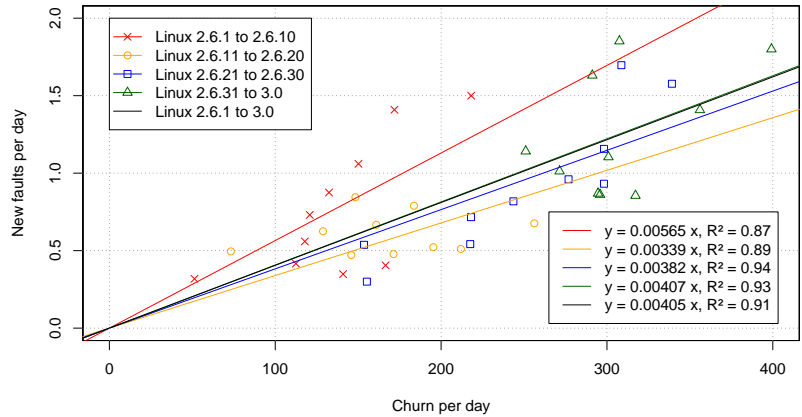}
%%  \includegraphics[width=.8\linewidth,keepaspectratio]{figures/faults_churn}

  %% \includegraphics[width=\linewidth,height=\textheight,keepaspectratio]{deaths_churn}

  %% \subfigure[Correlation between churn and fault rate]{
  %% \includegraphics[width=\linewidth,height=\textheight,keepaspectratio]{flt_rate_churn}
  %% \label{fig:faults_churn_rate}
  %% }

\caption{Churn vs.\ new faults}
\label{fig:faults_churn_overview}
\end{figure}

\citeN{Munson:ICSM98} observed that code churn, {\em i.e.}, the number
of times a file is modified, is a good predictor of fault rate.
\citeN{Nagappan:ICSE05} reached a similar conclusion in a study that
used metrics relating to the development of Windows Server 2003 to
predict Service Pack 1's fault rate.  Figure
\ref{fig:faults_churn_overview} shows the relationship between the
average churn per day preceding the release of each Linux version and
the number of the considered faults added per day in that version.
The relationship between churn and fault rate is similar.  There is an
overall tendency of high-churn versions to contain more new faults,
even if some high-churn versions have a smaller number of faults than
lower churn versions.  Recall, however, that in many versions, more
faults were eliminated than added; the churn includes the elimination
of faults as well.

% http://stat.ethz.ch/R-manual/R-patched/library/stats/html/summary.lm.html
\npnote*{Reviewer 2 requested trend lines.}{The solid lines in
  Figure~\ref{fig:faults_churn_overview} represent the linear
  regression of each group of versions and the overall trend.  The
  slope of the trend lines gives the rate of new faults per day as a
  function of the churn in a day.  The legend in the bottom right
  gives the equations used to produce these lines and the
  \textit{adjusted $R^2$}~\cite{R-summary-lm}, as computed by
  R~\cite{R-project}. Each formula is of the form $y = ax + b$, where
  $a$ is the slope, and $b$ is 0, representing the fact that the value
  at $x = 0$, where there is no churn and thus no faults are
  introduced, is naturally $0$.  The adjusted $R^2$ represents the
  goodness of the fit between the model and the actual data. Its value
  goes up to 1 for a perfect fit, and can be negative, as outliers
  introduce a penalty.

  The Linux versions are studied in four packets of ten Linux versions
  grouped by release order. For the first packet, from Linux 2.6.1 to
  2.6.10, the rate is the highest with an average churn of 138
  modifications per day. In the second packet, the rate has greatly
  improved and is indeed the lowest of the four ones. At the same
  time, the churn increased to 168 modifications per day. From Linux
  2.6.21 to 2.6.30, the third packet, the rate of new faults per day
  has increased but the average churn per day in those versions is
  also higher (251 modifications/day). Finally, \hl{in the last packet,
  the rate of new faults per day has slightly increased, but it is now
  almost identical to the overall rate between Linux 2.6.1 and
  3.0. However, the churn per day has further increased to 309
  modifications/day.}}

%% In the case of Linux, the git source code
%% manager \cite{git} makes it possible to collect the set of patches that
%% affect a given file between a pair of versions.\footnote{We have used the
%%   command \texttt{git log v2.6.12..v2.6.33 --oneline {\it{file}}}, for each
%%   {\it{file}}.}  This information, however, is only available from Linux
%% 2.6.12, when git was adopted as the version control system for Linux.

%% \begin{figure}[tb!]
%%   \centering
%%   \includegraphics[width=\linewidth,height=\textheight,keepaspectratio]{churn}
%%   \caption{Distribution of the file churns per directory}
%%   \label{fig:churn}
%% \end{figure}

%% Figure \ref{fig:churn} shows the amount of churn for the header files
%% and {\tt .c} files in each of the {\tt net}, {\tt fs}, {\tt drivers}
%% (including the {\tt staging} subdirectory), and {\tt arch} directories
%% of Linux 2.6.33.  Except for the {\tt net} directory, we observe that
%% over 50\% of the files have been modified 10 or fewer times between
%% Linux 2.6.12 and Linux 2.6.33.  Indeed, although drivers are often
%% considered to be unreliable, and are {\em e.g.}, reported by Chou {\em
%%   et al.} to harbor up to 7 times as many Block faults as files in
%% other directories, we observe that there is less churn in the {\tt
%%   drivers} directory than in {\tt fs} and {\tt net}.

\paragraph*{File age and function size}

\begin{figure}[t!p]
  \centering
  \subfigure[Fault rate by file age in Linux 3.0]{
    \includegraphics[width=.9\linewidth,keepaspectratio]{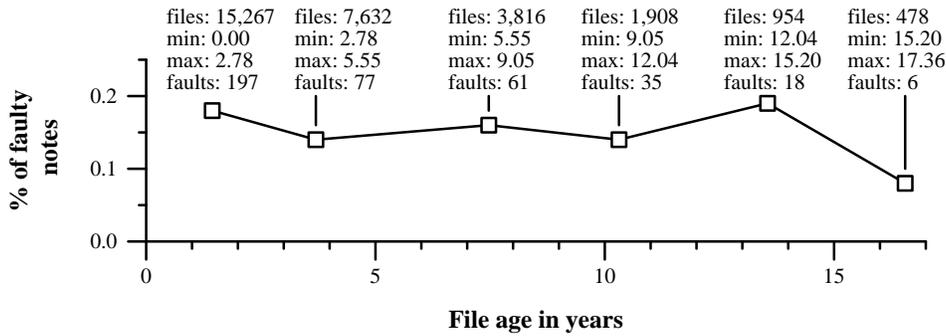}
  \label{fig:rate-by-age}
  }

  \subfigure[Fault rate by function size in Linux 3.0]{
    \includegraphics[width=.9\linewidth,keepaspectratio]{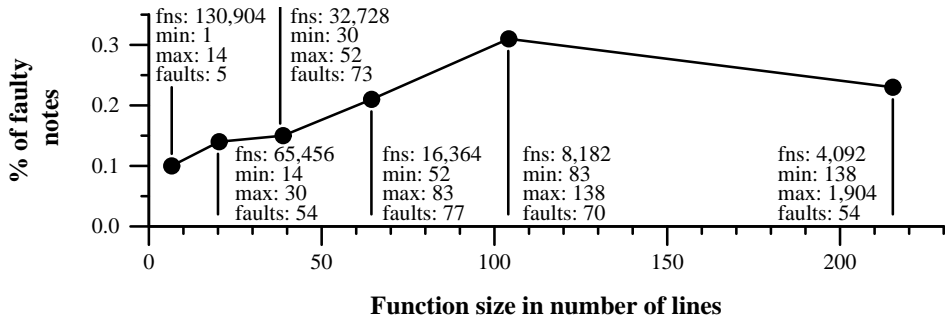}
  \label{fig:err-by-fct-size}
  }

%   \subfigure[Fault rate by file age or function size in 2.6.33]{
%   \includegraphics[width=\linewidth,height=\textheight,keepaspectratio]{figures/error-rate-by-X}
%   \label{fig:rate-by-X}
%   }

  \caption{Impact of file age and function size.  Each point
    represents twice as many files or functions as the next point.  At
    each point, ``files'' and ``fns'' refer respectively to the number
    of files and functions considered to compute the point, ``min''
    and ``max'' refer to the minimal or maximal age or size
    represented by the point, and faults indicates the number of
    faults that occur in the code in that range.}
\label{fig:err-by-age-size}
\end{figure}

One may expect that as a file ages the number of faults would
decrease. One may also expect that large functions would be more
complicated and thus tend to harbor more faults.  Indeed, Chou
\emph{et al.} found these trends in Linux 2.4.1, to a varying degree
for the different fault kinds.  Figure \ref{fig:err-by-age-size}
considers the relationship between file age or function size and fault
rate in Linux 3.0.  Files and functions are organized by increasing
age or size, respectively, then collected into buckets containing an
exponentially decreasing number of elements, from the smallest age or
size to the largest ({\em cf.}\ Section~\ref{buckets}). This strategy
permits a fine degree of granularity for the files with greater ages
or sizes, respectively, without degrading the results for the smaller
ages and sizes. \npnote*{Add explanation about buckets.}{Note that we
  change the bucket strategy with respect to the file age, compared to
  Figure~\ref{fig:rate-by-age-4-241}, page
  \pageref{fig:rate-by-age-4-241}. We thus reflect the changes in the
  file age distribution, as shown in
  Table~\ref{tab:file-age-distrib-30}, and the observations of
  Figure~\ref{fig:line-age}, page~\pageref{fig:line-age}.} Each graph
shows the average age or size of the files or functions in each bucket
and their average fault rate.

\begin{table}[h!tp]
\tbl{File age distribution for Linux 3.0}
{  \centering
  \begin{tabular}{| l |c|c|c|c|c|c|c|c|c|} \hline
    Age (in years) & 0 & 1 & 2 & 3 & 4 & 5 & 6 & 7 & 8  \\\hline
    Number of files & 
    5,934 & 3,826 & 7,989 & 2,220 & 1,943 &  1,522 & 858 &   638  &  1,487\\ \hline \hline

    Age (in years) & 9 & 10 & 11 & 12 & 13 & 14 & 15 & 16 & 17 \\ \hline
    Number of files & 
    1,276 & 265 & 662 & 301 & 311 & 251 & 208 & 123 &  241 \\ \hline
\end{tabular}
}
  \label{tab:file-age-distrib-30}
\end{table}

Figure \ref{fig:rate-by-age} shows that in Linux 3.0, the
youngest half of the files, represented by the leftmost point, has a
fault rate about twice that of the next youngest quarter of the files.
For older files, however, the relation between age and fault rate is
less clear, as the rate first increases and then decreases as the file
age increases.
%
% Fcts: 257 726
%
On the other hand, the average fault rate clearly increases as the
function size increases (Figure \ref{fig:err-by-fct-size}).  Indeed,
\hl{the bucket with the smallest functions (up to 14 lines) has a
significantly lower fault rate than the next buckets.  There is also
an increase up to 138 lines.  This suggests that large functions, with
about a hundred lines of code, need more attention than others, even
if they represent only 3\% of the functions.} 3\% of the functions in
Linux 3.0, however, amounts to about 8,000 functions, making them
difficult to check exhaustively.  Tool support is thus crucial.

For Linux 2.4.1, we observed in
Figure~\ref{fig:err-by-fct-size-4-241-log6b-all_fcts}
(page~\pageref{fig:err-by-fct-size-4-241-log6b-all_fcts}) that the
functions found in the fourth bucket had the highest fault rate at
1.1\% of the faulty notes. \hl{Compared to Linux 2.4.1, the maximum is for
larger functions in Linux 3.0, and in the fifth bucket. This maximum
has dropped to 0.3\% of the faulty notes. Finally, as observed for
Linux 2.4.1, the largest functions, represented by the rightmost
bucket, have a slightly smaller fault rate than those of around 100
lines.}

\paragraph*{Kernel configuration}
\label{sec:kernel-config}

In \emph{``The Cathedral and the Bazaar''} \cite{cathedral-bazaar}, Eric
S. Raymond formalized Linus' law: ``given enough eyeballs, all bugs are
shallow.'' But, in practice, code that is frequently executed, or at least
frequently compiled, is more likely to be reviewed than the rest.  When
code is frequently executed, many users are likely to encounter any faults
and some may fix the faults themselves or submit a request that the faults
be fixed by a kernel maintainer.  When code is frequently compiled, even if
it is not frequently executed, it can easily be submitted to fault-finding
tools that are integrated with the kernel compilation process.
``Eyeballs'' may also focus on fixing faults in code that they are able to
compile, as even standard compilers such as {\tt gcc} perform some sanity
checks that provide some confidence that a fix has not {\em e.g.},
introduced a typographical error, even if the code
cannot be tested.

% select standardized_name, allyes_compiled, count(report_id) as faults
% from full_reports join files using(file_id)
% where full_reports.version_name = 'linux-3.0' and
% (standardized_name = 'Intr' or standardized_name ='Lock')
% group by standardized_name, allyes_compiled;

% select standardized_name , allyes_compiled, count(note_id) as notes
% from full_notes
% where (standardized_name = 'Lock' OR standardized_name = 'Intr')
% and version_name = 'linux-3.0'
% group by allyes_compiled, standardized_name;

% From 'rates' for Lock
% FALSE    2     2898    0.06915629322268326418
% TRUE   13   14332    0.09070611219648339380

% allyes_compiled	standardized_name	rate
% FALSE	Lock	0.07320644216691068814
% TRUE	Lock	0.09070611219648339380

\begin{figure}[t!p]
     \centering
  %% \subfigure[Number of faults for 3.0]{
  %%   \includegraphics[width=.95\linewidth,height=\textheight,keepaspectratio]{count-4-30-x86}
  %%   \label{fig:where-count-30-x86}
  %% }

  %% \subfigure[Rate of faults for 3.0]{
     \includegraphics[width=.9\linewidth,height=\textheight,keepaspectratio]{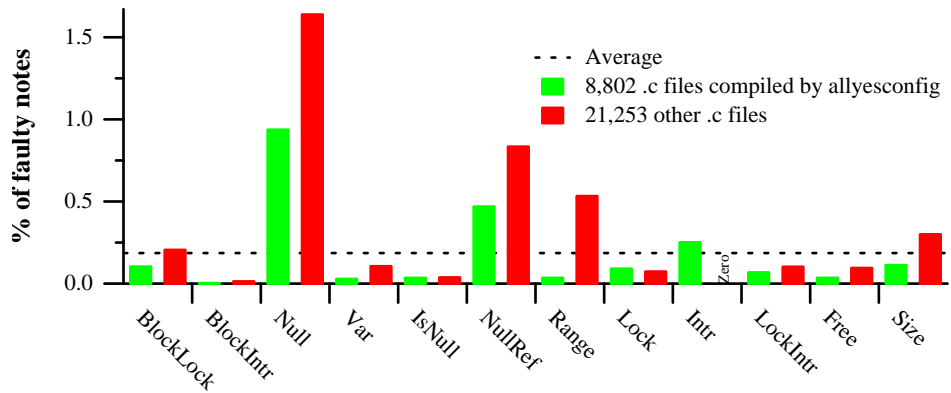}
    %%\label{fig:where-rate-30-x86}
  %%}
\caption{Fault rate compared between configurations (Linux 3.0)}
\label{fig:where-dft-x86}
\end{figure}

Figure~\ref{fig:where-dft-x86} compares the number of faults found in
the {\tt .c} files that are compiled using the configuration generated
on an x86 architecture by the Linux Makefile argument {\tt
  all\-yesconfig} to the number of faults found in the {\tt .c} files
found in the rest of the Linux kernel.\footnote{\scriptsize An x86\_64
  machine was used for the configuration. The compilation was done
  with \texttt{make} 3.82 and \texttt{gcc} 4.5.4.}  The Makefile
argument {\tt allyesconfig} creates a configuration file for the given
architecture that includes as many options as possible without causing
a conflict and without including {\tt drivers/staging}.  Thus, it can
be assumed to trigger the compilation of a set of well-tested files
and a superset of what is normally included with a Linux distribution,
and thus what is executed by ordinary users.  \hl{In most cases, we do
find that the {\tt allyesconfig} files have a lower fault rate than
the other files. } The only exceptions are for \textbf{Lock} and {\bf
  Intr} faults, but there is only a single fault in the case of
\textbf{Intr}. For \textbf{Lock}, 13 out of 15 \textbf{Lock} faults
are in files included in the \texttt{allyesconfig} configuration, and
this configuration contains only 5 times more \textbf{Lock} notes than the
other files. The fault rate of \textbf{Lock} notes in
\texttt{allyesconfig} files is thus higher than that in other
files.

%% \jlnote{In Figure 29, does the above average green bar for Intr really
%%   represent only one fault? Are there very few notes?}
%% \npnote{Checked in orgfile/db the 2013-12-2: single Intr fault for
%%   1820 notes in 3.0}

%%% Local Variables:
%%% mode: LaTeX
%%% TeX-master: "faults-in-linux-2.6-tocs"
%%% coding: utf-8
%%% TeX-PDF-mode: t
%%% ispell-local-dictionary: "american"
%%% End:

\section{Towards new code versions}
\label{sec:towards-new-code}

% \npwarning{RCU was introduced in the 2.5.43 release (16-Oct-2002). It
%   is one year before 2.6.0}
% commit 1477a825d7e6486a077608c7baf6abbb6f27ed95
% Author: Dipankar Sarma <dipankar@in.ibm.com>
% Date:   Tue Oct 15 05:40:46 2002 -0700

%     [PATCH] Read-Copy Update infrastructure

In this section, we first consider some new checkers related to the use of
the RCU locking API, which was introduced in 2.5.43, about one year before
the release of 2.6.0. Then, we consider how our experimental protocol eases
the extension of the results to new Linux versions.

\subsection{How does the fault rate in new APIs compare}

\begin{figure}[t!p]
  \centering
  \includegraphics[width=0.9\linewidth,keepaspectratio]{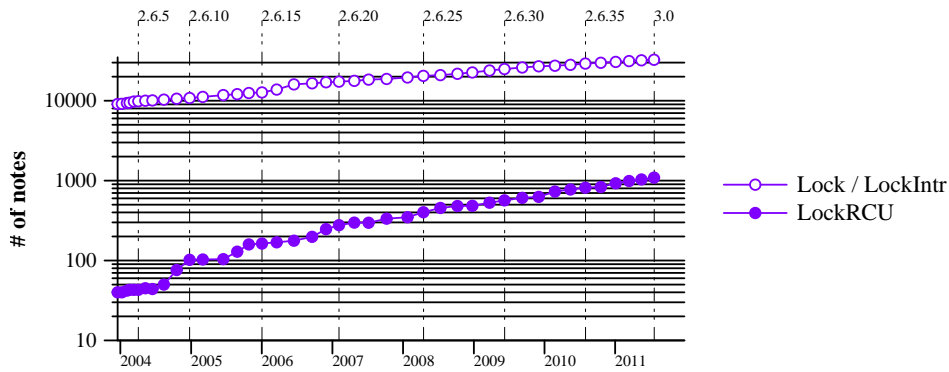}
  \caption{The use of spinlock and mutex locking functions and
    the use of RCU locking functions}
\label{fig:rcu-notes-evol}
\end{figure}

The fault kinds that we have considered until now involve code structures
that are either common to all C code or that have been present in Linux for
a long time.  To give an alternate perspective on the robustness of Linux
2.6 code, we consider the use of a more recent and specialized locking API:
the one that implements the Read-Copy-Update (RCU) mechanism
\cite{McKenney:osr08}.  RCU is a lightweight synchronization mechanism that
protects readers against writers. Reads are wait-free with very low
overhead while writes are more expensive. Therefore, RCU particularly
favors workloads that mostly read shared data rather than updating it.  RCU
has been increasingly used in Linux 2.6, but is still used less often
than spinlocks and mutexes, as shown in Figure~\ref{fig:rcu-notes-evol}.

% graph contains Lock + LockIntr vs RCULock notes

The main functions that we consider in the RCU API are the locking
functions: {\tt rcu\_\-read\_\-lock}, {\tt srcu\_\-read\_\-lock}, {\tt
  rcu\_\-read\_\-lock\_\-bh}, {\tt rcu\_\-read\_\-lock\_\-sched}, and {\tt
  rcu\_\-read\_\-lock\_\-sched\_\-notrace}.  We identify as faults any
cases where a blocking function is called while a RCU lock is held ({\bf
  BlockRCU}) and where a RCU lock is taken but not released ({\bf
  LockRCU}).  These are analogous to the previously considered {\bf
  BlockLock} and {\bf Lock} fault kinds, respectively.  Double-acquiring an
RCU lock is allowed, and thus this is not considered to be a fault.
Finally, we identify as a fault any case where shared data is accessed
using {\tt rcu\_\-dereference} when an RCU lock is not held ({\bf
  DerefRCU}).

% SELECT "version_name", count("note_id")
% FROM "public"."full_notes"
% WHERE "standardized_name" = 'LockRCU'
% GROUP BY "version_name", release_date
% ORDER BY release_date

%% SELECT "version_name", study_dirname, count("note_id")
%% FROM full_notes
%% WHERE "standardized_name" = 'LockRCU'
%% GROUP BY version_name, release_date, study_dirname
%% ORDER BY release_date

\begin{figure}[t!p]
  \centering
  \includegraphics[width=0.9\linewidth,keepaspectratio]{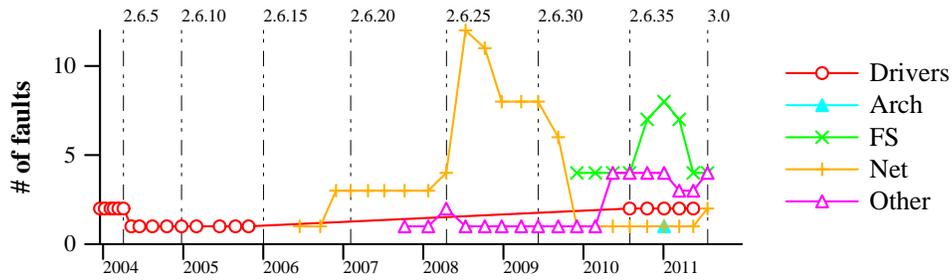}
  \caption{RCU faults per directory}
\label{fig:rcu-faults-per-dir}
\end{figure}

\begin{figure}[t!p]
  \centering
  \includegraphics[width=0.9\linewidth,keepaspectratio]{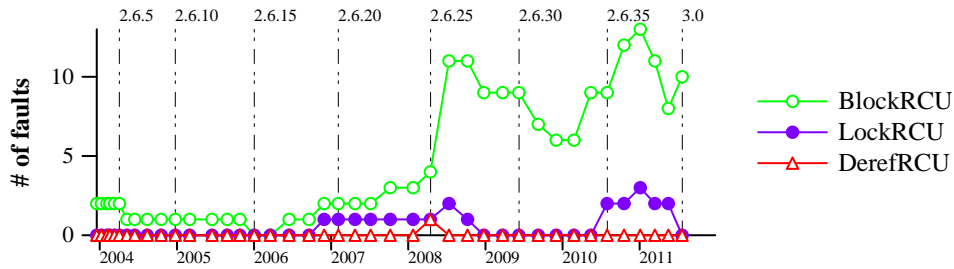}
  \caption{RCU faults through time}
\label{fig:rcu-count-evol}
\end{figure}

The largest number of uses of the RCU lock functions is in {\tt net},
followed by {\tt drivers} in versions earlier than 2007, and {\it
  other} in later versions, with {\tt kernel}, {\tt security} and
\texttt{mm} being the main directories in the \textit{other} category
where it is used.  Correspondingly, as shown in
Figure~\ref{fig:rcu-faults-per-dir}, most faults are found in {\tt
  net}, and the few remaining faults are found in {\it other}.  The
average lifespan of the {\tt net} faults is around one year, while
that of the {\it other} faults is a few months longer.  As shown in
Figure~\ref{fig:rcu-count-evol}, most of the faults are {\bf BlockRCU}
faults, and the largest number of faults per version is only 13, in
Linux 2.6.37.  Thus, overall, the fault rates are substantially below
the fault rates observed for any of the previously considered fault
kinds, and in particular far below the rates observed for the various
locking faults ({\bf Lock}, {\bf Intr}, and {\bf LockIntr}).  This may
be explained by the fact that the developers who have added calls to
the various RCU locking functions are relatively experienced, having
at the median 123 patches accepted between Linux 2.6.12 and Linux
2.6.33.  Still, these results suggest that \hl{Linux developers can be
successful at adopting a new API and using it correctly.}
%\npfatal*{To update the 123 value for 2.6.0 to 3.0. How was it measured?}

% \begin{figure}[h]
%   \centering
% \includegraphics[width=0.9\linewidth,keepaspectratio]{figures/avg-ages-per-dir-rcu}
%   \caption{Average age for RCU}
%   \label{fig:avg-ages-rcu}
% \end{figure}

% \begin{figure}[h]
%   \centering
%   \includegraphics[width=0.9\linewidth,keepaspectratio]{figures/birth-and-death-rcu}
%   \caption{Birth and Death for RCU}
% \label{fig:birth-death-rcu}
% \end{figure}

% \begin{figure}[h]
%   \centering
%   \includegraphics[width=0.9\linewidth,keepaspectratio]{count-evol-lock-vs-lock-rcu}
%   \caption{Lock versus RCU locks faults through time}
% \label{fig:rcu-count-evol-lock-vs}
% \end{figure}

% \begin{figure}[h]
%   \centering
%   \includegraphics[width=0.9\linewidth,keepaspectratio]{rate-4-2633-x86-rcu}
%   \caption{RCU fault rate for x86}
% \label{fig:rcu-rate-2633-x86}
% \end{figure}

% \begin{figure}[h]
%   \centering
%   \includegraphics[width=0.9\linewidth,keepaspectratio]{cmp-rate-per-dir-4-2633-rcu}
%   \caption{Comparison of RCU fault rate per dir. for 2.6.33}
% \label{fig:cmp-rcu-rate-2633}
% \end{figure}

%%% Local Variables:
%%% mode: LaTeX
%%% TeX-master: "faults-in-linux-2.6-tocs"
%%% coding: utf-8
%%% TeX-PDF-mode: t
%%% ispell-local-dictionary: "american"
%%% End:

\subsection{Processing New Versions}
\label{sec:tools}

% Numbers verified 29/10/2013

\begin{figure}
  \centering
  \includegraphics[width=.9\linewidth,keepaspectratio]{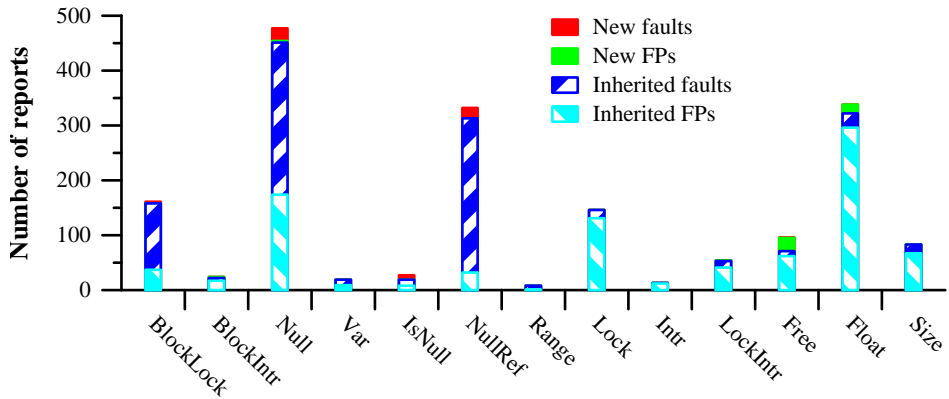}
  \caption{Reports generated for Linux 3.0}
  \label{fig:reports-linux-30}
\end{figure}

As presented in Section \ref{meth}, the automatic report correlations
provided by Herodotos make it easy to extend an existing set of
results to the next version of Linux.  As an experiment, we have studied
the extent to which the annotated reports for Linux 2.6.39 can help
evaluate the reports for Linux 3.0. For Linux 3.0, as shown in
Figure \ref{fig:reports-linux-30}, the \textbf{BlockLock} checker
produces 161 reports of potential faults of which 121 are faults
inherited from Linux 2.6.39, and 37 are false positives also present
in that version.  Herodotos would annotate these reports
automatically, leaving only three reports to be annotated by the user.
In this case, the three reports are faults.  Overall, for Linux 3.0
there are 100 new reports to consider.  Most of these can be dealt
with in a few minutes each.

% % For inherited
% SELECT standardized_name, status, count(correlation_id)
% FROM full_correlations
% WHERE linux_le(correlation_birth_version, 'linux-2.6.39')
% AND correlation_death_version is null
% AND status in ('BUG', 'FP')
% GROUP BY standardized_name, status
% ORDER BY standardized_name, status

% % For new
% SELECT standardized_name, status, count(correlation_id)
% FROM full_correlations
% WHERE correlation_birth_version = 'linux-3.0'
% AND status in ('BUG', 'FP')
% GROUP BY standardized_name, status
% ORDER BY standardized_name, status

%%% Local Variables:
%%% mode: LaTeX
%%% TeX-master: "faults-in-linux-2.6-tocs"
%%% coding: utf-8
%%% TeX-PDF-mode: t
%%% ispell-local-dictionary: "american"
%%% End:

%\newpage
\section{Limitations}
\label{sec:limitations}

The main limitations of our work are in the choice of faults considered and
the definition of the checkers.  We have focused on the same kinds of
faults as Chou {\em et al.}, to be able to assess the changes in Linux
since their work.  Current fault finding tools, including
Coccinelle, are able to find other kinds of faults, such as memory leaks.
The considered set of faults also does not include concurrency faults,
which are becoming increasingly important with the increasing prevalence of multicore
architectures.  A recent study of concurrency faults in infrastructure
software, however, has shown that over 20\% of deadlocks are caused by a
thread reacquiring a resource it already holds~\cite{Lu:ASPLOS08},
amounting to a double lock, as detected by our {\bf Lock} checker.

Our checkers could also be improved to reduce the number of false
positives. For instance, the {\bf Float} checker generates a large
number of false positives, as shown for Linux 3.0 in
Figure~\ref{fig:reports-linux-30} in the previous section. Some kinds
of false positives are due to recurring patterns specific to certain
Linux files.  Taking these patterns into account in the checkers would
avoid generating large numbers of trivial false positive reports.
Finally, we have tried to be conservative in our identification of
real faults, and this may have lead to an underestimation of their
number.  By making our results available in a public archive, we hope
to benefit from feedback from the Linux community to improve our
classification strategies.

%%% Local Variables:
%%% mode: LaTeX
%%% TeX-master: "faults-in-linux-2.6-tocs"
%%% coding: utf-8
%%% TeX-PDF-mode: t
%%% ispell-local-dictionary: "american"
%%% End:

\section{Related Work}
\label{sec:related}

\citeN{Engler:empirical:SOSP01} briefly considered OpenBSD, as well as
Linux. Because of its wider use and more active development, we have
focused only on Linux instead, comparing the properties of old and new
versions.

In previous work, Palix {\em et al.}~have used Coccinelle to conduct a
study of faults in versions of Linux and several other open-source projects
released between 2005 and 2009~\cite{palix:aosd10}.  They proposed
Herodotos, in order to correlate the faults between releases.  They did not
consider fault kinds requiring an interprocedural analysis, nor did they
consider lock-related faults.  As shown in Figure~\ref{fig:evol}, these are
among the most prevalent and also have high impact.  \citeN{Lawall:spe13}
proposed a methodology for finding interprocedural faults in Linux code,
but consider only a single Linux version.

\citeN{Israeli:JSS10} have studied 810 versions of the Linux kernel,
from Linux 1.0 to Linux 2.6.25.  They considered traditional
source-code metrics to measure complexity~\cite{McCabe:76} and
maintainability~\cite{Oman:JSS94}, rather than actual numbers of
faults.  They found that the complexity per Linux function is
decreasing, and the maintainability is increasing.  Nevertheless, they
did find that {\tt arch} and {\tt drivers} contain some very high
complexity functions, typically interrupt handlers or {\tt ioctl}
functions, and that {\tt arch} and {\tt drivers} code is somewhat less
maintainable than the code in other parts of the kernel.  Their work
is complementary to ours, and reaches some of the same conclusions.

\citeN{Song:DSN2010} have studied the reasons for software hangs in
open-source infrastructure software, such as MySQL and Apache.  They
focused on bug reports rather than analysis of the source code.  They
found that most types of concurrency faults are fixed on average
within 100 days.  Nevertheless, they did not take into account the
time elapsed between the introduction of the fault and the time when
it was first detected, and so the actual fault lifetime may be more in
line with what we have observed.  \citeN{Lu:ASPLOS08} also considered
concurrency faults in infrastructure software, primarily focusing on
the kinds of tools that would be helpful to address them.

\textsc{undertaker}
\cite{tartler2012revealing,DBLP:conf/msr/NadiDTHL13} finds
inconsistencies between the Linux configuration files and the
associated conditional compilation directives ({\em e.g.},
\texttt{\#ifdef}) present in the code. The faults that they consider
are thus orthogonal to the ones considered here. Using Herodotos, they
found an average fault lifespan of 6 Linux releases
\cite{DBLP:conf/msr/NadiDTHL13}. However, they do not seem to have
used the Kaplan-Meier estimator, and thus the censoring of the last
version studied does not seem to be correctly handled. As
\textsc{undertaker} focuses on configuration faults and Coccinelle
focuses on code faults, it would be interesting to integrate the
results obtained by the tool in the future, to obtain a more complete
view of the quality of Linux code.

Hector~\cite{saha2013hector} is a tool that checks protocol usage and
reports missing function calls in error-handling code. It thus finds a
narrower range of faults than those considered here, but it includes a
dataflow analysis that allows it to collect more precise information than
Coccinelle.  It could be interesting to conduct a similar study of the
history of the kinds of faults found using Hector.

We have noted that file systems have a higher fault rate than drivers in
some recent Linux versions (Figure~\ref{fig:evol-dir}, in
Section~\ref{where}).  Some recent work has focused on issues in file
systems.  \citeN{Rubio-Gonzalez:PLDI09} study the use of error codes in
file systems. \citeN{LuEtAl13-fsstudy-login} study the patches applied to
six major file systems, to understand the file systems' evolution.  Part of
this study includes categorization of the kinds of faults that are fixed by
these patches; some faults correspond to the kinds that we have considered,
however they consider only the faults that have been fixed, while our focus
is on those that remain in the code.  \citeN{Yang:2006} applied model
checking to find a small number of faults in file system code, including
some severe ones.  More recently, \citeN{Keller:PLOS13} have proposed a
methodology for designing a file system that is formally proved correct.
Our work is largely complementary to these studies.

%% A tale of four kernels:
%% http://doi.acm.org/10.1145/1368088.1368140
%% \jlfatal*{Fixme}{From 2008}

%% Evolution in Open Source Software: A Case Study:
%% http://www.st.cs.uni-saarland.de/edu/empirical-se/2006/PDFs/godfrey00.pdf
%% \jlfatal*{Fixme}{This paper actually dates from 2000, so it is not clear that we have
%%   something to say about it.}

%% The evolution of FreeBSD and linux:
%% http://doi.acm.org/10.1145/1159733.1159765
%% \jlfatal*{Fixme}{From 2006, abstract suggests it is mostly focused on code size.}

%% Put a sentence about churn and Microsoft.

%%% Local Variables:
%%% mode: LaTeX
%%% TeX-master: "faults-in-linux-2.6-tocs"
%%% coding: utf-8
%%% TeX-PDF-mode: t
%%% ispell-local-dictionary: "american"
%%% End:

\section{Conclusion}
\label{sec:concl}

During the last 10 years, much of the research in operating system
reliability has been predicated on the assumption that drivers are the
main problem.  The first major result of our study is that while {\tt
  drivers} still has the largest number of faults in absolute terms,
it no longer has the highest fault rate in Linux kernel code, having
been supplanted by the HAL.  The second major result of our study is
that even though faults are continually being introduced, the overall
code quality is improving.  Our work thus shows
  the importance of being able to periodically repeat the study of
  faults in source code in order to revise research priorities as the
  fault patterns change in response to research and development
  efforts for improving the quality of Linux code.  Because the
priorities of individuals and institutions change over time, the need
to repeat such studies implies that the tools and other data required
must be available in an open archival repository.

Our study also shows that tools, while used, are under-exploited.
Tools are indeed available to find all of the fault kinds considered
in this paper.  The fact that these kinds of faults remain and have a
relatively long lifespan suggests that research is needed on how to
design tools that are better integrated into the Linux development
process.  Another potential problem is the reactiveness of
maintainers. Indeed, some services have no maintainer, but remain in
the kernel source tree.  This may somewhat artificially increase the
number of faults.  Still, any such faults can impact anyone who uses
the affected code.

Our study has identified {\nbfaultsinlast} faults in Linux 3.0,
including RCU faults, some of which have not yet been corrected in the
current developer snapshot, {\tt linux-next}, as of November 5, 2013.
We have submitted a number of patches based on our results.

%%% Local Variables:
%%% mode: LaTeX
%%% TeX-master: "faults-in-linux-2.6-tocs"
%%% coding: utf-8
%%% TeX-PDF-mode: t
%%% ispell-local-dictionary: "american"
%%% End:

% Acknowledgments
\begin{acks}
  We thank the anonymous reviewers for their feedback. We also thank
  Emmanuel Cecchet, Willy Zwaenepoel and Frans Kaashoek for comments
  on an earlier version of this paper. Finally, we thank Nicolas
  Glade, Adeline Leclercq Samson and Frédérique Letué for their
  feedback on the Kaplan-Meier estimator.
\end{acks}

% Bibliography
\bibliographystyle{ACM-Reference-Format-Journals}
%\bibliography{../bib/cocci,../bib/proceedings,faults-in-linux-2.6-tocs}
\bibliography{faults-in-linux-2.6-tocs-full}


\begin{thebibliography}{00}

%%% the Web address.

\ifx \showCODEN    \undefined \def \showCODEN     #1{\unskip}     \fi
\ifx \showDOI      \undefined \def \showDOI       #1{{\tt DOI:}\penalty0{#1}\ }
  \fi
\ifx \showISBNx    \undefined \def \showISBNx     #1{\unskip}     \fi
\ifx \showISBNxiii \undefined \def \showISBNxiii  #1{\unskip}     \fi
\ifx \showISSN     \undefined \def \showISSN      #1{\unskip}     \fi
\ifx \showLCCN     \undefined \def \showLCCN      #1{\unskip}     \fi
\ifx \shownote     \undefined \def \shownote      #1{#1}          \fi
\ifx \showarticletitle \undefined \def \showarticletitle #1{#1}   \fi
\ifx \showURL      \undefined \def \showURL       #1{#1}          \fi

\bibitem[\protect\citeauthoryear{Aiken, Bugrara, Dillig, Dillig, Hackett, and
  Hawkins}{Aiken et~al\mbox{.}}{2007}]%
        {Aiken:paste07}
{Alex Aiken}, {Suhabe Bugrara}, {Isil Dillig}, {Thomas Dillig}, {Brian
  Hackett}, {and} {Peter Hawkins}. 2007.
\newblock \showarticletitle{An overview of the {Saturn} project}. In {\em
  Proceedings of the 7th ACM SIGPLAN-SIGSOFT Workshop on Program Analysis for
  Software Tools and Engineering, PASTE'07}. ACM, San Diego, CA, 43--48.
\newblock


\bibitem[\protect\citeauthoryear{Bird, Nagappan, Devanbu, Gall, and
  Murphy}{Bird et~al\mbox{.}}{2009}]%
        {bird2009does-icse}
{Christian Bird}, {Nachiappan Nagappan}, {Premkumar Devanbu}, {Harald Gall},
  {and} {Brendan Murphy}. 2009.
\newblock \showarticletitle{Does distributed development affect software
  quality? an empirical case study of windows vista}. In {\em 31st
  International Conference on Software Engineering}. IEEE, Vancouver, British
  Columbia, Canada, 518--528.
\newblock


\bibitem[\protect\citeauthoryear{Brunel, Doligez, Hansen, Lawall, and
  Muller}{Brunel et~al\mbox{.}}{2009}]%
        {Brunel:POPL09}
{Julien Brunel}, {Damien Doligez}, {Ren\'e~Rydhof Hansen}, {Julia Lawall},
  {and} {Gilles Muller}. 2009.
\newblock \showarticletitle{A foundation for flow-based program matching using
  temporal logic and model checking}. In {\em The 36th annual ACM
  SIGPLAN-SIGACT Symposium on Principles of Programming Languages}. ACM,
  Savannah, GA, USA, 114--126.
\newblock


\bibitem[\protect\citeauthoryear{CDHSE}{CDHSE}{2009}]%
        {coverity-dhs-end}
CDHSE 2009.
\newblock How is the department of homeland security involved?
\newblock   (2009).
\newblock
\newblock
\shownote{\fname{http://scan.coverity.com/faq.html\#how-department-homeland-security-involved}.}


\bibitem[\protect\citeauthoryear{Checkpatch}{Checkpatch}{2006}]%
        {Checkpatch}
Checkpatch 2006.
\newblock Checkpatch.
\newblock   (2006).
\newblock
\newblock
\shownote{\fname{http://www.codemonkey.org.uk/projects/checkpatch/}.}


\bibitem[\protect\citeauthoryear{Chou, Yang, Chelf, Hallem, and Engler}{Chou
  et~al\mbox{.}}{2001}]%
        {Engler:empirical:SOSP01}
{Andy Chou}, {Junfeng Yang}, {Benjamin Chelf}, {Seth Hallem}, {and} {Dawson
  Engler}. 2001.
\newblock \showarticletitle{An empirical study of operating systems errors}. In
  {\em Proceedings of the 18th ACM Symposium on Operating System Principles}.
  ACM, Banff, Canada, 73--88.
\newblock
\showURL{%
\url{http://www.stanford.edu/~engler/}}


\bibitem[\protect\citeauthoryear{Comedi}{Comedi}{1999}]%
        {comedi-drv}
Comedi 1999.
\newblock Comedi: {L}inux control and measurement device interface.
\newblock   (1999).
\newblock
\newblock
\shownote{\fname{http://www.comedi.org/}.}


\bibitem[\protect\citeauthoryear{Corbet}{Corbet}{2010a}]%
        {lwn374622}
{Jonathan Corbet}. 2010a.
\newblock The age of kernel code in various subsystems.
\newblock   (Feb. 2010).
\newblock
\newblock
\shownote{\fname{http://lwn.net/Articles/374622/}.}


\bibitem[\protect\citeauthoryear{Corbet}{Corbet}{2010b}]%
        {lwn374574}
{Jonathan Corbet}. 2010b.
\newblock How old is our kernel?
\newblock   (Feb. 2010).
\newblock
\newblock
\shownote{\fname{http://lwn.net/Articles/374574/}.}


\bibitem[\protect\citeauthoryear{Coverity}{Coverity}{2008}]%
        {coverity}
Coverity 2008.
\newblock Static source code analysis, static analysis, software quality tools
  by {Coverity Inc.}
\newblock \fname{http://www.coverity.com/}.   (2008).
\newblock


\bibitem[\protect\citeauthoryear{Depoutovitch and Stumm}{Depoutovitch and
  Stumm}{2010}]%
        {Depoutovitch:EuroSys10}
{Alex Depoutovitch} {and} {Michael Stumm}. 2010.
\newblock \showarticletitle{Otherworld -- giving applications a chance to
  survive {OS} kernel crashes}. In {\em EuroSys}. ACM, Paris, France, 181--194.
\newblock


\bibitem[\protect\citeauthoryear{Fedora}{Fedora}{2013}]%
        {Fedora}
Fedora 2013.
\newblock Fedora project.
\newblock   (2013).
\newblock
\newblock
\shownote{\fname{http://fedoraproject.org/}.}


\bibitem[\protect\citeauthoryear{Herder, Bos, Gras, Homburg, and
  Tanenbaum}{Herder et~al\mbox{.}}{2009}]%
        {Herder:DSN09}
{Jorrit~N. Herder}, {Herbert Bos}, {Ben Gras}, {Philip Homburg}, {and}
  {Andrew~S. Tanenbaum}. 2009.
\newblock \showarticletitle{Fault isolation for device drivers}. In {\em The
  39th Annual International Conference on Dependable Systems and Networks
  (DSN)}. IEEE/IFIP, Estoril, Portugal, 33--42.
\newblock


\bibitem[\protect\citeauthoryear{IEEE Std 982.2}{IEEE Std 982.2}{1988}]%
        {ieee_guide}
IEEE Std 982.2 1988.
\newblock {IEEE} guide for the use of {IEEE} standard dictionary of measures to
  produce reliable software.
\newblock   (1988).
\newblock


\bibitem[\protect\citeauthoryear{Israeli and Feitelson}{Israeli and
  Feitelson}{2010}]%
        {Israeli:JSS10}
{Ayelet Israeli} {and} {Dror~G. Feitelson}. 2010.
\newblock \showarticletitle{The {Linux} kernel as a case study in software
  evolution}.
\newblock {\em Journal of Systems and Software\/} {83}, 3 (2010), 485--501.
\newblock


\bibitem[\protect\citeauthoryear{Jiang, Adams, and Germ{\'a}n}{Jiang
  et~al\mbox{.}}{2013}]%
        {DBLP:conf/msr/JiangAG13}
{Yujuan Jiang}, {Bram Adams}, {and} {Daniel~M. Germ{\'a}n}. 2013.
\newblock \showarticletitle{Will my patch make it? and how fast?: case study on
  the {Linux} kernel}. In {\em Proceedings of the 10th Working Conference on
  Mining Software Repositories, MSR '13}. IEEE/ACM, San Francisco, CA, USA,
  101--110.
\newblock
\showISBNx{978-1-4673-2936-1}


\bibitem[\protect\citeauthoryear{Kaplan and Meier}{Kaplan and Meier}{1958}]%
        {kaplan1958nonparametric}
{Edward~L Kaplan} {and} {Paul Meier}. 1958.
\newblock \showarticletitle{Nonparametric estimation from incomplete
  observations}.
\newblock {\em Journal of the American statistical association\/} {53}, 282
  (1958), 457--481.
\newblock


\bibitem[\protect\citeauthoryear{Keller, Murray, Amani, O'Connor, Chen, Ryzhyk,
  Klein, and Heiser}{Keller et~al\mbox{.}}{2013}]%
        {Keller:PLOS13}
{Gabriele Keller}, {Toby Murray}, {Sidney Amani}, {Liam O'Connor}, {Zilin
  Chen}, {Leonid Ryzhyk}, {Gerwin Klein}, {and} {Gernot Heiser}. 2013.
\newblock \showarticletitle{File systems deserve verification too!}. In {\em
  7th Workshop on Programming Languages and Operating Systems (PLOS 2013)}.
  Nemacolin Woodlands Resort, PA, USA, 7.
\newblock


\bibitem[\protect\citeauthoryear{Lawall, Brunel, Palix, Hansen, Stuart, and
  Muller}{Lawall et~al\mbox{.}}{2013}]%
        {Lawall:spe13}
{Julia~L. Lawall}, {Julien Brunel}, {Nicolas Palix}, {Ren\'e~Rydhof Hansen},
  {Henrik Stuart}, {and} {Gilles Muller}. 2013.
\newblock \showarticletitle{{WYSIWIB}: exploiting fine-grained program
  structure in a scriptable {API}-usage protocol-finding process}.
\newblock {\em Software: Practice and Experience\/} {43}, 1 (Jan. 2013),
  67--92.
\newblock


\bibitem[\protect\citeauthoryear{Li and Zhou}{Li and Zhou}{2005}]%
        {Li:fse05}
{Zhenmin Li} {and} {Yuanyuan Zhou}. 2005.
\newblock \showarticletitle{{PR-Miner}: automatically extracting implicit
  programming rules and detecting violations in large software code}. In {\em
  Proceedings of the 10th European Software Engineering Conference held jointly
  with 13th ACM SIGSOFT International Symposium on Foundations of Software
  Engineering}. ACM, Lisbon, Portugal, 306--315.
\newblock


\bibitem[\protect\citeauthoryear{LKML}{LKML}{2013}]%
        {lkml}
LKML 2013.
\newblock Lkml: The {Linux} kernel mailing list.
\newblock \url{http://lkml.org/}.   (2013).
\newblock


\bibitem[\protect\citeauthoryear{Lu, Arpaci-Dusseau, Arpaci-Dusseau, and Lu}{Lu
  et~al\mbox{.}}{2013}]%
        {LuEtAl13-fsstudy-login}
{Lanyue Lu}, {Andrea~C. Arpaci-Dusseau}, {Remzi~H. Arpaci-Dusseau}, {and} {Shan
  Lu}. 2013.
\newblock \showarticletitle{A study of {Linux} file system evolution}.
\newblock {\em ;login: The USENIX Magazine\/} {38}, 3 (June 2013), 10--17.
\newblock


\bibitem[\protect\citeauthoryear{Lu, Park, Seo, and Zhou}{Lu
  et~al\mbox{.}}{2008}]%
        {Lu:ASPLOS08}
{Shan Lu}, {Soyeon Park}, {Eunsoo Seo}, {and} {Yuanyuan Zhou}. 2008.
\newblock \showarticletitle{Learning from mistakes: a comprehensive study on
  real world concurrency bug characteristics}. In {\em Architectural Support
  for Programming Languages and Operating Systems (ASPLOS)}. ACM, Seattle, WA,
  USA, 329--339.
\newblock


\bibitem[\protect\citeauthoryear{McCabe}{McCabe}{1976}]%
        {McCabe:76}
{T.~J. McCabe}. 1976.
\newblock \showarticletitle{A complexity measure}.
\newblock {\em IEEE Transactions on Software Engineering\/} {2}, 4 (July 1976),
  308--320.
\newblock


\bibitem[\protect\citeauthoryear{McKenney and Walpole}{McKenney and
  Walpole}{2008}]%
        {McKenney:osr08}
{Paul~E. McKenney} {and} {Jonathan Walpole}. 2008.
\newblock \showarticletitle{Introducing technology into the {Linux} kernel: a
  case study}.
\newblock {\em ACM SIGOPS Operating Systems Review\/} {42}, 5 (2008), 4--17.
\newblock


\bibitem[\protect\citeauthoryear{Munson and Elbaum}{Munson and Elbaum}{1998}]%
        {Munson:ICSM98}
{John~C. Munson} {and} {Sebastian~G. Elbaum}. 1998.
\newblock \showarticletitle{Code churn: A measure for estimating the impact of
  code change}. In {\em International Conference Software Maintenance (ICSM)}.
  IEEE, Bethesda, Maryland, USA, 24--31.
\newblock


\bibitem[\protect\citeauthoryear{Nadi, Dietrich, Tartler, Holt, and
  Lohmann}{Nadi et~al\mbox{.}}{2013}]%
        {DBLP:conf/msr/NadiDTHL13}
{Sarah Nadi}, {Christian Dietrich}, {Reinhard Tartler}, {Richard~C. Holt},
  {and} {Daniel Lohmann}. 2013.
\newblock \showarticletitle{Linux variability anomalies: what causes them and
  how do they get fixed?}. In {\em Proceedings of the 10th Working Conference
  on Mining Software Repositories, MSR '13}. IEEE/ACM, San Francisco, CA, USA,
  111--120.
\newblock
\showISBNx{978-1-4673-2936-1}


\bibitem[\protect\citeauthoryear{Nagappan and Ball}{Nagappan and Ball}{2005}]%
        {Nagappan:ICSE05}
{Nachiappan Nagappan} {and} {Thomas Ball}. 2005.
\newblock \showarticletitle{Use of relative code churn measures to predict
  system defect density}. In {\em 27th International Conference on Software
  Engineering (ICSE)}. ACM, St. Louis, Missouri, USA, 284--292.
\newblock


\bibitem[\protect\citeauthoryear{Oman and Hagemeister}{Oman and
  Hagemeister}{1994}]%
        {Oman:JSS94}
{Paul Oman} {and} {Jack Hagemeister}. 1994.
\newblock \showarticletitle{Construction and testing of polynomials predicting
  software maintainability}.
\newblock {\em Journal of Systems and Software\/} {24}, 3 (1994), 251--266.
\newblock


\bibitem[\protect\citeauthoryear{Org}{Org}{2013}]%
        {org-mode}
Org 2013.
\newblock Org-mode homepage.
\newblock \url{http://orgmode.org/}.   (2013).
\newblock


\bibitem[\protect\citeauthoryear{Padioleau, Lawall, Hansen, and
  Muller}{Padioleau et~al\mbox{.}}{2008}]%
        {Padioleau:eurosys08}
{Yoann Padioleau}, {Julia Lawall}, {Ren\'e~Rydhof Hansen}, {and} {Gilles
  Muller}. 2008.
\newblock \showarticletitle{Documenting and automating collateral evolutions in
  {Linux} device drivers}. In {\em EuroSys 2008}. ACM, Glasgow, Scotland,
  247--260.
\newblock


\bibitem[\protect\citeauthoryear{Palix, Lawall, and Muller}{Palix
  et~al\mbox{.}}{2010a}]%
        {palix:aosd10}
{Nicolas Palix}, {Julia Lawall}, {and} {Gilles Muller}. 2010a.
\newblock \showarticletitle{Tracking code patterns over multiple software
  versions with {Herodotos}}. In {\em Proc. of the ACM International Conference
  on Aspect-Oriented Software Development, AOSD'10}. ACM, Rennes and Saint
  Malo, France, 169--180.
\newblock
\showISBNx{978-1-60558-958-9}
\showDOI{%
\url{http://dx.doi.org/10.1145/1739230.1739250}}


\bibitem[\protect\citeauthoryear{{P}alix, {S}aha, {T}homas, {C}alv{\`e}s,
  {L}awall, and {M}uller}{{P}alix et~al\mbox{.}}{2010b}]%
        {10years_web}
{{N}icolas {P}alix}, {{S}uman {S}aha}, {{G}a{\"e}l {T}homas}, {{C}hristophe
  {C}alv{\`e}s}, {{J}ulia {L}awall}, {and} {{G}illes {M}uller}. 2010b.
\newblock Website of {F}aults in {L}inux: Ten years later.
\newblock   (Dec. 2010).
\newblock
\newblock
\shownote{\fname{http://faultlinux.lip6.fr/}.}


\bibitem[\protect\citeauthoryear{Palix, Saha, Thomas, Calv{\`e}s, Lawall, and
  Muller}{Palix et~al\mbox{.}}{2010c}]%
        {palix:inria-00509256}
{Nicolas Palix}, {Suman Saha}, {Ga{\"e}l Thomas}, {Christophe Calv{\`e}s}, {L.
  Lawall, Julia}, {and} {Gilles Muller}. 2010c.
\newblock {\em {Faults in Linux: Ten Years Later}}.
\newblock Research Report RR-7357. INRIA. 21 pages.
\newblock
\showURL{%
\url{http://hal.inria.fr/inria-00509256}}


\bibitem[\protect\citeauthoryear{Palix, Thomas, Saha, Calv\`es, Lawall, and
  Muller}{Palix et~al\mbox{.}}{2011}]%
        {Palix:asplos11}
{Nicolas Palix}, {Ga{\"e}l Thomas}, {Suman Saha}, {Christophe Calv\`es}, {Julia
  Lawall}, {and} {Gilles Muller}. 2011.
\newblock \showarticletitle{Faults in {Linux}: Ten years later}. In {\em
  Sixteenth International Conference on Architectural Support for Programming
  Languages and Operating Systems (ASPLOS 2011)}. ACM, Newport Beach, CA, USA,
  305--318.
\newblock


\bibitem[\protect\citeauthoryear{{R Core Team}}{{R Core Team}}{2014a}]%
        {R-project}
{{R Core Team}}. 2014a.
\newblock The {R} project for statistical computing.
\newblock   (April 2014).
\newblock
\newblock
\shownote{\fname{http://www.r-project.org/}.}


\bibitem[\protect\citeauthoryear{{R Core Team}}{{R Core Team}}{2014b}]%
        {R-summary-lm}
{{R Core Team}}. 2014b.
\newblock Summarizing linear model fits.
\newblock   (April 2014).
\newblock
\newblock
\shownote{\fname{http://stat.ethz.ch/R-manual/R-patched/library/stats/html/summary.lm.html}.}


\bibitem[\protect\citeauthoryear{Raymond}{Raymond}{2001}]%
        {cathedral-bazaar}
{Eric~S. Raymond}. 2001.
\newblock {\em The Cathedral and the Bazaar: Musings on Linux and Open Source
  by an Accidental Revolutionary}.
\newblock O'Reilly \& Associates, Inc.
\newblock
\showISBNx{0596001088}


\bibitem[\protect\citeauthoryear{Rubio-Gonz{\'a}lez, Gunawi, Liblit,
  Arpaci-Dusseau, and Arpaci-Dusseau}{Rubio-Gonz{\'a}lez et~al\mbox{.}}{2009}]%
        {Rubio-Gonzalez:PLDI09}
{Cindy Rubio-Gonz{\'a}lez}, {Haryadi~S. Gunawi}, {Ben Liblit}, {Remzi~H.
  Arpaci-Dusseau}, {and} {Andrea~C. Arpaci-Dusseau}. 2009.
\newblock \showarticletitle{Error propagation analysis for file systems}. In
  {\em PLDI}. ACM, Dublin, Ireland, 270--280.
\newblock


\bibitem[\protect\citeauthoryear{Saha, Lozi, Thomas, Lawall, and Muller}{Saha
  et~al\mbox{.}}{2013}]%
        {saha2013hector}
{Suman Saha}, {Jean-Pierre Lozi}, {Ga{\"e}l Thomas}, {Julia~L Lawall}, {and}
  {Gilles Muller}. 2013.
\newblock \showarticletitle{Hector: Detecting resource-release omission faults
  in error-handling code for systems software}. In {\em Dependable Systems and
  Networks (DSN), 2013 43rd Annual International Conference on}. IEEE/IFIP,
  Budapest, Hungary, 1--12.
\newblock


\bibitem[\protect\citeauthoryear{Searls}{Searls}{2004}]%
        {sparse}
{D. Searls}. 2004.
\newblock Sparse, {Linus} \& the {Lunatics}.
\newblock   (Nov. 2004).
\newblock
\newblock
\shownote{Available at \fname{http://www.linuxjournal.com/article/7272}.}


\bibitem[\protect\citeauthoryear{Song, Chen, and Zang}{Song
  et~al\mbox{.}}{2010}]%
        {Song:DSN2010}
{Xiang Song}, {Haibo Chen}, {and} {Binyu Zang}. 2010.
\newblock \showarticletitle{Why software hangs and what can be done with it}.
  In {\em International Conference on Dependable Systems and Networks (DSN
  2010)}. IEEE/IFIP, Chicago, IL, USA, 311--316.
\newblock


\bibitem[\protect\citeauthoryear{Sparse}{Sparse}{2003}]%
        {Sparse-web}
Sparse 2003.
\newblock Sparse.
\newblock   (2003).
\newblock
\newblock
\shownote{\fname{https://sparse.wiki.kernel.org/}.}


\bibitem[\protect\citeauthoryear{Spencer}{Spencer}{2009}]%
        {spencer-exploit}
{Brad Spencer}. 2009.
\newblock Local kernel exploit in /dev/net/tun.
\newblock   (July 2009).
\newblock
\newblock
\shownote{\fname{http://grsecurity.net/~spender/exploits/cheddar_bay.tgz}.}


\bibitem[\protect\citeauthoryear{Swift, Annamalai, Bershad, and Levy}{Swift
  et~al\mbox{.}}{2006}]%
        {Swift:TOCS06}
{Michael~M. Swift}, {Muthukaruppan Annamalai}, {Brian~N. Bershad}, {and}
  {Henry~M. Levy}. 2006.
\newblock \showarticletitle{Recovering device drivers}.
\newblock {\em ACM Transactions on Computer Systems\/} {24}, 4 (2006),
  333--360.
\newblock


\bibitem[\protect\citeauthoryear{Tartler, Sincero, Dietrich,
  Schr{\"o}der-Preikschat, and Lohmann}{Tartler et~al\mbox{.}}{2012}]%
        {tartler2012revealing}
{Reinhard Tartler}, {Julio Sincero}, {Christian Dietrich}, {Wolfgang
  Schr{\"o}der-Preikschat}, {and} {Daniel Lohmann}. 2012.
\newblock \showarticletitle{Revealing and repairing configuration
  inconsistencies in large-scale system software}.
\newblock {\em International Journal on Software Tools for Technology
  Transfer\/} {14}, 5 (2012), 531--551.
\newblock


\bibitem[\protect\citeauthoryear{{The Kernel Janitors}}{{The Kernel
  Janitors}}{2010}]%
        {smatch}
{{The Kernel Janitors}}. 2010.
\newblock Smatch, the source matcher.
\newblock   (2010).
\newblock
\newblock
\shownote{Available at \url{http://smatch.sourceforge.net}.}


\bibitem[\protect\citeauthoryear{Ubuntu}{Ubuntu}{2013}]%
        {Ubuntu}
Ubuntu 2013.
\newblock Ubuntu.
\newblock   (2013).
\newblock
\newblock
\shownote{\fname{http://www.ubuntu.com/}.}


\bibitem[\protect\citeauthoryear{Wang, Zeldovich, Kaashoek, and
  Solar-Lezama}{Wang et~al\mbox{.}}{2013}]%
        {Wang:2013:TOS:2517349.2522728}
{Xi Wang}, {Nickolai Zeldovich}, {M.~Frans Kaashoek}, {and} {Armando
  Solar-Lezama}. 2013.
\newblock \showarticletitle{Towards optimization-safe systems: analyzing the
  impact of undefined behavior}. In {\em Proceedings of the Twenty-Fourth ACM
  Symposium on Operating Systems Principles} {\em (SOSP '13)}. ACM, Farminton,
  Pennsylvania, 260--275.
\newblock
\showISBNx{978-1-4503-2388-8}


\bibitem[\protect\citeauthoryear{Wheeler}{Wheeler}{2006}]%
        {wheeler06:flawfinder}
{David Wheeler}. 2006.
\newblock Flawfinder home page.
\newblock Web page: \url{http://www.dwheeler.com/flawfinder/}.   (Oct. 2006).
\newblock
\showURL{%
\url{http://www.dwheeler.com/flawfinder/}}


\bibitem[\protect\citeauthoryear{Wheeler}{Wheeler}{2013}]%
        {sloccount}
{David~A. Wheeler}. 2013.
\newblock {SLOCC}ount.
\newblock   (2013).
\newblock
\newblock
\shownote{\fname{http://www.dwheeler.com/sloccount/}.}


\bibitem[\protect\citeauthoryear{Yang, Twohey, Engler, and Musuvathi}{Yang
  et~al\mbox{.}}{2006}]%
        {Yang:2006}
{Junfeng Yang}, {Paul Twohey}, {Dawson Engler}, {and} {Madanlal Musuvathi}.
  2006.
\newblock \showarticletitle{Using model checking to find serious file system
  errors}.
\newblock {\em ACM Trans. Comput. Syst.\/} {24}, 4 (Nov. 2006), 393--423.
\newblock
\showISSN{0734-2071}
\showDOI{%
\url{http://dx.doi.org/10.1145/1189256.1189259}}


\end{thebibliography}

% % History dates
% \received{February 2007}{March 2009}{June 2009}

% Appendix
\appendix

\section{Appendix}

\subsection{Lock and interrupt functions}
%\label{sec:fct-lock-intr}

\paragraph{Locking functions:}

\{mutex,spin,read,write\}\_lock, \{mutex,spin,read,write\}\_trylock

\paragraph{Interrupt disabling functions:}

 cli, local\_irq\_disable

\paragraph{Functions combining both:}

\{read,write,spin\}\_lock\_irq, \{read,write,spin\}\_lock\_irqsave,
local\_irq\_save, save\_and\_cli

\end{document}